\documentclass{article}[10pt]
\usepackage{amsmath}
\usepackage{epsfig}
\usepackage{graphicx}
\usepackage{amsfonts}
\usepackage{amssymb}
\numberwithin{equation}{section} 
\newcommand{\be}{\begin{equation}}
\newcommand{\ee}{\end{equation}}
\newcommand{\bea}{\begin{eqnarray}}
\newcommand{\eea}{\end{eqnarray}}

\begin{document}

\title{{\bf  Non-Equilibrium Large
$N$ Yukawa Dynamics: marching through the Landau pole}}
\author{Daniel Boyanovsky$^{(a,b)}$\footnote{boyan@pitt.edu}, Hector J. de Vega$^{\left(  b,a\right)  }$\footnote{devega@lpthe.jussieu.fr},
\and Richard Holman$^{(c)}\footnote{holman@cmuhep2.phys.cmu.edu}$, Matthew R. Martin$^{(c)}$\footnote{mmartin@cmu.edu}\\
$^{(a)}$ Department of Physics and Astronomy, University of\\Pittsburgh, Pittsburgh PA 15260,
\\$^{\left(  b\right)  }$ Universite Pierre et Marie Curie, Paris VI, Jussieu,\\Paris CEDEX, France\\
$^{(c)}$ Physics Department, Carnegie Mellon University, \\Pittsbugh, PA, 15213}
\maketitle

\begin{abstract}
The non-equilibrium dynamics of a Yukawa theory with N fermions
coupled to a scalar field is studied in the large N limit with
the goal of comparing the dynamics predicted from the renormalization
group improved effective potential to that obtained including the
fermionic backreaction. The effective potential is of the
Coleman-Weinberg type. Its renormalization group improvement is
unbounded from below and features a Landau pole.
When viewed self-consistently, the initial time singularity
does not arise. The different regimes
of the dynamics of the fully renormalized theory are studied both
analytically and numerically. Despite the existence of a
Landau pole in the model, the dynamics of the mean field
is smooth as it passes
the location of the pole. This is a consequence of a remarkable
cancellation between the effective potential and the dynamical chiral
condensate.  The asymptotic evolution is effectively described by a
\emph{quartic upright} effective potential. In all regimes, profuse
particle production results in the formation of a dense fermionic
plasma with occupation
numbers nearly saturated up to a scale of the order of the mean
field. This can be interpreted as a chemical potential. We discuss the
implications of these results for cosmological preheating.

\end{abstract}

\section{Introduction}

In the Standard model or its extensions, the Yukawa couplings of
Fermions to scalars (Higgs) play a fundamental role. Not only do
such couplings determine the masses of the fermionic degrees of
freedom, but in turn, it is through these couplings that the
fermionic sector influences the dynamics of the scalar fields. If
these Yukawa couplings are large enough, they can lead to negative
contributions to the beta functions of the running scalar
self-couplings and so to destabilizing the vacuum by large {\em
negative} radiative corrections to the scalar effective
potential\cite{vacuumstab}. This is the Coleman-Weinberg mechanism
of symmetry breaking by radiative
corrections\cite{colemanweinberg}.

While such a scenario has been ruled out within the Standard model
due to an unacceptably low value of the Higgs and the top quark
masses, large negative radiative corrections to the effective
potential from large Yukawa couplings could still be relevant in
extensions of the Standard model with more complicated
Higgs-Yukawa sectors\cite{nonstanhiggs}. Coleman Weinberg phase
transitions in extended Higgs models and their potential
cosmological implications have been studied by
Sher\cite{nonstanhiggs} who analyzed the effective potential of
this theory. While this study extracted bounds on the parameters
of extended Higgs sectors from vacuum stability and thermodynamic
considerations, these results are based on an {\em equilibrium}
description based on the effective potential, a purely {\em
static} quantity.

Detailed studies reveal  that the information extracted from a
static effective potential is restricted to situations very close
to equilibrium and that a deeper understanding of dynamical
processes requires a non-equilibrium treatment. The necessity for
a non-equilibrium description of quantum field theory has become
clear in cosmology where inflationary phase transitions require a
fully dynamical description\cite{noneqcosm}, in  heavy-ion
collisions where a transient quark-gluon plasma may be
formed\cite{lanllargeN,rhicstuff} and domain formation in phase
transitions\cite{domains1,domains2} which may have consequences in
cosmology as well as in heavy ion collisions.

In each of these fields,  non-equilibrium effects can give rise to
new phenomena which can differ from equilibrium behavior in
dramatic ways that cannot be captured by an effective potential
description.

From the point of view of cosmology and inflationary cosmology in
particular, non-equilibrium effects associated with particle
production via resonances and or instabilities in \emph{bosonic}
field theories have taken center stage. This is evident in the
theory of preheating\cite{preheating} as well as in the
classicalization of fluctuations during inflation\cite{ZMR}, where
spinodally unstable dynamical fluctuations about an homogeneous
condensate modify the long wavelength behavior of the theory.

While the non-equilibrium dynamics of fermionic fields has been
studied recently by several authors
\cite{baackefermions,ferminos,linde, Nilles} and while there has
been some remarkable progress in lattice simulations of fermionic
dynamics in low dimensional gauge theories\cite{smitfermions}, the
dynamical behavior of fermions has not received the same level of
attention as the bosonic case. This is mostly due to the argument
that Fermi-Dirac statistics preclude parametric amplification of
occupation numbers with non-perturbative particle production.

In order to have access to non-perturbative dynamics, in this work
we consider theories containing $N$ fermions with Yukawa couplings
to one scalar field $\Phi$ in leading order in the large $N$
limit. The large $N$ limit leads to a consistent, non-perturbative
approximation scheme that can be systematically improved. This
approximation has been applied to {\em bosonic} theories  and has
been studied both analytically to leading
order\cite{lanllargeN,reviews} and more recently numerically in
real time including corrections beyond leading order in the large
$N$\cite{aartsberges,cooper}. These studies reveal a wealth of new
phenomena not accessible via perturbative methods.

In the large $N$ limit we study here, the fermions serve to
suppress scalar field fluctuations to leading order in $1/N$. We
find that even at leading order a new and important aspect of
field theory comes into play: the wave function renormalization.
It should be noted that this does \emph{not} occur at the same
order in purely scalar theories. Our study reveals that this new
ingredient is responsible for dramatic new dynamical phenomena.

We obtain the equations of motion of the mean field, or
expectation value of the scalar field, taking into account the
non-equilibrium back reaction of the fermionic modes to leading
order in $1/N$. This analysis goes beyond the effective potential
approximation\cite{colemanweinberg} and affords us a window to
understanding the non-perturbative dynamics of this coupled
system.

We can summarize our main results as follows:
\begin{itemize}
\item{ The fully renormalized theory displays the Coleman-Weinberg mechanism of dimensional
transmutation through radiative corrections. It has an effective
potential that is unbounded from below at large values of the mean
field with two symmetric global maxima and a local minimum at the
origin. It also exhibits a Landau pole at an energy scale $E_{L}$
which is non-perturbative in the Yukawa coupling. The presence of
the Landau pole distinguishes {\em two} distinct regimes to be
studied depending on the relationship between the cutoff
$\Lambda$, which is required for the numerical analysis of the
theory, and  the position $E_{L}$ of the Landau pole. }

\item{ Suppose that we take $\Lambda<E_{L}$. Then, if the initial value of the mean field
is between the origin and the global maxima, the mean field
oscillates about the origin, and the fermionic quantum
fluctuations grow as a result of particle production in a
\emph{preferred} band of wave vectors. This is akin to what
happens in the bosonic case but here the production must saturate
due to Pauli blocking. The width of the band of wave vectors is
determined by the initial amplitude of the mean field and the mass
of the scalar field. If, however, the initial value of the mean
field is larger than the maxima, its amplitude runs away to the
cutoff scale at which point the evolution must be stopped since
the theory reaches the edge of its domain of validity. }

\item{ Our most noteworthy results are those for which $\Lambda >>
E_{L}$ and the initial value of the mean field is taken to be
larger than the position of the maxima of the effective potential.
In this case, an analysis of the dynamics based {\em solely} on
the static effective potential would lead to the conclusion that
the time evolution of the mean field would lead to a divergence or
discontinuity in the time derivatives when the amplitude reaches
the value of the Landau pole.  However a detailed analysis of the
{\em full} dynamics, including the fermionic fluctuations and
their backreaction on to the mean field, reveals that the
evolution of the mean field is {\em smooth}. When the amplitude of
the mean field becomes larger than the Landau pole, the dynamics
becomes {\em oscillatory} and asymptotically reaches a fixed point
described by a simple {\em quartic, upright effective potential}
with a quartic coupling of order one and with the mean field
oscillating with a large amplitude around the origin. This novel
dynamical behavior arises from a remarkable cancellation between
the fermionic fluctuations and the contribution from the
instantaneous effective potential that leads to smooth dynamics
through the Landau pole. When the initial value of the mean field
is larger than the maxima of the effective potential but much
smaller than the Landau pole, the ensuing non-equilibrium
evolution leads to fermion production in a band of wave-vectors up
to the scale of the Landau pole. These modes become populated with
almost Pauli blocking saturation at large times and describe a
very dense medium. We study this behavior numerically and confirm
that this phenomenon occurs for a wide range of parameters. We are
led  to {\em conjecture} that the theory is, in fact, sensibly
behaved beyond the Landau pole when studied both {\em dynamically
and non-perturbatively}, at least at the mean-field level. While
at this point this is merely a conjecture, these phenomena may
have some interesting phenomenological consequences. }

\item{ A consistent analysis of the renormalization during the dynamical evolution
reveals that the wave function renormalization builds up in time
over a time scale of ${\cal O}(1/\Lambda)$. The fully renormalized
equations of motion display a renormalized coupling at a scale
determined by the amplitude of the mean field, and which therefore
depends parametrically on time. This is a consequence of the
``running'' of the coupling constant with scale, which in the
dynamical evolution translates to a ``running'' with time.  }

\end{itemize}

The article is organized as follows: in section 2 we obtain the
renormalization group improved effective potential, discuss its
features including the presence of the Landau pole and the
potential singularities that would occur in an analysis based
solely on the effective potential. In section 3 we obtain the
equations of motion to leading order in the large $N$ limit. In
section 4 we address the renormalization of the equations of
motion and the energy density. We discuss and resolve the issue of
potential initial time singularities; in particular, we highlight
the fact that the wave function renormalization {\em builds up} on
time scales determined by the cutoff. In this section we also
establish contact between the non-equilibrium equations of motion
and the renormalization group improved effective potential,
emphasizing the emergence of {\em smooth} dynamics as the
mean-field approaches and passes the Landau pole. In section 5 we
provide a detailed and comprehensive numerical study of the
dynamics in several cases in a wide range of parameters. In
section 6 we provide an {\em exact proof} of the lack of unstable
bands for fermionic mode functions in the background of a scalar
field that oscillates, by showing that the Floquet indices are
purely real. This is the underlying reason for Pauli blocking at
the level of mode functions.  We also offer a perturbative
analysis of this important phenomenon, which provides the reason
for the existence of a preferred band of wave vectors for the
produced fermions. We summarize our findings and offer some
conjectures for potential implications of our results in cosmology
as well as for the phenomenology of theories with extended
Higgs-Yukawa sectors containing heavy fermions (and hence large
Yukawa couplings) that could feature a Landau pole in an energy
range of phenomenological interest.

\section{Static aspects: the effective potential and its RG
improvement}\label{sec:veff}

Before studying the dynamical aspects of the Yukawa theory in the
large $N$ limit it is illuminating to understand the static
aspects via the effective potential and its renormalization group
improvement.

The Yukawa model under consideration is described by the Lagrangian
density

\bea
 \mathcal{L} &=&\frac{1}{2} \left(\partial_{\mu} \Phi_{B} \right)^2-
 \frac{1}{2}m_{B}^{2}\,\Phi_{B}^{2}-\,\frac{\lambda
_{B}}{4!\,N}\,\Phi_{B}^{4} \nonumber \\ && +\sum_{i=1}^N
\bar{\psi}_{i}\left[i
\gamma^{\mu}\partial_{\mu}-\frac{y_{B}}{\sqrt{N}}\,\Phi_{B}
\right]\,\psi_{i}, \label{lagrangian}
 \eea

\noindent where the subscript $B$ denotes the bare fields,
anticipating the need for renormalization. The factors of $N$ in
the coupling constants are explicitly displayed so that both the
quartic self coupling and the Yukawa coupling are of ${\cal O}(1)$
in the large $N$ limit.

The effective potential is defined as the expectation value of the
Hamiltonian in a state that minimizes the energy subject to the
constraint that the field has a space-time independent expectation
value\cite{coleman}

\be
V_{eff}(\bar\Phi)= \mbox{min}\frac{\langle \bar\Phi|H| \bar\Phi\rangle}{\Omega}
\ee

\noindent with $\Omega$ the spatial volume and  $|\bar
\Phi\rangle$ a (coherent) state for which

\be \langle \bar \Phi|\Phi| \bar \Phi \rangle = \bar \Phi. \ee

To leading order in the large $N$ the effective potential is
obtained by replacing $\Phi(\vec x) \rightarrow \bar \Phi$ in the
Hamiltonian and neglecting the scalar field fluctuations, since
the energy will be dominated by the $N$ fermion fields. This is
the mean field approximation, which becomes exact in the large $N$
limit.

Hence

\bea && H[\bar \Phi] = \Omega\left(\frac{1}{2}m_{B}^{2}\,{\bar
\Phi}^{2}+\,\frac{\lambda _{B}}{4!\,N}\,{\bar\Phi}^{4}\right)
+\sum_{i=1}^N \bar{\psi}_{i}\left[-i\vec{\alpha}\cdot
\vec{\triangledown}+M \right]\,\psi_{i}, \label{meanfieldham} \\
&&M = \frac{y_B}{\sqrt{N}} \bar \Phi. \eea

The fermionic contribution to the Hamiltonian is simply that of
$N$ Dirac fermions of mass $M$, and can be diagonalized in terms
of creation and annihilation operators for particles and
antiparticles with dispersion relation $\omega_k =
\sqrt{k^2+M^2}$. The state that minimizes the expectation value of
the normal ordered Hamiltonian is the  vacuum state for particles
and antiparticles and corresponds to the Dirac sea completely
filled (with two spin states per wave vector). It is convenient to
rescale the expectation value of the scalar field to exhibit  the
large $N$ limit more clearly:

\be \bar \Phi = \sqrt{N}~ \delta_B \label{rescaleN} \ee

\noindent  in terms of which

\be V_{eff}(\delta)= N\left\{\frac{m_B^2}{2} \delta^2_B +
\frac{\lambda_B}{4!}\delta^4_B - 2 \int \frac{d^3k}{(2\pi)^3}
\sqrt{k^2+M^2} \right\}. \label{veffmeanfield} \ee

The integral in (\ref{veffmeanfield}) is the (negative)
contribution from the Dirac sea, and is calculated with an
ultraviolet cutoff $\Lambda$. A straightforward calculation,
subtracting the ``zero point energy'' $\propto \Lambda^4$ and
neglecting terms that vanish in the limit $\Lambda \rightarrow
\infty$, leads to

\bea \frac{V_{eff}(\delta)}{N} = &&\frac{\delta^2_B}{2}\left( m^2_B -
\frac{y^2_B \Lambda^2}{4\pi^2}
\right)+\frac{\delta^4_B}{4}\left(\frac{\lambda_B}{3!} +
\frac{y^4_B}{2\pi^2}\ln\left[\frac{2\Lambda}{\mu} \right]\right)  \nonumber\\
&& -\frac{y^4_B~
\delta^4_B}{8\pi^2}\ln\left[\left|\frac{y_B~\delta_B}{\mu}
\right|\right] -\frac{y^4_B \delta^4_B}{32\pi^2}
,\label{veffcutoff} \eea

\noindent where $\mu$ is an arbitrary renormalization scale. The
renormalization of the mass and quartic scalar coupling $\lambda$
can be gleaned directly from the form of the effective potential
above. However, to make contact with the dynamics in which the
equations of motion are obtained from the non-equilibrium
effective action, we also need the field or wave function
renormalization. The wave function renormalization cannot be
extracted from the effective potential since it is associated with
gradient terms in the effective action. To leading order in $1/N$
it can be obtained from the one-fermion loop self energy shown in
figure (\ref{fig:2fermionloop}) below, where the scalar field
lines have non-vanishing external momentum.

\begin{figure}[ht!]
\begin{center}
\includegraphics[height=3.0in,width=3.0in,keepaspectratio=true]{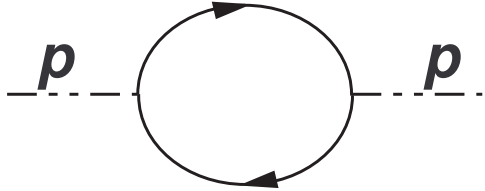}
\caption{Scalar self energy in leading order in large $N$.
External dashed lines correspond to the scalar field, internal
solid lines are fermions. } \label{fig:2fermionloop}
\end{center}
\end{figure}

This calculation leads to

\be Z[\mu] = \left(1 +
\frac{y^2_B}{4\pi^2}\ln\left[\frac{2\Lambda}{\mu}
\right]\right)^{-1}. \label{wavefun}\ee

The renormalization of field, mass and couplings is now achieved
by introducing

\bea \delta_R & = &
\frac{\delta_B}{\sqrt{Z}}, \label{fieldren} \\
m^2_R & = & Z\left(m^2_B-\frac{y^2_B
\Lambda^2}{2\pi^2}\right),\label{massren}\\
\frac{\lambda_R}{3!} & = & Z^2\left(\frac{\lambda_B}{3!} +
\frac{y^4_B}{2\pi^2}\ln\left[\frac{2\Lambda}{\mu} \right]
\right),\label{scalarcoupren} \\
y_R & = & y_B\sqrt{Z}, \label{yukaren}
 \eea
in terms of which the renormalized effective potential is given by

 \be
V_{eff}(\delta_R) = \frac{1}{2}m^2_R~\delta^2_R +
\frac{\delta^4_R}{4}\left(\frac{\lambda_R}{3!} -\frac{y^4_R
}{2\pi^2}\ln\left|\frac{y_R~\delta_R e^{\frac{1}{4}}}{\mu
 } \right| \right).\label{veffren}\ee

We will consider the case in which the renormalized mass of the
scalar field vanishes, since this case will highlight the
important feature of dimensional transmutation both at the level
of the static effective potential as well as the dynamics.
Therefore in what follows we set $m_R=0$.

The equation of motion, which will be the focus of next section,
requires $V'(\delta_R)=\partial V(\delta_R)/\partial\delta_R$.
This is given by

\be \label{vprime} V'(\delta_R)= \frac{y^4_R}{2\pi^2}\delta^3_R
\left[\frac{2\pi^2 \lambda_R}{3! y^4_R}-\ln\left[\frac{|M_R|}{\mu}
\right]-\frac{1}{2} \right],
 \ee

\noindent where we introduced the renormalized effective fermion
mass

\be \label{fermimass} M_R = y_R\delta_R . \ee

\noindent We see that the effective potential features an extremum
at $M_R = {\bar M}$ with ${\bar M}$ determined by \be
\label{dimtrans} \left[\frac{2\pi^2 \lambda_R}{3!
y^4_R}-\ln\left[\frac{|{\bar M}|}{\mu} \right]-\frac{1}{2}
\right]=0 .\ee

In terms of the scale ${\bar M}$ we find

\bea  V'_{eff}(\delta_R) &=& - \frac{y^4_R \delta^3_R}{2\pi^2}
\ln\left[\frac{|M_R|}{{\bar M}}
\right] ,\label{finvprime}\\
V_{eff}(\delta_R)&  = & -\frac{y^4_R \delta^4_R}{8\pi^2}
\ln\left[\frac{|M_R|e^{-\frac{1}{4}}}{{\bar M}}
\right].\label{veffdimtrans}\eea

Thus we see that the term $\lambda_R/3!$ inside the parenthesis in
(\ref{veffren}) above has been traded for a new scale ${\bar M}$ at
which the effective potential features a maximum. This is the
manifestation of dimensional transmutation\cite{colemanweinberg}.

Figure (\ref{fig:1veff}) below displays $V_{eff}/{\bar
M}^4$ vs. $\chi=\frac{M_R}{{\bar M} }$.
The $V_{eff}$ is given by (\ref{veffdimtrans}).

\begin{figure}[ht!]
\begin{center}
\includegraphics[height=3.0in,width=3.0in]{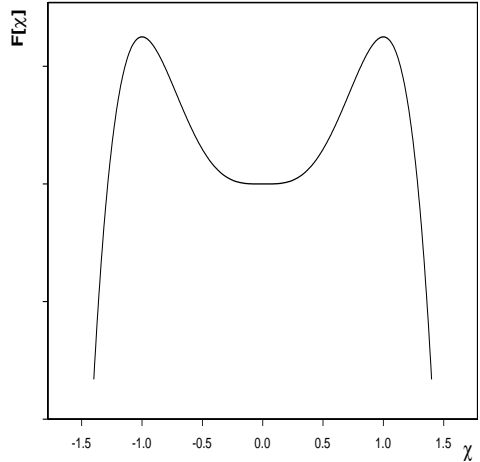}
\caption{$F(\chi)=-\chi^4 \ln(|\chi|e^{-\frac{1}{4}})$ vs.
$\chi=\frac{M_R}{\bar M}$} \label{fig:1veff}
\end{center}
\end{figure}

While there are alternative calculations of the effective
potential, the Hamiltonian formulation highlights many important
aspects that will be relevant to the discussion in the next
sections. In particular, it makes clear that the effective
potential is the expectation value of the Hamiltonian in a
``vacuum'' state in which the scalar field attains an expectation
value. Furthermore, this Hamiltonian interpretation immediately
provides the physical reason for the effective potential being
unbounded below in this approximation: it is completely determined
by the negative energy Dirac sea. For larger amplitudes of the
expectation value, the effective fermionic mass $M_R$ is larger,
thus the negative energy of a free fermion mode of momentum $k$ in
the Dirac sea decreases further. Thus for larger amplitudes of the
expectation value, the energy stored in the Dirac sea becomes more
negative. This particular point should be borne in mind when we
study the {\em dynamics} of the  mean field  below, since we will
find that the evolution of the scalar field ``feeds off'' the
negative energy Dirac sea.

\subsection{RG improvement:}\label{RGimpVeff}
The emergence of the  scale ${\bar M}$ is a consequence of the
renormalization scale $\mu$ introduced above. A change in this
scale is compensated for by a change in the couplings. The
effective action  \be \Gamma[\delta_R] = \int d^4x
\left[-V_{eff}[\delta_R]+{\cal Z}[\delta_R]\frac{1}{2}
(\partial_{\mu}\delta_R)^2+\cdots\right] \label{effaction}\ee

\noindent is invariant under a change of the renormalization
scale, and consequently under a change in the scale ${\bar M}$.
This invariance leads to a renormalization group equation for the
effective action\cite{colemanweinberg}.

Since our main focus is to study dynamical behavior, which
involves $V'_{eff}$, we now use the renormalization group to
improve the {\em derivative of the effective potential}.

While in principle we can study the full solution of the
renormalization group equation as in\cite{colemanweinberg}, the
large $N$ approximation simplifies the task. In this limit we need
only keep the one loop fermion contribution to renormalization.
Therefore, after trading the quartic self-coupling for the scale
${\bar M}$ via dimensional transmutation, the effective potential
(and its derivatives) are only functions of the Yukawa coupling.
Furthermore, from the renormalization conditions
(\ref{fieldren},\ref{yukaren}) the product $y_R\delta_R$ is {\em a
renormalization group invariant}, i.e, is constant under a change
of scales. With the purpose of comparing with the dynamics to be
studied in the next section, it is convenient to introduce the
coupling

\be g_R(\mu) = \frac{y^2_R(\mu)}{2\pi^2} \label{couplingg}, \ee

\noindent and to RG improve the product

\be y_R(\mu) V'_{eff}(\delta_R(\mu))= -g_R(\mu)M^3_R
\ln|\frac{M_R}{\bar M}| \label{prod}. \ee

The reason for studying this product is based on the idea that the
effective equation of motion of the scalar field is loosely of the
form

\be \ddot{\delta}_R(t)+V'_{eff}(\delta_R(t))=0. \ee

While $\delta$ is not invariant under a change of scale (i.e,
under a renormalization group transformation) in the large $N$
limit the product $M_R=y_R \delta_R$  \emph{is} a renormalization
group invariant. Thus one is led to consider the product $y_R
V'_{eff}$.

However, from the renormalization condition (\ref{yukaren}) of the
Yukawa coupling and the wave-function renormalization constant in
the large $N$ limit given by (\ref{wavefun}) we find

\be \label{gcoupren}
\frac{2}{g_R(\mu)}=\frac{2}{g_B}+\ln\left[\frac{2\Lambda}{\mu}
\right] ,\ee

\noindent which leads to the renormalization group running of this
coupling

\be\label{rgrun}
\frac{2}{g_R(\mu')}=\frac{2}{g_R(\mu)}+\ln\left[\frac{\mu}{\mu'}
\right]. \ee

Therefore choosing the new scale $\mu' \equiv M_R$ with the scale
of dimensional transmutation ${\bar M}$ fixed, the renormalization
group improvement of the product (\ref{prod}) leads to

\be \label{rgimproved} y_R(|M_R|) V'_{eff}(M_R)= -g_R(|M_R|)M^3_R
\ln\left|\frac{M_R}{\bar M}\right|= -\frac{g~ {\bar M}^3 \chi^3
~\ln|\chi|}{(1-\frac{g}{2}\ln|\chi|)}, \ee

with $g= g_R({\bar M})~,~ \chi= M_R/{\bar M}$.

The  expression (\ref{rgimproved}) features a {\em Landau pole} at
\be \label{LP}
M_R \sim {\bar M} e^{\frac{2}{g}} \ee

\noindent which, if interpreted in terms of the equation of motion
of the scalar field via the effective action (\ref{effaction})
would signal infinite time derivatives when the value of the
scalar field reaches the putative Landau pole. Since the large $N$
limit does not restrict the coupling to be weak, the value of $g$
can be ${\cal O}(1)$. Therefore if the dynamical evolution of the
expectation value of the scalar field is solely determined by
$V'_{eff}$ we would expect large derivatives and non-analytic
behavior of the dynamics as the scalar field approaches the
position of the Landau pole.

An important result of this work is that the Landau pole is {\em
not relevant} for the dynamical evolution of the scalar field and
contributions from particle production that {\em cannot be
captured by the effective potential} become very important. These
non-equilibrium contributions lead to smooth dynamics as the
expectation value of the scalar field nears the Landau pole.

\section{Large N Yukawa Dynamics}
Having studied the static aspects of the Yukawa theory in the
large $N$ limit via the effective potential and its
renormalization group improvement, we now focus our attention on
the dynamical aspects of this model.

As discussed in the introduction, we consider a system of $N$
fermions $\left\{  \psi_{i}\right\}  _{i=1}^{N}$ coupled to a
scalar field $\Phi$. The Lagrangian density in terms of bare
fields, mass and couplings is given by equation (\ref{lagrangian})
above. In order to study the dynamics as an initial value problem
we introduce an external ``magnetic field'' coupled to the scalar
field so that ${\cal L}\rightarrow {\cal L}+h(t)\Phi_B$ with
${\cal L}$ the Lagrangian density in (\ref{lagrangian}). The
external source $h(t)$ serves to generate a spatially homogeneous
expectation value for the scalar field

We will assume that the source was switched on adiabatically in
the infinite past and then slowly switched off at $t=0$. This
means that the scalar field evolves in the {\em absence} of this
external source for $t\geq 0$. This adiabatic switching on
procedure allows us to establish a connection with the effective
potential formalism of the previous section. If the initial state
as $t\rightarrow -\infty$ is the vacuum, an adiabatically switched
on source ensures that the state is the adiabatic vacuum; recall
that the (zero temperature) effective potential is referred to the
expectation value of the field in the {\em vacuum state}. We will
discuss this issue in greater detail below when we address the
renormalization aspects.

In order to extract the dynamics of the mean field, we expand the
scalar field as $\Phi_{B}\left(  \vec{x},t\right)
=\sqrt{N}\,\delta _{B}\left(  t\right)  +\chi\left(
\vec{x},t\right)  $ with $\left\langle \chi\left( \vec{x},t\right)
\right\rangle =0$. Implementing this last equation within the path
integral via the tadpole method\cite{ferminos}, we find that to
leading order in $1/N$, we arrive at the following equation of
motion for the mean field $\delta_{B}\left( t\right) $:
\begin{equation}
\ddot{\delta}_{B}\left(  t\right)  +m_{B}^{2}\,\delta_{B}\left(  t\right)
+\frac{\lambda_{B}}{3!}\delta_{B}^{3}\left(  t\right)  +\frac{y_{B}}{N}%
\sum_{i=1}^{N}\left\langle \bar{\psi}_{i}\,\psi_{i}\right\rangle =h(t).
\label{meanfieldeqn}%
\end{equation}

Note that this equation is non-perturbatively \emph{exact} in the
$N\rightarrow\infty$ limit. The Dirac equations for the $N$
species of  fermions $\psi_{i}$ are :
\begin{equation}
\left(  i\,\gamma^{\mu}\partial_{\mu}-M_{B}\left(  t\right)  \right)  \psi
_{i}=0\text{, where }M_{B}\left(  t\right)  =y_{B}\,\delta_{B}\left(
t\right)  \text{.} \label{DiracEqn}%
\end{equation}
The above equations are invariant under a permutation of the
fermion fields so that we need only deal with one of them, denoted
generically as $\psi$; therefore we make the replacement
\begin{equation}
\frac{y_{B}}{N}\sum_{i=1}^{N}\left\langle \bar{\psi}_{i}\,\psi_{i}%
\right\rangle \rightarrow y_{B}\left\langle \bar{\psi}\psi\right\rangle
\end{equation}
in eq.(\ref{meanfieldeqn}).

In order to proceed, we expand the spinor field operators in terms
of a complete set of mode functions solutions of the time
dependent Dirac equation into eq.(\ref{DiracEqn}):
\begin{equation}
\psi\left(  \vec{x},t\right)  =\sum_{\alpha=1}^{2}\int\frac{d^{3}p}{\left(
2\pi\right)  ^{3}}\left[  b_{\vec{p},\alpha}\,U_{\vec{p}}^{\left(
\alpha\right)  }\left(  t\right)  \,e^{i\,\vec{p}\cdot\vec{x}}+d_{\vec
{p},\alpha}^{\dagger}\,V_{\vec{p}}^{\left(  \alpha\right)  }\left(  t\right)
\,e^{-i\,\vec{p}\cdot\vec{x}}\right]  , \label{modedecomp}%
\end{equation}
where the creation/annihilation operators obey the usual
anticommutation relations
\begin{equation}
\left\{  b_{\vec{p},\alpha},b_{\vec{q},\beta}^{\dagger}\right\}  =\left\{
d_{\vec{p},\alpha},d_{\vec{q},\beta}^{\dagger}\right\}  =\left(  2\pi\right)
^{3}\,\delta_{\alpha\beta}\,\delta^{\left(  3\right)  }\left(  \vec{p}-\vec
{q}\right)  , \label{anticomrel}%
\end{equation}
and the Dirac spinors satisfy the completeness relation
\begin{equation}
\sum_{\alpha=1}^{2}\left[  U_{\vec{p}}^{\left(  \alpha\right)  }\left(
t\right)  _{a}\,U_{\vec{p}}^{\left(  \alpha\right)  \dagger}\left(  t\right)
_{b}+V_{-\vec{p}}^{\left(  \alpha\right)  }\left(  t\right)  _{a}\,V_{-\vec
{p}}^{\left(  \alpha\right)  \dagger}\left(  t\right)  _{b}\right]
=\delta_{ab}, \label{completeness}%
\end{equation}
with $a,b$ being Dirac space indices. Furthermore,
eq.(\ref{DiracEqn}) implies:%
\begin{equation}
\left(  i\,\gamma^{0}\partial_{0}-\vec{\gamma}\cdot\vec{p}-M_{B}\left(
t\right)  \right)  U_{\vec{p}}^{\left(  \alpha\right)  }\left(  t\right)
=0,\text{ }\left(  i\,\gamma^{0}\partial_{0}+\vec{\gamma}\cdot\vec{p}%
-M_{B}\left(  t\right)  \right)  V_{\vec{p}}^{\left(  \alpha\right)  }\left(
t\right)  =0. \label{modeequns}%
\end{equation}
Since the time derivative operator $\partial_{0}$ is singled out by our need
to consider the time evolution of the system, it is best to work in the basis
in which $\gamma^{0}$ is diagonal. Writing the four-component spinors
$U_{\vec{p}}^{\left(  \alpha\right)  }\left(  t\right)  $, $V_{\vec{p}%
}^{\left(  \alpha\right)  }\left(  t\right)  $ in terms of two-component
spinors $\chi_{i\vec{p}}^{\left(  \alpha\right)  },\xi_{i\vec{p}}^{\left(
\alpha\right)  }$ , $i=1,2$:
\begin{equation}
U_{\vec{p}}^{\left(  \alpha\right)  }\left(  t\right)  =\left(
\begin{array}
[c]{c}%
f_{1\,\vec{p}}\left(  t\right)  \,\chi_{1\vec{p}}^{\left(  \alpha\right)  }\\
f_{2\,\vec{p}}\left(  t\right)  \,\chi_{2\vec{p}}^{\left(  \alpha\right)  }%
\end{array}
\right)  ,\,V_{\vec{p}}^{\left(  \alpha\right)  }\left(  t\right)  =\left(
\begin{array}
[c]{c}%
g_{1\,\vec{p}}\left(  t\right)  \,\xi_{1\vec{p}}^{\left(  \alpha\right)  }\\
g_{2\,\vec{p}}\left(  t\right)  \,\xi_{2\vec{p}}^{\left(  \alpha\right)  }%
\end{array}
\right)  , \label{spinors}%
\end{equation}
we can use eq.(\ref{modeequns}) to find:
\begin{align}
\left(  i\,\partial_{0}-M_{B}\left(  t\right)  \right)  f_{1\,\vec{p}}\left(
t\right)  \,\chi_{1\vec{p}}^{\left(  \alpha\right)  }  &  =\left(  \vec
{\sigma}\cdot\vec{p}\right)  f_{2\,\vec{p}}\left(  t\right)  \,\chi_{2\vec{p}%
}^{\left(  \alpha\right)  }\label{fmodes}\\
\left(  i\,\partial_{0}+M_{B}\left(  t\right)  \right)  f_{2\,\vec{p}}\left(
t\right)  \,\chi_{2\vec{p}}^{\left(  \alpha\right)  }  &  =\left(  \vec
{\sigma}\cdot\vec{p}\right)  f_{1\,\vec{p}}\left(  t\right)  \,\chi_{1\vec{p}%
}^{\left(  \alpha\right)  },\nonumber
\end{align}%
\begin{align}
\left(  i\,\partial_{0}-M_{B}\left(  t\right)  \right)  g_{1\,\vec{p}}\left(
t\right)  \,\xi_{1\vec{p}}^{\left(  \alpha\right)  }  &  =-\left(  \vec
{\sigma}\cdot\vec{p}\right)  g_{2\,\vec{p}}\left(  t\right)  \,\,\xi_{2\vec
{p}}^{\left(  \alpha\right)  }\label{gmodes}\\
\left(  i\,\partial_{0}+M_{B}\left(  t\right)  \right)  g_{2\,\vec{p}}\left(
t\right)  \,\,\xi_{2\vec{p}}^{\left(  \alpha\right)  }  &  =-\left(
\vec{\sigma}\cdot\vec{p}\right)  g_{1\,\vec{p}}\left(  t\right)  \xi_{1\vec
{p}}^{\left(  \alpha\right)  }.\nonumber
\end{align}
We also impose the normalization conditions%
\begin{equation}
U_{\vec{p}}^{\left(  \alpha\right)  \dagger}\left(  t\right)  U_{\vec{p}%
}^{\left(  \beta\right)  }\left(  t\right)  =\delta^{\alpha\beta}=V_{\vec{p}%
}^{\left(  \alpha\right)  \dagger}\left(  t\right)  V_{\vec{p}}^{\left(
\beta\right)  }\left(  t\right)  , \label{normalizationconds}%
\end{equation}
which together with $\chi_{i\vec{p}}^{\left(  \alpha\right)  \dagger}%
\,\chi_{i\vec{p}}^{\left(  \alpha\right)  }=\delta^{\alpha\beta}$,
$i=1,2,$ imply that

\be\left|  f_{1\,\vec{p}}\left(  t\right)  \right| ^{2}+\left|
f_{2\,\vec{p}}\left(  t\right)  \right|  ^{2}=1\ee

\noindent which is a consequence of the conservation of
probability in the Dirac theory, which in turn is a consequence of
the fact that the Dirac equation is first order in time. This
constraint on the mode functions will have important consequences
in the dynamics and underlies the mechanism of Pauli blocking, as
will be analyzed in detail below.

We determine  the \emph{initial} state by demanding that
$U_{\vec{p}}^{\left( \alpha\right)
}\left(  t=0\right)  $ represent  positive energy states while $V_{\vec{p}%
}^{\left(  \alpha\right)  }\left(  t=0\right)  $ represent  negative
energy ones:
\begin{subequations}
\label{InState}%
\begin{gather}
i\,\partial_{0}U_{\vec{p}}^{\left(  \alpha\right)  }\left(  t=0\right)
=\omega_{p}\left(  t=0\right)  U_{\vec{p}}^{\left(  \alpha\right)
}\left(
t<0\right)  ,\,\label{initialstateU}\\
i\,\partial_{0}V_{\vec{p}}^{\left(  \alpha\right)  }\left(  t=0\right)
=-\omega_{p}\left(  t=0\right)  V_{\vec{p}}^{\left(  \alpha\right)
}\left(
t=0\right)  \,, \label{initialstateV}%
\end{gather}
where $\omega_{p}\left(  t\right)  =\sqrt{p^{2}+M_{B}\left( t
\right) ^{2}}$ are the mode frequencies. Considering an initial
state for which the time derivative of the scalar field vanishes
at $t=0$ (which can
always be achieved by a shift in the time variable) and then evaluating eqs.(\ref{fmodes}%
,\ref{gmodes}) at $t=0$ and using eqs.(\ref{InState}), we find:
\end{subequations}
\begin{align}
\left(  \omega_{0}-M_{0}\right)  \,f_{1\,\vec{p}}\left(  0\right)
\,\chi_{1\vec{p}}^{\left(  \alpha\right)  }  &  =\left(  \vec{\sigma}\cdot
\vec{p}\right)  \,f_{2\,\vec{p}}\left(  0\right)  \,\chi_{2\vec{p}}^{\left(
\alpha\right)  }\label{initialfmodes}\\
\left(  \omega_{0}+M_{0}\right)  \,f_{2\,\vec{p}}\left(  0\right)
\,\chi_{2\vec{p}}^{\left(  \alpha\right)  }  &  =\left(  \vec{\sigma}\cdot
\vec{p}\right)  \,f_{1\,\vec{p}}\left(  0\right)  \,\chi_{1\vec{p}}^{\left(
\alpha\right)  },
\end{align}
where $\omega_{0}=\omega_{p}\left(  0\right)  $, $M_{0}=M_{B}\left(  0\right)
$. This leads to the relations:
\begin{equation}
f_{1\,\vec{p}}\left(  0\right)  =\sqrt{\frac{\omega_{0}+M_{0}}{2\,\omega_{0}}%
},\,f_{2\,\vec{p}}\left(  0\right)  =\sqrt{\frac{\omega_{0}-M_{0}}%
{2\,\omega_{0}}},\,\chi_{2\vec{p}}^{\left(  \alpha\right)  }=\left(
\vec{\sigma}\cdot\hat{p}\right)  \chi_{1\vec{p}}^{\left(  \alpha\right)  }.
\label{modeIC's}%
\end{equation}
A similar analysis for the negative energy modes shows that in fact
$g_{1\,\vec{p}}\left(  t\right)  =f_{2\,\vec{p}}^{\ast}\left(  t\right)
,$ $g_{2\,\vec{p}}\left(  t\right)  =f_{1\,\vec{p}}^{\ast}\left(
t\right) ,\,\xi_{2\vec{p}}^{\left(  \alpha\right)  }=\left(
\vec{\sigma}\cdot\hat {p}\right)  \,\xi_{1\vec{p}}^{\left(
\alpha\right)  }$ so that we only need to solve for the positive energy
modes. Using the relation between the spinors in (\ref{modeIC's}),
equations (\ref{fmodes}) reduce to:
\begin{align}
\left( i\,\partial_{0}-M_{B}\left(  t\right)  \right)  f_{1\,\vec{p}}\left(
t\right)  &  = p f_{2\,\vec{p}} \label{modof}\\
\left(  i\,\partial_{0}+M_{B}\left(  t\right)  \right)  f_{2\,\vec{p}}\left(
t\right) &  = p f_{1\,\vec{p}} \; .\nonumber
\end{align}
These can be seperated into two second order equations:
\begin{equation}
\left(  \partial_{0}^{2}+p^2+M^2_B(t)\pm i\,\dot{M}%
_{B}\left(  t\right)  \right)  \,f_{1,2\ \vec{p}}\left(  t\right)  =0.
\label{fmodeeqns}%
\end{equation}
The second order equations have twice as many solutions as
the first order equations, but since the initial conditions for these
equations are determined from the first order equations, the correct,
physical, pair of solutions will be found.
The second order equations are of interest because they are more amenable
to the WKB expansion in the next section.

The quantity $\left\langle \bar{\psi}\psi\right\rangle $ appears in the mean
field equation eq.(\ref{meanfieldeqn}); we need to calculate it in terms of
the mode functions $\left\{  f_{1\,\vec{p}}\left(  t\right)  ,\,f_{2\,\vec{p}%
}\left(  t\right)  \right\}  $. This is easily done for our state when we note
that $\left\langle b_{\vec{q},\alpha}^{\dagger}\,b_{\vec{p},\beta
}\right\rangle =0$ $\,$while $\left\langle d_{\vec{p},\alpha}\,d_{\vec
{q},\alpha}^{\dagger}\,\right\rangle =\left\langle \left\{  d_{\vec{p},\alpha
},\,d_{\vec{q},\beta}^{\dagger}\right\}  \,\right\rangle =\left(  2\pi\right)
^{3}\,\delta_{\alpha\beta}\,\delta^{\left(  3\right)  }\left(  \vec{p}-\vec
{q}\right)  $ and use $g_{1\,\vec{p}}\left(  t\right)  =f_{2\,\vec{p}}^{\ast
}\left(  t\right)  ,$ $g_{2\,\vec{p}}\left(  t\right)  =f_{1\,\vec{p}}^{\ast
}\left(  t\right)  $:
\begin{equation}
\left\langle \bar{\psi}\psi\right\rangle =2\,\int\frac{d^{3}p}{\left(
2\pi\right)  ^{3}}\left(  \left|  \,f_{2\,\vec{p}}\left(  t\right)  \right|
^{2}-\left|  \,f_{1\,\vec{p}}\left(  t\right)  \right|  ^{2}\right)  .
\label{psibarpsi}%
\end{equation}

Another important quantity for us is the energy density $\rho_{\text{f}}$ in
the fermionic fluctuations:
\begin{equation}
\rho_{\text{f}}\equiv\left\langle T_{0}^{0}\right\rangle =\frac{i}%
{2}\left\langle \psi^{\dagger}\,\dot{\psi}-\dot{\psi}^{\dagger}\,\psi
\,\right\rangle =2\,\int\frac{d^{3}p}{\left(  2\pi\right)  ^{3}}\Im m\left(
f_{1\,\vec{p}}^{\ast}\left(  t\right)  \,\dot{f}_{1\,\vec{p}}\left(  t\right)
+f_{2\,\vec{p}}^{\ast}\left(  t\right)  \,\dot{f}_{2\,\vec{p}}\left(
t\right)  \right)  , \label{energydensity}%
\end{equation}
where $T_{0}^{0}$ is the indicated component of the fermionic stress energy
tensor operator.

At this point it is important to highlight the connection with the
{\em static} effective potential studied in section \ref{sec:veff}
above.

The equation of motion for the mean field eqn.
(\ref{meanfieldeqn}) suggests the identification of the fermionic
contribution, the last term of (\ref{meanfieldeqn}), with the
derivative of the fermionic contribution to the effective
potential (divided by $N$) given by eqn. (\ref{veffmeanfield}).
Such an identification, however, could only hold for a time
independent mean field. Indeed in this case, when the scalar field
is independent of time, the solutions of the mode equations with
the initial conditions given in eqn. (\ref{modeIC's}) are given by
$f_{1,{\vec p}}(t)=f_{1,{\vec p}}(0);f_{2,{\vec p}}(t)=f_{2,{\vec
p}}(0)$ leading to

\be \langle {\bar \psi}\psi \rangle|_{\text{static}} = -2 \int
\frac{d^3k}{(2\pi)^3} \frac{M_B}{\sqrt{k^2+M^2_B}}, \ee

\noindent which is the derivative of the fermionic contribution to
the effective potential with respect to the effective mass $M_B$.
Thus  the connection with the effective potential for a static
mean field is manifest. As it will be explained in detail below,
for a {\em dynamically} evolving mean field there are
non-equilibrium contributions that cannot be captured by the {\em
static} effective potential and that are responsible for a wealth
of new phenomena associated with the non-equilibrium dynamics.

\section{WKB Expansion and Renormalization}

While the renormalization aspects of the Yukawa theory to leading
order in the large $N$ approximation have been studied previously
\cite{baackefermions,ferminos}, here we provide an alternative
method, based on the WKB expansion, that makes contact with the
effective potential formalism studied in section (\ref{sec:veff})
above.

\subsection{WKB Solution of the Mode Equations}

Both the chiral condensate $\left\langle
\bar{\psi}\psi\right\rangle $ and the energy density
$\rho_{\text{f}}$ are divergent. In order to construct the
renormalized equations of motion, we need to extract their
divergent parts so that they can be absorbed by appropriate
counterterms. This can be done by finding solutions to
eq.(\ref{fmodeeqns}) that allow for a large momentum expansion,
i.e. a WKB expansion. We will concentrate on $f_{2\,\vec{p}}\left(
t\right)  $ since the results for $f_{1\,\vec{p}}\left(  t\right)
$ can be then be obtained via the replacement $M_{B}\left(
t\right)  \leftrightarrow-M_{B}\left( t\right)  $ throughout,
including in the initial conditions.

The WKB \textit{ansatz} for $f_{2\,\vec{p}}\left(  t\right)  $ is:
\begin{equation}
f_{2}^{\left(  I\right)  }\left(  t\right)  =A\left(  t\right)  \,\exp
\,i\,\int_{0}^{t}du\,\Omega\left(  u\right)  \,, \label{WKBAnsatzI}%
\end{equation}
where from now on we omit the $\vec{p}$ index on the mode functions. Insert
this into eq.(\ref{fmodeeqns}) and take the real and imaginary parts to find:%
\begin{equation}
\frac{\ddot{A}\left(  t\right)  }{A\left(  t\right)  }+\omega_{\vec{p}}%
^{2}\left(  t\right)  =\Omega^{2}\left(  t\right)  ;\,\,2\frac{\dot{A}\left(
t\right)  }{A\left(  t\right)  }+\frac{\dot{\Omega}\left(  t\right)  }%
{\Omega\left(  t\right)  }=\frac{\dot{M}_{B}\left(  t\right)  }{\Omega\left(
t\right)  }. \label{RealImWKB}%
\end{equation}
We can solve these equations:
\begin{equation}
f_{2}^{\left(  I\right)  }\left(  t\right)  =\frac{1}{\sqrt{2\,\Omega\left(
t\right)  }}\,\exp\,\int_{0}^{t}du\,\left(  i\Omega\left(  u\right)
+\frac{\dot{M}_{B}\left(  u\right)  }{2\,\Omega\left(  u\right)  }\right)
,\text{ where } \label{f2One}%
\end{equation}%
\begin{gather}
\Omega^{2}\left(  t\right)  =\omega_{\vec{p}}^{2}\left(  t\right)  +\left(
\frac{3}{4}\left(  \frac{\dot{\Omega}\left(  t\right)  }{\Omega\left(
t\right)  }\right)  ^{2}-\frac{1}{2}\left(  \frac{\ddot{\Omega}\left(
t\right)  }{\Omega\left(  t\right)  }\right)  \right)  +\label{WKBfreq}\\
\left(  \frac{1}{2}\left(  \frac{\ddot{M}_{B}\left(  t\right)  }{\Omega\left(
t\right)  }\right)  +\frac{1}{4}\left(  \frac{\dot{M}_{B}\left(  t\right)
}{\Omega\left(  t\right)  }\right)  ^{2}-\frac{\dot{M}_{B}\left(  t\right)
\,\dot{\Omega}\left(  t\right)  }{\Omega\left(  t\right)  ^{2}}\right)
.\nonumber
\end{gather}

Given $f_{2}^{\left(  I\right)  }\left(  t\right) $, the
second,linearly independent, solution $f_{2}^{\left(  II\right)  }\left(  t\right)$ is:%
\begin{align}
f_{2}^{\left(  II\right)  }\left(  t\right)   &  =\,f_{2}^{\left(  I\right)
}\left(  t\right)  \,F\left(  t\right)  ,\label{f2Two}\\
F\left(  t\right)   &  =-i\int_{0}^{t}\frac{du\,}{f_{2}^{\left(  I\right)
}\left(  u\right)  ^{2}}.\nonumber
\end{align}
Without loss of generality we can take the lower limit of integration in
$F\left(  t\right)  $ to be $t=0$, since including an arbitrary constant in
$F\left(  t\right)  $ will only give rise to a part proportional to
$f_{2}^{\left(  I\right)  }\left(  t\right)  $ in $f_{2}^{\left(  II\right)
}\left(  t\right)  $. This in turn can be absorbed when constructing the
appropriate linear combination to satisfy the initial conditions. The overall
factor of $-i$ was chosen so that the Wronskian of $f_{2}^{\left(  I\right)
}\left(  t\right)  $, $f_{2}^{\left(  II\right)  }\left(  t\right)  $ would
coincide with its value in the equilibrium case.

Write $f_{2}\left(  t\right)  =A_{I}\,f_{2}^{\left(  I\right)  }\left(
t\right)  +A_{II}\,f_{2}^{\left(  II\right)  }\left(  t\right)  $, and
impose the conditions

\begin{equation}
f_{2}\left(  0 \right) =
\sqrt{\frac{\omega_{0}-M_{0}}{2\,\omega_{0}}}\quad \quad ,\quad
\quad i\,\dot{f}_{2}\left( 0\right)     = \omega_{0}f_{2}\left(
0\right)
 \label{initialcondsf2}
 \end{equation}

 to find:

\begin{equation}
f_{2}\left(  t\right)  =f_{2}\left(  0\right)  \left\{  \frac{f_{2}^{\left(
I\right)  }\left(  t\right)  }{f_{2}^{\left(  I\right)  }\left(  0\right)
}\left[  1+\left(  \omega_{0}+\Omega\left(  0\right)  -i\frac{\dot{M}\left(
0\right)  -\dot{\Omega}\left(  0\right)  }{2\Omega\left(  0\right)  }\right)
f_{2}^{\left(  I\right)  }\left(  0\right)  ^{2}F\left(  t\right)  \right]
\right\}  . \label{finalWKBf2}%
\end{equation}

\subsection{Renormalizing $\left\langle \bar{\psi}\psi\right\rangle $}

We begin the renormalization program by first studying the divergences
in the fermion (chiral) condensate $\left\langle \bar{\psi}%
\psi\right\rangle $. The ultraviolet divergences of this
expectation value can be extracted by  performing a high momentum
expansion of $\left|  f_{2}\left(  t\right)  \right|  ^{2}-$
$\left| f_{1}\left( t\right)  \right|  ^{2}$; the only terms that
will be relevant are those proportional to $1/k^{p}$, $0\leq
p\leq3$ for large $k$. An alternative formulation can be found in
references\cite{coomot,baackefermions}. Furthermore, since we can
obtain $f_{1}\left( t\right) $ from $f_{2}\left(  t\right) $ by
making the replacement $M_{B}\left( t\right)
\leftrightarrow-M_{B}\left( t\right) $, $\left| f_{2}\left(
t\right) \right|  ^{2}-$ $\left| f_{1}\left( t\right) \right|
^{2}$ is odd in $M_{B}\left(  t\right) $.

We first solve for the WKB frequencies by iterating eq.(\ref{WKBfreq}) once
and keeping the appropriate power of the momentum in the large momentum limit:%
\begin{align}
\Omega^{\left(  0\right)  }\left(  t\right)   &  =\omega\left(  t\right)
\label{WKB0}\\
\Omega^{\left(  2\right)  }\left(  t\right)   &  =\omega\left(  t\right)
\left(  1+\frac{\ddot{M}_{B}\left(  t\right)  }{4\omega\left(  t\right)  ^{3}%
}\right)  . \label{WKB2}%
\end{align}
We will also need the high momentum behavior of $F\left(  t\right)  $ as
defined in eq.(\ref{f2Two}). This is most easily obtained by integrating by
parts a sufficient number of times to extract the relevant part. Doing this
for $\left\langle \bar{\psi}\psi\right\rangle $, we find%
\begin{gather}
F\left(  t\right)  =-1-i\frac{\dot{M}_{B}\left(  0\right)  }{2\Omega\left(
0\right)  ^{2}}-\frac{\ddot{M}_{B}\left(  0\right)  }{4\Omega\left(  0\right)
^{3}}+\label{Fexp}\\
\left(  1+i\frac{\dot{M}_{B}\left(  t\right)  }{2\Omega\left(  t\right)  ^{2}%
}+\frac{\ddot{M}_{B}\left(  t\right)  }{4\Omega\left(  t\right)  ^{3}}\right)
\exp-2\int_{0}^{t}dt^{\prime}\left(  i\Omega\left(  t^{\prime}\right)
+\frac{\dot{M}_{B}\left(  t^{\prime}\right)  }{2\Omega\left(  t^{\prime
}\right)  }\right)  +\mathcal{O}\left(  \omega^{-4}\right) \nonumber
\end{gather}
After some algebra we find:%
\begin{gather}
\left(  \left|  f_{2}\left(  t\right)  \right|  ^{2}-\left|  f_{1}\left(
t\right)  \right|  ^{2}\right)  _{\text{div}}=-\frac{M_{B}\left(  t\right)
}{\omega\left(  t\right)  }+\frac{\ddot{M}_{B}\left(  t\right)  }%
{4\omega ^{3}\left(  t\right) }-\frac{\ddot{M}_{B}\left(  0\right)  }%
{4\omega_{0}^{2}\omega\left(  t\right)  }\cos\left[2\int_{0}^{t}dt^{\prime}%
\omega\left(  t^{\prime}\right)\right]  +\label{divergentpiece}\\
\frac{\dot{M}_{B}\left(  0\right)  }{2\omega_{0}\omega\left(  t\right)  }%
\sin\left[2\int_{0}^{t}dt^{\prime}\omega\left(
t^{\prime}\right)\right] .\nonumber
\end{gather}
Note that
\begin{gather}\label{statrel}
\left|  f_{2}\left(  0\right)  \right|  ^{2}-\left|  f_{1}\left(  0\right)
\right|  ^{2}=\left(  \sqrt{\frac{\omega_{0}-M_{0}}{2\,\omega_{0}}}\right)
^{2}-\left(  \sqrt{\frac{\omega_{0}+M_{0}}{2\,\omega_{0}}}\right)
^{2}=-\frac{M_{0}}{\omega_{0}}\nonumber\\
=\left(  \left|  f_{2}\left(  0\right)  \right|  ^{2}-\left|  f_{1}\left(
0\right)  \right|  ^{2}\right)  _{\text{div}},
\end{gather}

There are several important aspects of the renormalization of the
condensate $\langle {\bar \psi}\psi \rangle$ that should be
emphasized at this point:

\begin{itemize}

\item{The momentum integral of the first term in eqn.(\ref{divergentpiece})
yields the derivative of the effective potential. This would be
the {\em only} contribution in an adiabatic limit in which the
derivatives of the expectation value of the scalar field vanish.
This observation allows us to make a first contact with the
preparation of the initial value problem via the external source
term $h(t)$. Switching this source on adiabatically from the
infinite past up to the initial time $t=0$ leads to the first term
in eqn. (\ref{divergentpiece}) for $t < 0$ only. }

\item{Integrating the second term in  (\ref{divergentpiece}) in momentum up to
an upper momentum cutoff $\Lambda$ leads to a contribution of the
form $\propto \ddot{\delta}_B(t)\ln(\Lambda)$; this should be
identified with the wave function renormalization. }

\item{By choosing $\dot{M}_{B}\left(  0\right) = 0$ we are able to dispense
with the fourth term in (\ref{divergentpiece}).}

\item{The third term in (\ref{divergentpiece}), proportional to
$\ddot{M}_{B}\left(  0\right)$ , has a logarithmic UV divergence
at $t=0$ which can potentially give rise to initial time
singularities in the equations of
motion\cite{coomot,baackeinitialtime,allinitial}, for which there
are no counter terms.

However, in the infinite momentum cutoff limit, the contributions
that give an initial time singularity, are actually {\em finite}
for any $t>0$. This is because the integrand
is averaged out by the strong
oscillations\cite{allinitial}\footnote{This can be seen simply by
considering the contribution to the momentum integral in the limit
of $k>>M_B$. The contribution of the large $k$-modes can be
estimated by taking the ultraviolet cutoff to infinity but
introducing a lower momentum cutoff $\mu$, leading to
$\int^{\infty}_{\mu}\cos(2kt)\frac{dk}{k} = -Ci(2\mu t)$ with
$Ci(x)$ the cosine integral function which is finite for $t>0$ and
diverges logarithmically as $t\rightarrow 0$. }. At $t=0$ the
contributions from the second and third terms cancel exactly. For
finite but large cutoff $\Lambda$ it is a straightforward exercise
to show that the combination of the second and third terms
(proportional to $\ddot{M}$) is finite and small for $t \leq
1/\Lambda$. The UV logarithmic divergence in the combined second
and third terms
begins to develop on a time scale $t \geq 1/\Lambda$. Thus, for
$t\gg 1/\Lambda$, when the third term in (\ref{divergentpiece}) is
{\em finite}, we can write

\be\label{splittlarge}
\left\langle \bar{\psi}\,\psi\right\rangle _{B}(t)=\left\langle \bar{\psi}%
\,\psi\right\rangle_{\text{div2}}^{>}(t)+\left\langle
\bar{\psi}\,\psi\right\rangle^{>} _{R}(t) \ee with $\left\langle
\bar{\psi}\,\psi\right\rangle _{\text{div2}}^{>}(t)$ given as the
momentum integral of the first two terms in
eq.(\ref{divergentpiece}).

Using an upper momentum cutoff $\Lambda$ and dropping terms of
order $(\frac{M^2}{\Lambda^2})$ we find
\begin{equation}
\left\langle \bar{\psi}\,\psi\right\rangle _{\text{div2}}^{>}(t)=
\frac{1}{2\pi^{2}}\left[  -M_{R}^{3}\left(  t\right)  \left(  \ln\left|
\frac{M_{R}\left(  t\right)  }{2\Lambda}\right|
+\frac{1}{2}\right)-M_{R}\left( t\right) \Lambda ^{2}
-\frac{\ddot{M}_{R}\left(  t\right) }{2}\left(  \ln\left|
\frac{M_{R}\left( t\right)  }{2\Lambda}\right|  +1\right)  \right]  , \label{div2psibarpsi}%
\end{equation}
\noindent and $\left\langle \bar{\psi}\,\psi\right\rangle^{>}
_{R}(t)$ contains the third term of (\ref{divergentpiece}) and is
{\em finite} for $t\gg 1/\Lambda$.  In
section(\ref{sec:fullfermionicdynamics}) we will show that this is also
finite for t=0 since $\ddot{M}_{B}(0)$ is proportional to $1/ln(\Lambda)$.

Alternatively we can also write

\be\label{splittsmall}
\left\langle \bar{\psi}\,\psi\right\rangle _{B}=\left\langle \bar{\psi}%
\,\psi\right\rangle_{\text{div1}}(t)+\left\langle
\bar{\psi}\,\psi\right\rangle_{AF} (t)\ee
 with $\left\langle
\bar{\psi}\,\psi\right\rangle _{\text{div1}}(t)$ given as the
momentum integral of only the {\em first} term in
eq.(\ref{divergentpiece}):

\begin{equation}
\left\langle \bar{\psi}\,\psi\right\rangle _{\text{div1}}=
\frac{1}{2\pi^{2}}\left[  -M_{R}^{3}\left(  t\right)  \left(  \ln\left|
\frac{M_{R}\left(  t\right)  }{2\Lambda}\right|  +\frac{1}{2}\right)
-M_{R}\left( t\right)  \Lambda
^{2}\right]  . \label{div1psibarpsi}%
\end{equation}

\noindent Now $\left\langle \bar{\psi}\,\psi\right\rangle_{AF}(t)$
includes both the second and third term in
eq.(\ref{divergentpiece}). It is clear from the discussion above,
that $\left\langle \bar{\psi}\,\psi\right\rangle_{AF}(t)$ is
actually {\em finite} for $t\leq 1/\Lambda$; it vanishes
identically at $t=0$  and does not have any initial time
singularity. This quantity will, however, develop a logarithmic
divergence due to the second term in eq.(\ref{divergentpiece})
associated with wave function renormalization for $t\gg
1/\Lambda$ (hence AF for ``almost finite'').

Several important features of the above expressions must be
highlighted. First, the argument of the logarithms contains the
full time dependent mass, unlike a renormalization scheme that
only extracts the logarithmic divergences in terms of the initial
mass(\cite{baackefermions,ferminos,baackeinitialtime,allinitial});
note that these schemes differ only by finite terms. However, as
will become clear below, keeping the full time dependent
frequencies in the denominators will lead to the {\em
instantaneous effective potential}, i.e, the static effective
potential but now as a function of the time dependent mean field.
Furthermore, taking the full time dependent frequencies will lead
to the identification of the effective coupling that ``runs'' with
the amplitude of the mean field. This identification will allow us
to establish a direct correspondence with the RG improved
effective potential. In this manner we will be able to clearly
separate the contribution from an adiabatic or instantaneous
generalization of the static effective potential, from important
non-equilibrium and fully dynamical effects which can only be
described in terms of the time dependent mean field. In
particular, we seek to clearly separate the effect of particle
production and its concomitant contribution to the dynamical
evolution.

Second, the term $ \frac{1}{2\pi^{2}}\left[  -M_{R}^{3}\left(
t\right) \left(  \ln\left| \frac{M_{R}\left(  t\right)
}{2\Lambda}\right| +\frac{1}{2}\right)  -M_{R}\left( t\right)
\Lambda ^{2}\right] $ in both $\left\langle
\bar{\psi}\,\psi\right\rangle _{\text{div2}}^{>}(t)$ and
$\left\langle \bar{\psi}\,\psi\right\rangle _{\text{div1}}(t)$
will lead to the effective potential. This is implied because this
term does not depend on the derivatives of $M_B(t)$.

Finally, while $\left\langle \bar{\psi}\,\psi\right\rangle^{>}
_{R}(t)$ is finite for $t \gg 1/\Lambda$ but features an initial
time singularity at $t=0$, $\left\langle
\bar{\psi}\,\psi\right\rangle_{AF}(t)$ vanishes identically at
$t=0$, it is finite for $t\leq 1/\Lambda$ but exhibits a
logarithmic ultraviolet divergence $\propto \ddot{\delta}
\ln(\Lambda)$ associated with wave function renormalization for
$t\gg 1/\Lambda$.

If we insist in using the split (\ref{splittlarge}), and thus
extract the wave function renormalization divergent term at {\em
all} times, including $t\leq 1/\Lambda$ the quantity $\langle
{\bar \psi}\psi\rangle^{>}_R$ will contain an {\em initial time
singularity} given by the short time limit $t\leq 1/\Lambda$ of
the third term in eq.(\ref{divergentpiece}).

In references\cite{coomot,baackeinitialtime,allinitial} this
initial time divergence is dealt with  in several ways: by
choosing the initial conditions on the mode functions to a high
(fourth) order in the adiabatic expansion\cite{coomot}, or by
performing an appropriately chosen Bogoliubov transformation of
the initial state\cite{baackeinitialtime}, or equivalently, by
including a counterterm in the external ``magnetic field''
$h\left( t\right) $ \cite{allinitial} so as to cancel this
singularity at $t=0$. All of these methods are equivalent and lead
to a set of equations that conserve energy and are free of initial
time singularities\cite{coomot,baackeinitialtime,allinitial}.

However, these methods all suffer from the drawback that they do
not lead to an interpretation of the dynamics in terms of the
effective potential. This is evident in the Bogoliubov approach
advocated in references\cite{baackeinitialtime,allinitial} since
the Bogoliubov transformation involves the derivatives of the mean
field, and the Bogoliubov coefficients multiply the terms that
lead to the effective potential, thereby mixing terms that depend
on the derivatives of the mean field with terms that arise from
the adiabatic effective potential. We refer the reader to
reference\cite{baackeinitialtime} for a thorough exposition of the
Bogoliubov method. As discussed in detail
in\cite{coomot,baackeinitialtime,allinitial} any approach to
regulating the initial time singularity of $\left\langle
\bar{\psi}\,\psi\right\rangle^{>} _{R}(t)$ (when extrapolated to
$t=0$) is tantamount to a {\em redefinition of the state} at
$t=0$.

Instead of seeking a regularization of this initial time
singularity, we recognize that this singularity arises from trying
to extract a wave function renormalization from very early time
{\em even when there is no such divergence}. We interpret the fact
that the logarithmic divergence associated with wave function
renormalization emerges at time scales $t\gg 1/\Lambda$ as the
{\em build up} of the wave function renormalization over this time
scale. This is consistent with the adiabatic hypothesis of
preparation of the state at $t=0$ via an external current. At this
point the following question could be raised: in the formulation
presented above, where is the adiabatic assumption explicit?
The answer to this important question is in the initial conditions
on the fermionic mode functions given by
eqn.(\ref{initialcondsf2})(up to an overall trivial phase). These
initial conditions determine that the state at $t=0$ is the {\em
vacuum} state for free Dirac spinors of mass $M_B(0)$ which is
determined by the value of the mean field at $t=0$. Obviously this
is the state obtained by adiabatically displacing the mean field
from the trivial vacuum. As we will see in detail below,
recognizing that $\left\langle
\bar{\psi}\,\psi\right\rangle_{AF}(t)$ vanishes identically at
$t=0$ and is {\em finite} over a time interval $t \leq 1/\Lambda$
will allow us to solve the initial value problem self-consistently
with the result that the potential initial time singularity is
simply not there and the evolution is continuous throughout. }

\end{itemize}

Thus in summary, the above discussion highlights the very
important {\em dynamical} aspect that the wave function
renormalization {\em builds up} on time scales $t > 1/\Lambda$ and
eqn.(\ref{splittlarge}) or (\ref{splittsmall}) must be used
according to the time scale studied in the evolution. However, as
will be discussed in detail below, when we study the full
equations of motion, we will see that eqn. (\ref{splittsmall}) is
far more convenient for numerical studies. Furthermore we will
find in section(\ref{sec:fullfermionicdynamics}) below that a
self-consistent analysis reveals that in fact {\em there is no
initial time singularity}, i.e,  $\left\langle
\bar{\psi}\,\psi\right\rangle^{>} _{R}(0)$ is actually {\em
finite}.

\bigskip

\subsection{The Renormalized Equations of
Motion}\label{sec:reneqnofmotion}

We now have all the ingredients to  obtain the renormalized
equations of motion. We recall our system of equations:

\bea &&\left(\frac{d^2}{dt^2}+p^2+M^2_B(t)\pm i\,\dot {M}_{B}\left(
t\right) \right) \,f_{1,2\ \vec{p}}\left(  t\right)
=0.\label{equations}\\
&&\ddot{\delta}_{B}\left(  t\right)  +m_{B}^{2}\,\delta_{B}\left(
t\right) +\frac{\lambda_{B}}{3!}\delta_{B}^{3}\left(  t\right)
+y_{B}\left\langle \bar{\psi}\,\psi\right\rangle _{B}=0 \eea \noindent
where we have set $h(t>0)=0$.

Starting with the fermionic mode equations, we see that since
there are no operators present to absorb the divergences in
$M_{B}\left( t\right) $, we need to impose the condition:
$M_{B}\left( t\right) =M_{R}\left( t\right) $ or
$y_{B}\delta_{B}\left( t\right) =y_{R}\delta_{R}\left( t\right) $.
When coupled with the wavefunction renormalization conditions
below, this will relate the bare and renormalized Yukawa
couplings. This condition also implies $\omega_{p}\left(  t\right)
_{B}=\omega_{p}\left(  t\right) _{R}$ .

To study the dynamics for time scales $t \gg 1/\Lambda$ we use
eqn.(\ref{splittlarge}) to separate the divergences, leading to
the following  equation for the mean field
\begin{gather}
\left(  1+\frac{y_{B}^{2}}{4\pi^{2}}\ln\frac{2\Lambda}{\mu}\right)
\ddot{\delta}_{B}+\left(  m_{B}^{2}-\frac{y_{B}^{2}\Lambda^{2}}{2\pi^{2}%
}\right)  \delta_{B}+\left(  \frac{\lambda_{B}}{3!}+\frac{y_{B}^{4}}{2\pi^{2}%
}\ln\frac{2\Lambda}{\mu}\right)  \delta_{B}^{3}\left(  t\right)  +\nonumber\\
y_{B}\left\langle \bar{\psi}\,\psi\right\rangle _{R}^{>}-\frac{y_{B}}%
{2\pi^{2}}\left\{  M_{R}^{3}\left(  t\right)  \left(  \ln\left|
\frac{M_{R}\left(  t\right)  }{\mu}\right|  +\frac{1}{2}\right)
+\frac{\ddot{M}_{R}\left(  t\right)  }{2}\left(  \ln\left|  \frac{M_{R}\left(
t\right)  }{\mu}\right|  +1\right)  \right\}  =0,
\end{gather}
where $\mu$ is a renormalization point. Set $\delta_{B}\left(  t\right)
=\sqrt{Z}\delta_{R}\left(  t\right)  $, so that $M_{B}\left(  t\right)
=M_{R}\left(  t\right)  $ implies $y_{B}=y_{R}/\sqrt{Z}$, and choose the
coefficient of $\ddot{\delta}_{R}$ to be unity. This yields the
renormalization conditions:
\begin{subequations}
\label{renormconds}%
\begin{gather}
Z\left(  1+\frac{y_{B}^{2}}{4\pi^{2}}\ln\frac{2\Lambda}{\mu}\right)
=1,\label{wavefunction}\\
Z\left(  m_{B}^{2}-\frac{y_{B}^{2}\Lambda^{2}}{2\pi^{2}}\right)  =m_{R}%
^{2}\left(  \mu\right)  ,\label{massrenorm}\\
Z^{2}\left(  \frac{\lambda_{B}}{3!}+\frac{y_{B}^{4}}{2\pi^{2}}\ln
\frac{2\Lambda}{\mu}\right)  =\frac{\lambda_{R}\left(  \mu\right)  }%
{3!},\label{couplingrenorm}\\
\sqrt{Z}y_{B}=y_{R}\left(  \mu\right)  , \label{yukawarenorm}%
\end{gather}

These are exactly the same renormalization conditions obtained
from the renormalization of the {\em static} effective potential,
together with the wave function renormalization obtained from the
one loop fermionic self energy calculated in section
(\ref{sec:veff}) above. Furthermore, the renormalization of the
Yukawa coupling (\ref{yukawarenorm}) guarantees that the fermionic
mode equations are {\em renormalization group invariant} since
they depend only on the product $y_B\delta_B=y_R\delta_R$.

The renormalized equation of motion for the mean field now becomes
\end{subequations}
\begin{gather}
\left(  1-\frac{g_{R}\left(  \mu\right)  }{2}\left(  \ln\left|  \frac{eM_{R}%
\left(  t\right)  }{\mu}\right|  \right)  \right)  \ddot{M}_{R}+m_{R}%
^{2}\left(  \mu\right)  M_{R}+\label{renormEOM}\\
g_{R}\left(  \mu\right)  \left(  \left(  \frac{1}{12\pi^{2}}\frac{\lambda
_{R}\left(  \mu\right)  }{g_{R}^{2}\left(  \mu\right)  }-\ln\left|
\frac{M_{R}\left(  t\right)  }{\mu}\right|  -\frac{1}{2}\right)  M_{R}%
^{3}\left(  t\right)  +2\pi^{2}\left\langle \bar{\psi}\,\psi\right\rangle
_{R}^{>}\right)  =0,\nonumber
\end{gather}
where $g_{R}\left(  \mu\right)  \equiv y_{R}^{2}\left(  \mu\right)
/2\pi^{2}$, as defined previously, and we have multiplied through
by $y_R$ to write the
equation of motion for the {\em renormalization group invariant}
product $y_R \delta_R$.  We can rewrite this as:%
\begin{equation}
\ddot{M}_{R}+\frac{1}{D\left(  \mu\right)  }\left(  \frac{2m_{R}^{2}\left(
\mu\right)  }{g_{R}\left(  \mu\right)  }M_{R}+2\left(  \frac{1}{12\pi^{2}%
}\frac{\lambda_{R}\left(  \mu\right)  }{g_{R}^{2}\left(  \mu\right)  }%
-\ln\left|  \frac{M_{R}\left(  t\right)  }{\mu}\right|  -\frac{1}{2}\right)
M_{R}^{3}\left(  t\right)  +4\pi^{2}\left\langle \bar{\psi}\,\psi\right\rangle
_{R}^{>}\right)  =0, \label{RGInvEOM}%
\end{equation}
where $D\left(  \mu\right)  =\left(  2/g_{R}\left(  \mu\right)  -\ln\left|
eM_{R}\left(  t\right)  /\mu\right|  \right)  $. Using the following relations
we can see that all the terms in eq.(\ref{RGInvEOM}) are $\mu$ independent:
\begin{subequations}
\label{RGeqns}%
\begin{gather}
\frac{m_{R}^{2}\left(  \mu\right)  }{g_{R}\left(  \mu\right)  }=\frac{m_{B}%
^{2}}{g_{B}}-\Lambda^{2}\label{massRG}\\
\frac{2}{g_{R}\left(  \mu\right)  }=\frac{2}{g_{B}}+\ln\frac{2\Lambda}{\mu
}\label{gRG}\\
\frac{1}{12\pi^{2}}\frac{\lambda_{R}\left(  \mu\right)  }{g_{R}^{2}\left(
\mu\right)  }=\frac{1}{12\pi^{2}}\frac{\lambda_{B}}{g_{B}^{2}{}}%
+\ln\frac{2\Lambda}{\mu} \label{lambdaRG}%
\end{gather}
In particular,
\end{subequations}
\begin{equation}
D\left(  \mu\right)  \equiv\frac{2}{g_{R}\left(  \mu\right)  }-\ln\left|
\frac{eM_{R}\left(  t\right)  }{\mu}\right|  =\frac{2}{g_{R}\left(
eM_{R}\left(  t\right)  \right)  }. \label{Landau1}%
\end{equation}
Since $g_{R}\left(  \mu\right)  \propto y_{R}^{2}\left( \mu\right)
\geq0$, unitarity appears to require that the range of validity of
the theory be restricted to be below the Landau pole at
$E_{L}=e^{-1}\mu\exp\left( 2/g_{R}\left(  \mu\right)  \right) $.
As discussed in the Introduction, however, this may not actually
be necessary under all circumstances; we will explore this issue
further in the next section.

We can now compare to the static case. Taking $m_R=0$ (as in the
static case) we see that the term proportional to $M_{R}^{3}$ in
(\ref{RGInvEOM}) is precisely the derivative of the {\em static}
effective potential given by eqn. (\ref{vprime}) but now in terms
of the time dependent renormalized mean field. Therefore, just as
with the static effective potential, we introduce the
dimensional transmutation scale $\bar{M}$ by demanding that the
{\em instantaneous or adiabatic} effective potential, i.e, the
static effective potential in terms of the time dependent mean
field, have a maximum at this scale when $m_R=0$:
\begin{equation}
\frac{1}{12\pi^{2}}\frac{\lambda_{R}\left(  \mu\right)  }{g_{R}^{2}\left(
\mu\right)  }-\ln\frac{\bar{M}}{\mu}-\frac{1}{2}=0. \label{Mbar}%
\end{equation}
We also define $\chi\equiv M_{R}\left(  t\right)
/\bar{M},g=g_{R}\left( \bar {M}\right)
,q=k/\bar{M},\tilde{m}^{2}=m_{R}^{2}\left( \mu=\bar{M}\right)
/\bar{M}^{2},\tau=\bar{M}t.$ Then eq.(\ref{RGInvEOM}) can be written as:%
\begin{equation}
\chi^{\prime\prime}\left(  \tau\right)  +g_{R}\left(  e\chi\left(
\tau\right)  \right)  \left(  \frac{\tilde{m}^{2}}{g}\chi\left(  \tau\right)
-\chi^{3}\left(  \tau\right)  \ln\left|  \chi\left(  \tau\right)  \right|
+2\pi^{2}\frac{\left\langle \bar{\psi}\,\psi\right\rangle _{R}^{>}(\tau)}%
{\bar{M}^{3}}\right)  =0, \label{NoDimEOM}%
\end{equation}
where primes denote $\tau$ derivatives.

For $m_R=0$ we see immediately that the combination
$-g_R(e\chi(\tau))\chi^{3}\left(  \tau\right)  \ln\left|
\chi\left( \tau\right)  \right|$ is the derivative of the
renormalization group improved effective potential (see equation
(\ref{rgimproved}))  in terms of the running coupling constant at
a scale $\chi(\tau)$ (up to a finite term).

Thus this form of the mean field equation of motion splits off the
effects due to the RG improved effective potential ( the first two
terms) from those due to the time evolution of the fermionic
fluctuations which are represented by $\left\langle
\bar{\psi}\,\psi\right\rangle _{R}^{>}$. We also see that the
effects of wavefunction renormalization are encoded in the prefactor
$g_{R}\left(  e\chi\left( \tau\right)  \right)  $ multiplying the
potential and fluctuation terms.

A remarkable aspect of equation (\ref{NoDimEOM}) is that the
effective coupling {\em depends on time}. In a well defined sense,
this is a {\em dynamical renormalization} much in the same way as
that explored in real time in references\cite{drg}. As mentioned
before we could have renormalized simply by absorbing the
divergences with the frequencies at the initial time. However,
doing this would still leave ``large logarithms'' arising from
when the amplitude of the mean field becomes large.  As the mean
field evolves in time, the fermion fields probe different energy
scales.  Since the coupling ``runs'' with energy scale, it is then
natural that it will ``run'' as a parametric function of time
through the evolution of the mean field. This physical picture is
manifest in eq.(\ref{NoDimEOM}). This important aspect of our study
is a novel result which only becomes
manifest in a real time non-equilibrium framework that allows one
to study the dynamics of the fully renormalized fields and
couplings.

It would appear from this equation that the dynamics of
$\chi$ could be singular as\thinspace$\chi$ crosses $E_{L}/\bar{M}$ since%
\begin{equation}
g_{R}\left(  e\chi\left(  \tau\right)  \right)
=\frac{2}{2/g-\ln\left| e\chi\left(  \tau\right)  \right|
}=\frac{2}{\ln\left[\frac{\bar{E}}{\left|  \chi\left( \tau\right)
\right|}\right]  },
\label{Landau2}%
\end{equation}
\noindent  where
\be
 \bar{E}=\frac{E_{L}}{\bar{M}}=e^{\frac{2}{g}-1}. \label{Lanpole}
 \ee
This would certainly be the case if the equation of motion only
involved the derivative of the {\em static} effective potential.
However, we will see below that this is  {\em not} the case
when the fluctuations are taken into account.

We conclude this section by collecting together our renormalized
equations of motion and initial conditions, now in terms of
dimensionless variables, valid in principle for $t \gg 1/\Lambda$ when
$\langle \bar{\psi}\,\psi \rangle_{R}^{>}(\tau)$  is free of potential
initial time singularities:
\begin{eqnarray}
&& \left(  \frac{d^2}{d\tau^2}+q^2+\chi^2\left(  \tau\right)  \pm
i\ \chi^{\prime} \left(  \tau\right)  \right)  \,f_{1,2\
\vec{q}}\left( \tau\right)
=0,\label{dimlessfermode}\\
&& \chi^{\prime\prime}\left(  \tau\right)  +g_{R}\left(  e\chi\left(
\tau\right)  \right)  \left(  \frac{\tilde{m}^{2}}{g}\chi\left(
\tau\right) -\chi^{3}\left(  \tau\right)  \ln\left|  \chi\left(
\tau\right)  \right| +\mathcal{F}\left(  \tau\right)  \right)
=0,\label{dimlessmeanfield}\\
&&\mathcal{F}\left(  \tau\right)  \equiv
2\pi^{2}\frac{\left\langle \bar{\psi }\,\psi\right\rangle
_{R}^{>}(\tau)}{\bar{M}^{3}}= 2\,\int_{0}^{\infty}dqq^{2}\left(
\left|  \,f_{2\,\vec{q}}\left( \tau\right)  \right|^{2}-\left|
\,f_{1\vec{q}}\left(  \tau\right) \right|^{2}-\left\{
-\frac{\chi\left(  \tau\right) }{\omega_{q}\left( \tau\right)
}+\frac{\chi^{\prime\prime}\left( \tau\right)
}{4\omega_{q}^{3}\left( \tau\right)  }\right\}  \right),
\label{dimlesspsibarpsi}
\end{eqnarray}

\noindent with the initial conditions
\begin{eqnarray}
&& f_{1,2\ \vec{q}}\left(  0\right)  =\sqrt{\frac{\omega_{q}\left(
0\right) \pm\chi\left(  0\right)  }{2\omega_{q}\left(  0\right)
}}~,~i\dot{f} _{1,2\ \vec{q}}\left(  0\right)  =\omega_{q}\left(
0\right)  f_{1,2\ \vec{q}
}\left(  0\right) \label{dimlessfermodeIC}\\
&&\chi\left(  0\right)  =\chi_{0}~,~\dot{\chi}\left(  0\right)
=0\label{dimlesschiIC}\\
&&\left\langle \bar{\psi}\,\psi\right\rangle _{AF}\left(  0\right)
=0 . \label{dimlesspsibarpsiIC}
\end{eqnarray}

The quantity ${\cal F}(\tau)$ is the fermionic condensate $\langle
\bar{\psi}\psi \rangle $ after subtracting the contributions that
lead to the effective potential and the wave function
renormalization.  It therefore represents the {\em purely
dynamical} fluctuations in the fermion field associated with
particle production.

\subsection{RG Improved Effective Potential}

Since we will focus our study on the effects of the dynamics, we
start by highlighting the dynamics that would ensue from a
consideration of the RG improved effective potential {\em
alone}\cite{colemanweinberg}. A comparison between the
renormalization group improved effective potential eq.
(\ref{rgimproved}) and the renormalized equation of motion
(\ref{dimlessmeanfield}) clearly indicates that we can extract the
dynamics that would ensue solely from the effective potential by
neglecting the dynamical contribution of the fermionic
fluctuations encoded in ${\cal F}(\tau)$ in the equation of motion
(\ref{dimlessmeanfield}) leading to

\begin{equation}
\chi^{\prime\prime}\left(  \tau\right)  +\frac{2}{2/g-\ln\left|
e\chi\left( \tau\right)  \right|  }\left(
\frac{\tilde{m}^{2}}{g}\chi\left(  \tau\right) -\chi^{3}\left(
\tau\right)  \ln\left|  \chi\left(  \tau\right)  \right|
\right)  =0. \label{EffPotTrunc}%
\end{equation}
Notice that the denominator in the second term comes solely from
wavefunction renormalization considerations and that it has a zero at
the Landau pole, as discussed above. This truncated form of the equation of
motion displays clearly the connection with the renormalization group
improved effective potential,  as obtained in section
(\ref{RGimpVeff}), eqn. (\ref{rgimproved})). Obviously for generic
values of $\tilde{m}^{2}$ and $g$, the numerator will \emph{not} vanish when
the denominator does, leading to infinite acceleration at the time when
the mean field reaches the Landau pole. Thus, an investigation of the
above equation would lead to unphysical behavior of the mean field as
it approaches the Landau pole and one would then conclude that the
existence of a Landau pole precludes a sensible interpretation of the
theory when the amplitude of the mean field is comparable to the
position of the Landau pole.

We bring this discussion to the fore because it is one of the important
points of this study that the effect of the fluctuations is very
dramatic and completely changes the picture extracted from the
effective potential.

\subsection{Full dynamics: fermionic fluctuations}\label{sec:fullfermionicdynamics}

Having established the connection with the renormalization group
improved effective potential, we now study the evolution of the full
equation of motion (\ref{dimlessmeanfield}) including the fermionic
fluctuations ${\cal F}(\tau)$.

However at this point we face two problems: i) the renormalized
equation of motion (\ref{dimlessmeanfield}) is only valid for
$t\gg 1/\Lambda$ and cannot be extrapolated to the initial time
$t=0$ because of the potential initial time singularity in ${\cal
F}(0)$ discussed above. This in turn, entails a potential problem
with the  initialization of the dynamical evolution. ii) Equation
(\ref{dimlessmeanfield}) cannot be numerically implemented accurately
because ${\cal F}(\tau)$ requires a subtraction that
involves the second derivative of the mean field at the same time
as the update (see eqn. (\ref{dimlesspsibarpsi}). Both problems
can be circumvented at once by invoking the split
(\ref{splittsmall}) or equivalently by introducing the
dimensionless quantity ${\cal F}_{AF}(\tau)$ as

\bea \label{almostfinite} &&\mathcal{F}\left(  t\right)
=\mathcal{F}_{\text{AF}}\left( \tau\right) +\left(
-\frac{1}{2}\int_{0}^{\Lambda}dqq^{2}\frac{1}{\omega_{q}^{3}\left(
\tau\right)  }\right)  \chi^{\prime\prime}\left(  \tau\right)  ,\text{
where}\label{almostfiniteflucts}\\
&&\mathcal{F}_{\text{AF}}\left(  \tau\right)  =2\,\int_{0}^{\Lambda}%
dqq^{2}\left(  \left|  \,f_{2\,\vec{q}}\left(  \tau\right)  \right|
^{2}-\left|  \,f_{1\vec{q}}\left(  \tau\right)  \right|  ^{2}-\left\{
-\frac{\chi\left(  \tau\right)  }{\omega_{q}\left(  \tau\right)
}\right\} \right)  \label{FAFdef}
 \eea

 We call $\mathcal{F}_{\text{AF}}\left(  \tau\right)  $ the
``almost finite'' fluctuation, since it only has the part of the
divergences proportional to $\chi\left(  \tau\right)  $ subtracted.
Furthermore, as discussed above, $\mathcal{F}_{\text{AF}}\left(
0\right) =0 $.

We emphasize that we have \emph{not} changed the equation but have
merely redistributed the terms, just as the two alternative forms
of writing the fermionic condensate given by (\ref{splittlarge})
and (\ref{splittsmall}).

The integral in eq.(\ref{almostfiniteflucts}) can be done
explicitly. We can then combine the terms proportional to
$\chi^{\prime\prime}\left( \tau\right)  $ and rewrite
eq.(\ref{dimlessmeanfield}) as

\begin{gather}
\chi^{\prime\prime}\left(  \tau\right)  +\frac{2}{Z_{\Lambda}\left(
\tau\right)  }\left(  \frac{\tilde{m}^{2}}{g}\chi\left(  \tau\right)
-\chi^{3}\left(  \tau \right)  \ln\left|  \chi\left(  \tau\right)
\right| +\mathcal{F}_{\text{AF}}\left(  \tau\right)  \right)  =0
\label{cutoffEOM}\\
Z_{\Lambda}\left(  \tau\right)  =2/g-1+\frac{\Lambda}{\sqrt{\Lambda^{2}%
+\chi^{2}\left(  \tau\right)  }}-\ln\left(
\Lambda+\sqrt{\Lambda^{2}+\chi ^{2}\left(  \tau\right)  }\right)
.\nonumber
\end{gather}
We have kept the full expressions for the integral for the sake of
completeness. However, if $\chi\left(  \tau\right)  $ approaches the
cutoff, fermion modes with momenta of the order of the cutoff will be
excited indicating that we are approaching the limit of validity of our
numerical approximation. This means that we should restrict $\chi\left(
\tau\right)  $ to be much less than $\Lambda$. Doing so allows us to
simplify the expression
for $Z_{\Lambda}\left(  \tau\right)  $:%
\begin{equation}
Z_{\Lambda}\left(  \tau\right)  \sim2/g-\ln2\Lambda=\ln\frac{\bar{E}}%
{2e^{-1}\Lambda}\equiv\ln\frac{\bar{E}}{\bar{\Lambda}}. \label{ZLambdaSimple}%
\end{equation}

The important point to note here is that $Z_{\Lambda}\left(
\tau\right) $ is independent of time. In particular, in this
formulation the Landau pole does not give rise to any singular
behavior.

From the renormalization of the coupling $g$ given by eqn.
(\ref{gRG}) we identify

\be \frac{2}{Z_{\Lambda}} = g_B \ee

\noindent i.e, the coupling at the scale $\Lambda$ or the ``bare
coupling''.

We reiterate that we are still solving the renormalized equations;
all we have done is reformulate them for the purposes of numerical
analysis. The fact that $Z_{\Lambda}\left( \tau\right)  $ is
non-vanishing provides a strong hint that there should be no
problems as $\chi$ crosses the Landau pole. We shall see below
that indeed this is borne out by the numerics.

The equation of motion (\ref{cutoffEOM}) is now in a form that can
be studied numerically. In particular, it can be initialized for
any large but finite cutoff. Furthermore, the form of the equation
of motion given by eqn. (\ref{cutoffEOM}) above suggests that the
dynamics is {\em smooth} even at time scales $t\gg 1/\Lambda$ when
${\cal F}_{AF}(\tau)$ develops an ultraviolet logarithmic
divergence. This is so because the  logarithmic divergence in
${\cal F}_{AF}(\tau)$ will be compensated by the logarithm in
$Z_{\Lambda}$ in the denominator. Hence we conclude that the
equation of motion (\ref{cutoffEOM}) has a well defined
initialization and the dynamics is smooth without logarithmic
divergences or discontinuities  as $\chi(\tau)$ approaches and
passes the Landau pole. This in fact will become clear from the
detailed numerical analysis provided in the next section. The
equivalence of the equations of motion in terms of ${\cal F}_{AF}$
or ${\cal F}$ and the observation that the equation of motion
(\ref{cutoffEOM}) does not feature any initial divergence
suggests that the equation of motion (\ref{dimlessmeanfield}) is
free from initial time singularities and the numerical evolution
is indeed smooth. That this is indeed the case  can be seen as
follows. From the fact that $\mathcal{F}_{\text{AF}}\left(
0\right) =0 $ we now find that

\be \chi^{\prime\prime}\left(  0 \right)  = -\frac{2}{Z_{\Lambda}}
\left(  \frac{\tilde{m}^{2}}{g}\chi\left( 0\right) -\chi^{3}\left(
0\right)  \ln\left|  \chi\left(  0\right) \right| \right)\ee

Furthermore, from the relation between ${\cal F}$ and ${\cal F}_{AF}$
given by eqn. (\ref{almostfinite}) we find that

\be {\cal F}(0) = \frac{1}{2}\chi^{\prime\prime}\left(  0
\right)\left[1-\ln\left(\frac{2\Lambda}{\chi(0)}\right) \right] =
\frac{\ln\left[\frac{{\bar \Lambda}}{ \chi(0)}
\right]}{\ln\left[\frac{{\bar E}}{{\bar \Lambda}} \right] }\left(
\frac{\tilde{m}^{2}}{g}\chi\left( 0\right) -\chi^{3}\left(
0\right)  \ln\left|  \chi\left(  0\right) \right|
\right).\label{selfconsini}
 \ee
In the large cutoff limit, the above expression becomes cutoff
independent and we conclude that the equation of motion in terms
of the renormalized fermionic condensate is also free of initial
time divergences. Thus either formulation can now be used for a
numerical study with well defined initialization and smooth
evolution throughout with no cutoff dependence in the limit when
the cutoff is taken to infinity.

The resolution of the initial time singularity is now clear: from
equation (\ref{divergentpiece}), it is clear that the initial time
singularity is completely determined by $\chi^{\prime \prime}(0)$
which in turn must be obtained {\em self-consistently} from the
renormalized equation of motion in terms of ${\cal F}$. The
formulation of the equation of motion in terms of ${\cal F}_{AF}$
which vanishes at $t=0$ allows us to find the value of the second
derivative at the origin, the logarithmic singularity is now
encoded in $Z_{\Lambda}$ which leads to an initial value of the
second derivative of the mean field  that is vanishingly small in
the limit of large cutoff.

The solution (\ref{selfconsini}) has a remarkable aspect that
explains how the dynamics manages to be smooth when the amplitude
of the mean field approaches the Landau pole. Consider the initial
value problem in which $\chi(0)$ is {\em very near the Landau
pole}, i.e, $\chi(0)\sim {\bar E}$. In this case we find from
(\ref{selfconsini}) that  \be \mathcal{F}(0) = -\left(
\frac{\tilde{m}^{2}}{g}\chi\left( 0\right) -\chi^{3}\left(
0\right)  \ln\left|  \chi\left(  0\right) \right| \right)
\label{rightonLP} \ee which leads to the conclusion that the
coefficient of the coupling $g_R(e\chi(0))$ in
(\ref{dimlessmeanfield}) actually vanishes. That is to say, the
potential singularity  at the Landau pole is actually rendered
finite by an exact cancellation between the fermionic fluctuation
and the adiabatic effective potential. We will see numerically
below that this remarkable feature is borne out by the dynamics in
all cases.

In the cases below we can solve the mean field equation in the
form given by  eq.(\ref{cutoffEOM}) together with the fermionic
mode equations, simply because the update does not require the
specification of the second derivative and is therefore more
accurate. However after each step in the iteration we have
constructed $\mathcal{F}$ and checked that the value of the second
derivative obtained from both formulations coincide, thus
providing a numerical check of consistency.

Standard numerical techniques (fourth order predictor-corrector
Runge-Kutta ODE solver together with a fourth order Simpson's rule
integrator) are used.  We also compute the fermion occupation
number relative to the initial vacuum state in
each momentum mode as a function of time given by\cite{baackefermions}%
\begin{equation}
n_{q}\left(  \tau\right)  =\left|  \,f_{2\,\vec{q}}\left(  \tau\right)
\,f_{1\vec{q}}^{\ast}\left(  0\right)  -\,f_{1\,\vec{q}}\left(
\tau\right)
\,f_{2\,\vec{q}}^{\ast}\left(  0\right)  \right|  ^{2} \label{fermionnumber}%
\end{equation}

In the  section (\ref{sec:soleqn}) below, we  analyze the
equations governing the system for various values of the
parameters $\tilde{m}^{2},g,\chi_{0}$.

\subsection{Renormalizing $\varrho_{\text{f}}$}\label{sec:renenergy}

Before proceeding to the numerical study of the renormalized equations
of motion we now  turn our attention to the renormalization of the
energy density $\varrho_{\text{f}}$. From eq.(\ref{energydensity}), we
see that a \emph{time derivative} of the mode functions is involved in
computing $\varrho_{\text{f} }$. This has the effect of bringing down
one more power of momentum into the integrand and implies that we need
the WKB expansion of the mode functions to order $k^{-4}$. This entails
both one more iteration of the WKB frequency, eq.(\ref{WKBfreq}), as
well as one more integration by parts on the function $F\left(
t\right)  $. Doing this yields
\begin{gather}
\Im m\left(  f_{1\,\vec{p}}^{\ast}\left(  t\right)  \,\dot{f}_{1\,\vec{p}%
}\left(  t\right)  +f_{2\,\vec{p}}^{\ast}\left(  t\right)  \,\dot{f}%
_{2\,\vec{p}}\left(  t\right)  \right)  _{\text{div}}=-\omega\left(
t\right) -\frac{\dot{M}_{B}^{2}\left(  t\right)  -\dot{M}_{B}^{2}\left(
0\right)
}{8\omega\left(  t\right)  ^{3}}+\label{divergentrho}\\
\frac{\dot{M}_{B}\left(  t\right)  }{4\omega\left(  t\right)
^{3}}\left(
\dot{M}_{B}\left(  t\right)  -\dot{M}_{B}\left(  0\right)  \cos2\int_{0}%
^{t}dt^{\prime}\omega\left(  t^{\prime}\right)  \right)  -\nonumber\\
\frac{\dot{M}_{B}\left(  0\right)  \left(  \dot{M}_{B}\left(  t\right)
-\dot{M}_{B}\left(  0\right)  \right)  }{8\omega\left(  t\right)  ^{3}}%
\sin2\int_{0}^{t}dt^{\prime}\omega\left(  t^{\prime}\right)  .\nonumber
\end{gather}
When the factor of two in eq.(\ref{energydensity}) is included,
the first term above gives the contribution of the zero point
energy densities for a four component Dirac fermion. This would be
the one-loop approximation to the effective
potential\cite{colemanweinberg} if the frequencies were constant.
In this case, we can consider this piece to be an adiabatic
approximation to the effective potential as described above. The
other terms have been grouped so that their contribution vanishes
at the initial time. While the fully renormalized energy density
has a lengthy but not very illuminating expression, we just
highlight its main features. After mass, coupling, wavefunction
renormalization and a substraction of the zero point energy  (time
independent and proportional to the fourth power of the cutoff)
the energy density is finite and {\em conserved} by use of the
fully renormalized equations of motion for the scalar field and
the mode functions. There is no unambiguous separation between the
fermionic and scalar energy density because of the wave function
renormalization, which arises from the fermionic fluctuations but
contributes to the {\em kinetic} energy of the scalar field. The
total energy density is renormalization group invariant, finite
and conserved. Furthermore we have checked numerically in all
cases that the energy density is constant throughout the evolution
to the accuracy required in the numerical implementation, thus
providing an alternative check of the reliability of the numerical
calculation.

\bigskip

\section{Solving the Equations of Motion}\label{sec:soleqn}

We now turn to a discussion of the actual time evolution of the
coupled scalar-fermion system. At this stage we summarize the
discussion of the previous section on renormalization to be able
to focus on the important aspects to be gleaned from the numerical
study:

\begin{itemize}
\item{ The fully renormalized equations of motion for the fermion
field modes (\ref{dimlessfermode})
and for the mean field either in the form given by
(\ref{dimlessmeanfield}) or by (\ref{cutoffEOM}) are free of
initial time singularities or ultraviolet divergences. They can be
consistently initialized and lead to smooth evolution. The two
forms of the mean field equation  are completely equivalent as
they are obtained one from the other by a rearrangement of terms.
While eqn. (\ref{dimlessmeanfield}) seems to suggest a singular
behavior when $\chi$ reaches the Landau pole, the equivalent form
(\ref{cutoffEOM}) suggests smooth and continuous evolution. }

\item{Mass, coupling and wave function renormalization and a
renormalization of the zero point energy (time independent)
renders the energy density finite and conserved as a consequence
of the equations of motion.}
\end{itemize}

We now study in detail different cases to bring the role of the
fermionic fluctuations to the fore.

\subsection{Case 1: $\tilde{m}^{2}\neq 0$ , $\Lambda<\bar{E}$ and
$\left|  \chi_{0}\right|  <\chi_{\text{max}}$}

For this case we will use $\tilde{m}^{2}/g=1,g=0.261$ and look at
two different values of $\chi_{0}:\chi_{0}=0.75,0.2$. For
reference, the location of the maxima of the  effective potential
are at $\left| \chi_{\text{max}}\right|  =1.53$ so for these
values of $\chi_{0}$ we expect $\chi\left(  \tau\right)  $ to
oscillate about the origin. Figure (\ref{fig:chipsivst}) below
displays $\chi(\tau)$ as a function of $\tau$ for $\chi_{0}=0.75$.

\begin{figure}[ht!]
\begin{center}
\epsfig{file=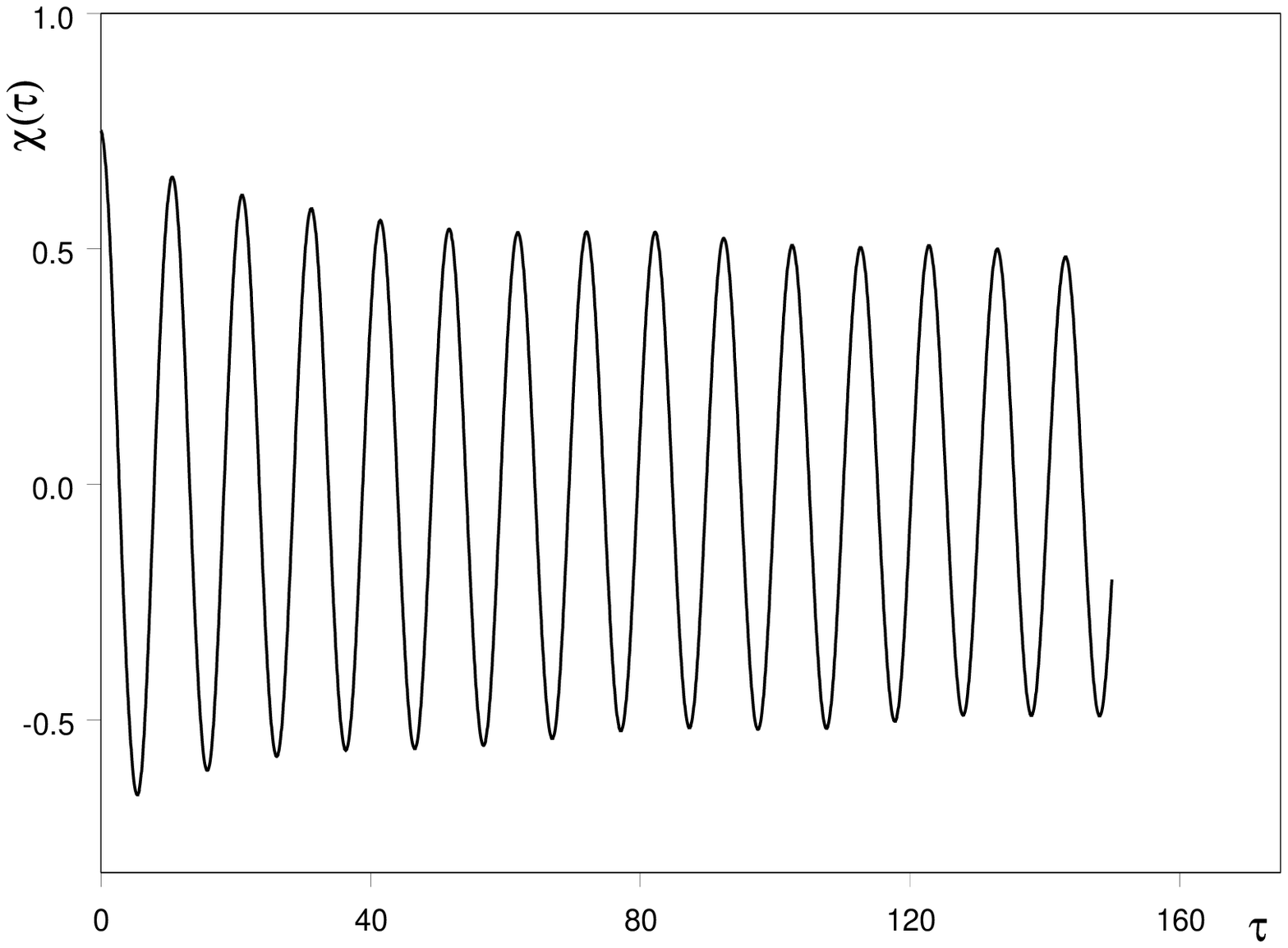,height=3.0in,width=3.0in,keepaspectratio=true}
\epsfig{file=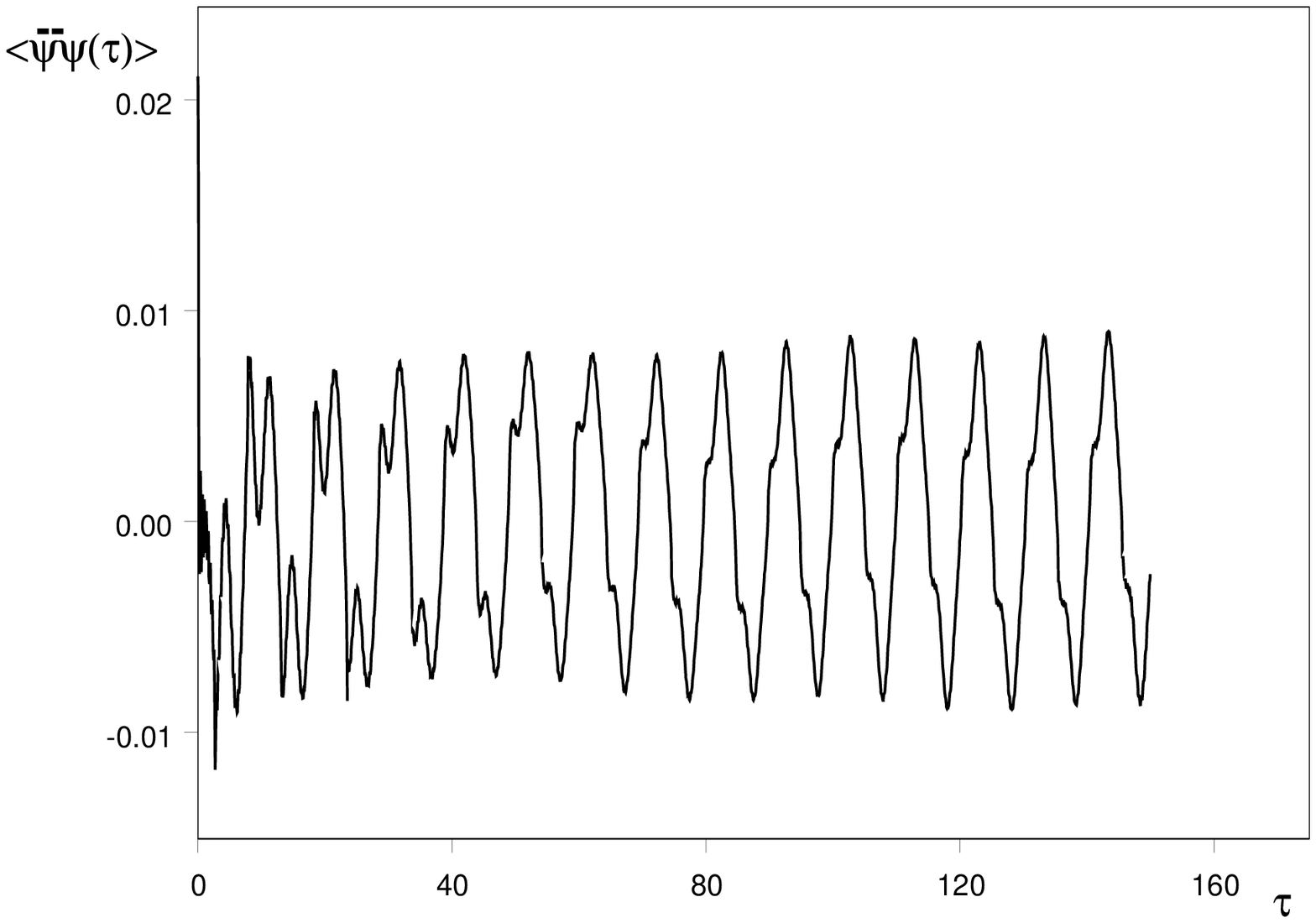,height=3.0in,width=3.0in,keepaspectratio=true}
\caption{$\chi(\tau)$ and $\langle\bar{\psi}\psi(\tau)\rangle_R^{>}$
vs. $\tau$. $\chi(0)=0.75$} \label{fig:chipsivst}
 \end{center}
\end{figure}

\begin{figure}[ht!]
\begin{center}
\includegraphics[height=4.0in,width=4.0in,keepaspectratio=true]{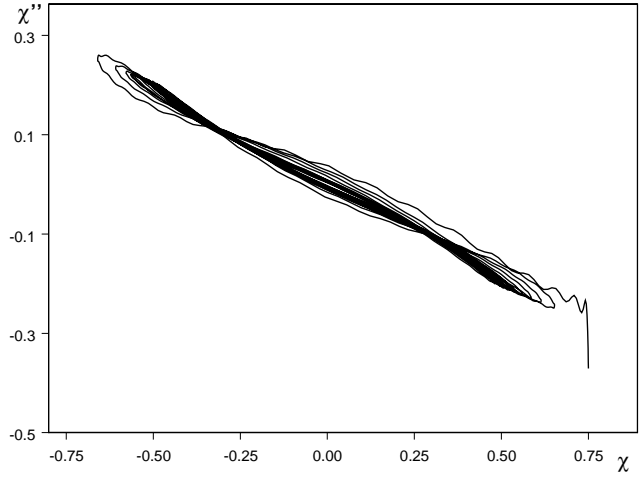}
\caption{$\chi''(\tau)$ vs. $\chi(\tau)$} \label{fig:chippvschi}
\end{center}
\end{figure}

The first figure reveals that the amplitude of the mean field decays
as would be expected; there is energy transfer to fermion particle
production.  These two figures taken together
exhibit an interesting feature.
There is a remarkable similarity at later times
between the mean field and the
oscillations in $\mathcal{F}$. A comparison of the two figures
suggests that the amplitude of the fluctuations is
\emph{proportional} to $\chi$.  Since the mean field has a
decreasing envelope while the fluctuations increase, if such a
proportionality exists, it must involve a coefficient that is
slowly increasing in time. We can actually extract more
information from a parametric plot of $\chi^{\prime\prime}$ versus
$\chi$, which is shown in figure (\ref{fig:chippvschi}).

The fact that such a tight curve is produced is indicative of an
underlying  relationship. In fact, we found that this curve could
be well fit to $a  \chi\left(  \tau\right)  +b \chi^{3}\left(
\tau\right) \ln\left| \chi\left(  \tau\right)  \right|  $ with
$a;b$ very slowly varying over the time scale of the oscillations.
This in turn implies that the fermionic fluctuations
$\mathcal{F}\left( t\right) $ can also be fit to this form, with
coefficients that are slowly varying function of time indicating
that the growth of the fermionic fluctuations results in  a time
dependent correction to the mass term and quartic coupling of the
mean field. This, we believe, is a noteworthy aspect of the
dynamics: the non-equilibrium fermionic fluctuations, those that
were {\em not} accounted for by the adiabatic effective potential,
introduce a slow time dependent  renormalization of the parameters
of the {\em effective potential}, mass and quartic coupling.

There is a new time scale emerging from the dynamics that is
associated with the (slow) time evolution of this renormalization
and the decay of the mean field. A full analysis of these time
scales is beyond the scope of this article, but we expect to use
the methods of dynamical renormalization group \cite{drg} to
investigate the relaxation of the mean field in future work.

We next consider the behavior of the fermion occupation numbers as
a function of time. Figure (\ref{fig:nofk}) below gives three
snapshots of $n\left(  k\right)  $ versus $k$ at different times.

\begin{figure}[ht!]
\begin{center}
\epsfig{file=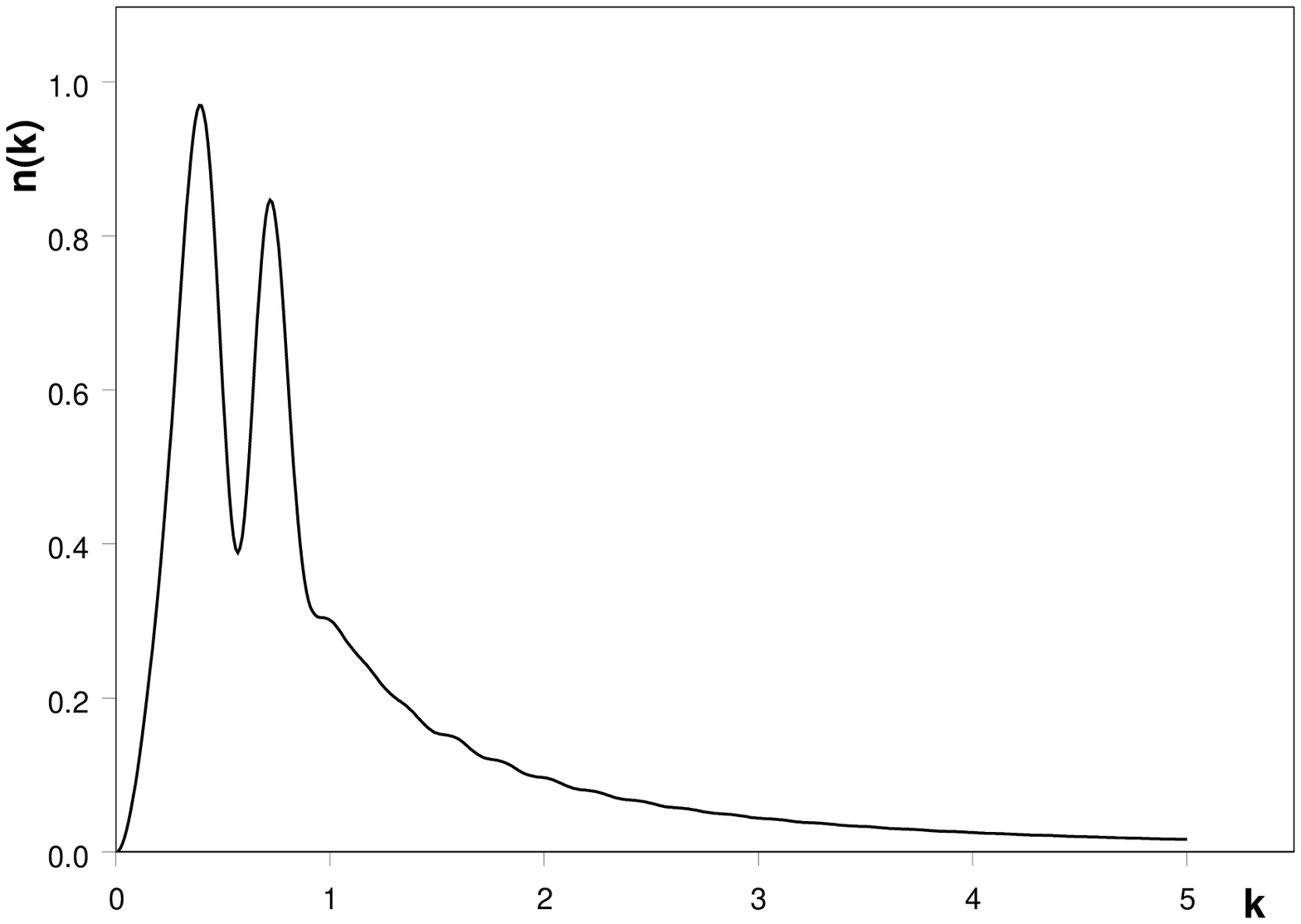,height=2.5in,width=2.5in,keepaspectratio=true}
\epsfig{file=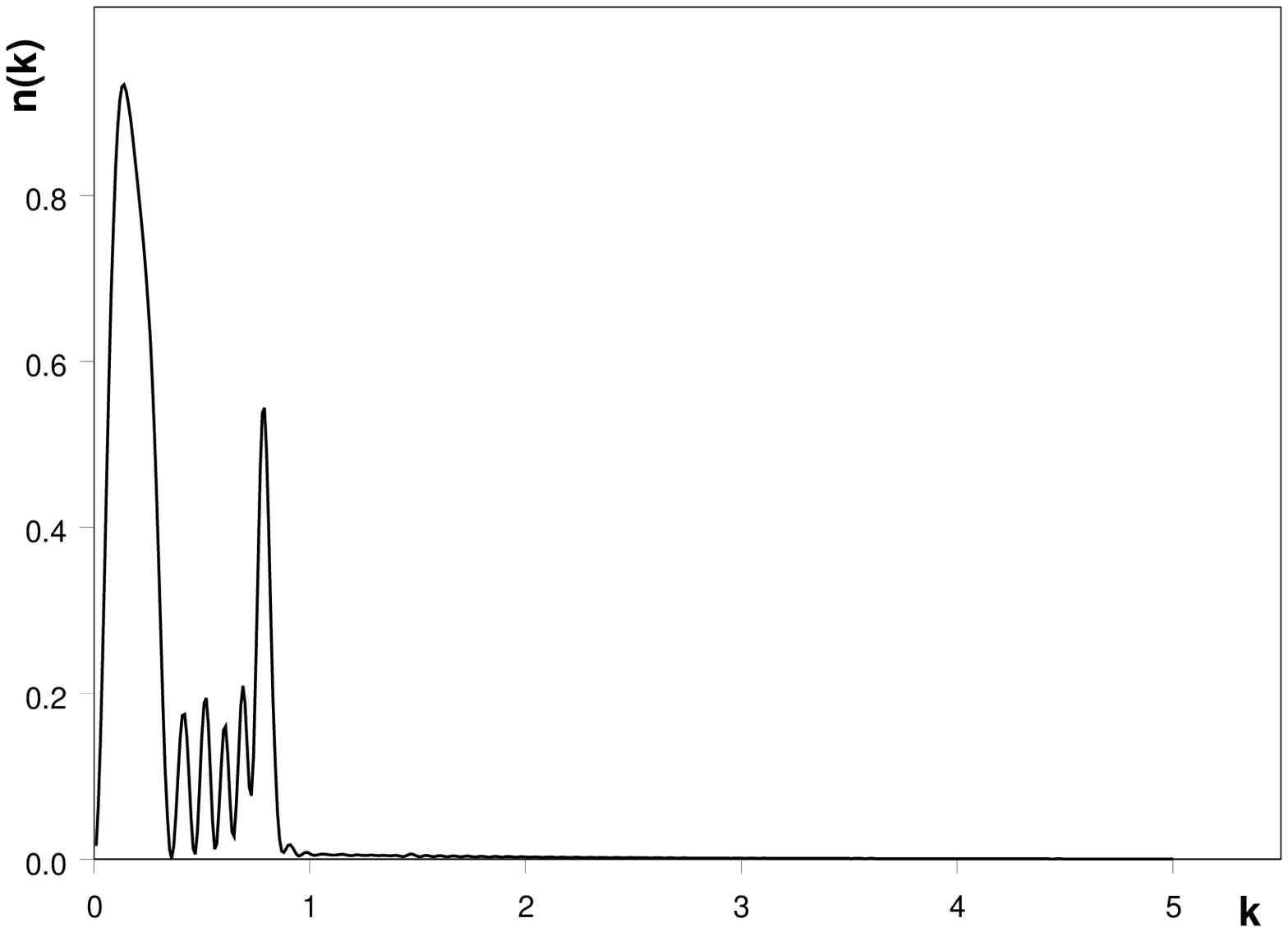,height=2.5in,width=2.5in,keepaspectratio=true}
\epsfig{file=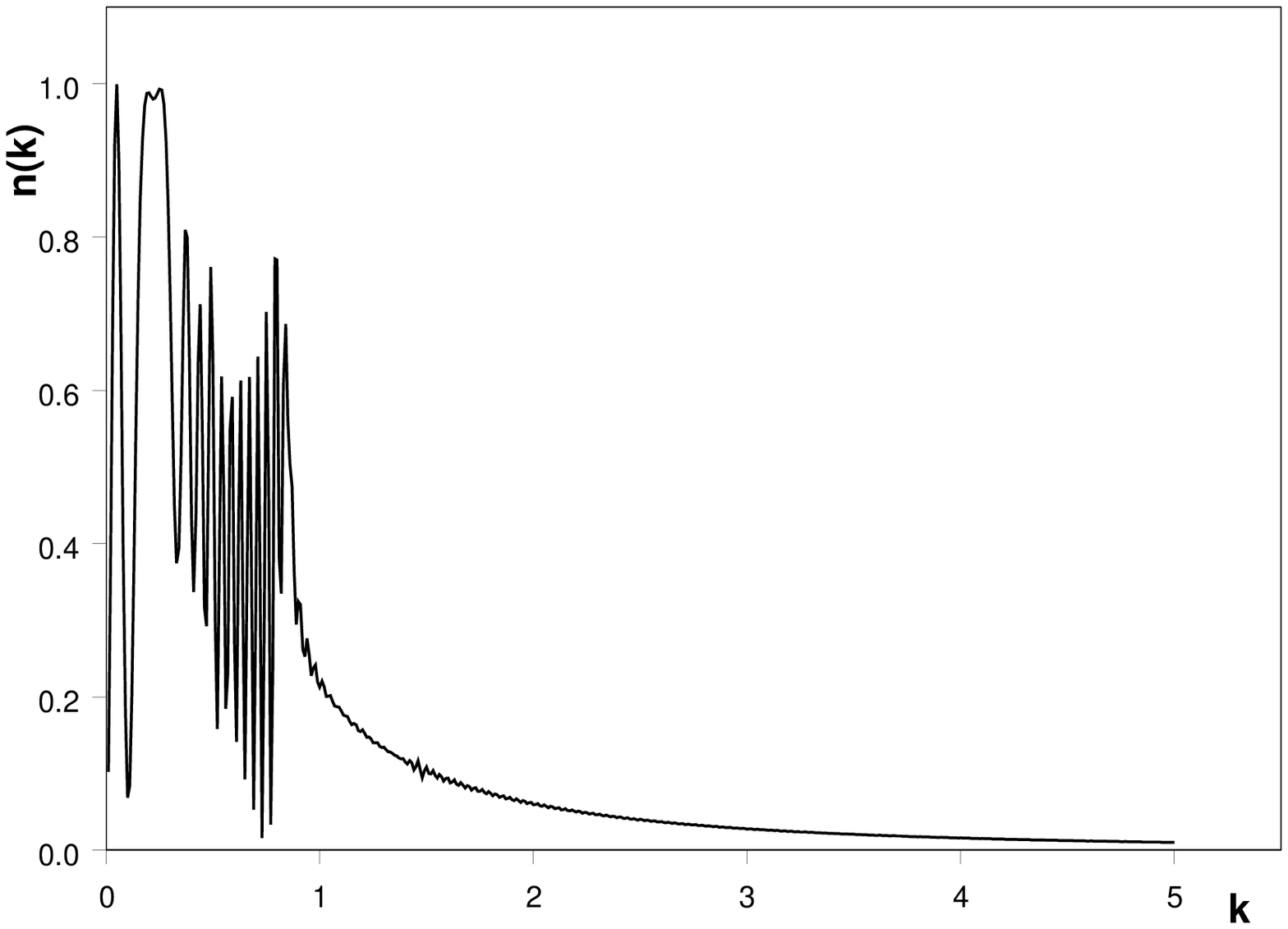,height=2.5in,width=2.5in,keepaspectratio=true}
\caption{$n(k)$ vs. $k$ at times $\tau=15.01;51.05;99.09$
respectively.} \label{fig:nofk}
\end{center}
\end{figure}

What should be apparent from these pictures is the existence of
band structure in the fermion occupation numbers. In fact, the
produced fermions seem to have wavenumbers that lie within a
region spanning $k=0$ to $k=k_{c}$ where $k_{c}\propto \chi_{0}$.
While we are used to band structure in bosonic theories with
parametric resonance, it is unexpected to encounter this structure
in fermionic theories. Pauli blocking prevents exponential
particle production since the occupation numbers can at most
become unity. In section (\ref{sec:pauli}) we study in detail  the
issue of parametric amplification versus Pauli blocking.

Now consider the case where $\chi_{0}=0.20$.

\begin{figure}[ht!]
\begin{center}
\epsfig{file=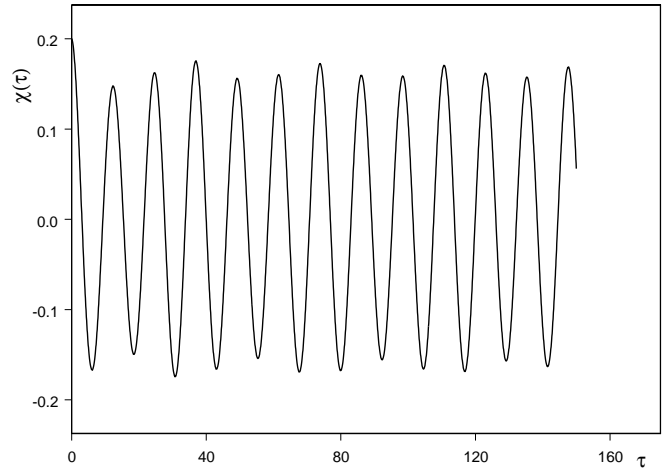,height=4.0in,width=4.0in,keepaspectratio=true}
\caption{$\chi(\tau)$ vs. $\tau$ for $\chi(0)=0.2$. }
\label{fig:chioft}
\end{center}
\end{figure}

\begin{figure}[ht!]
\begin{center}
\epsfig{file=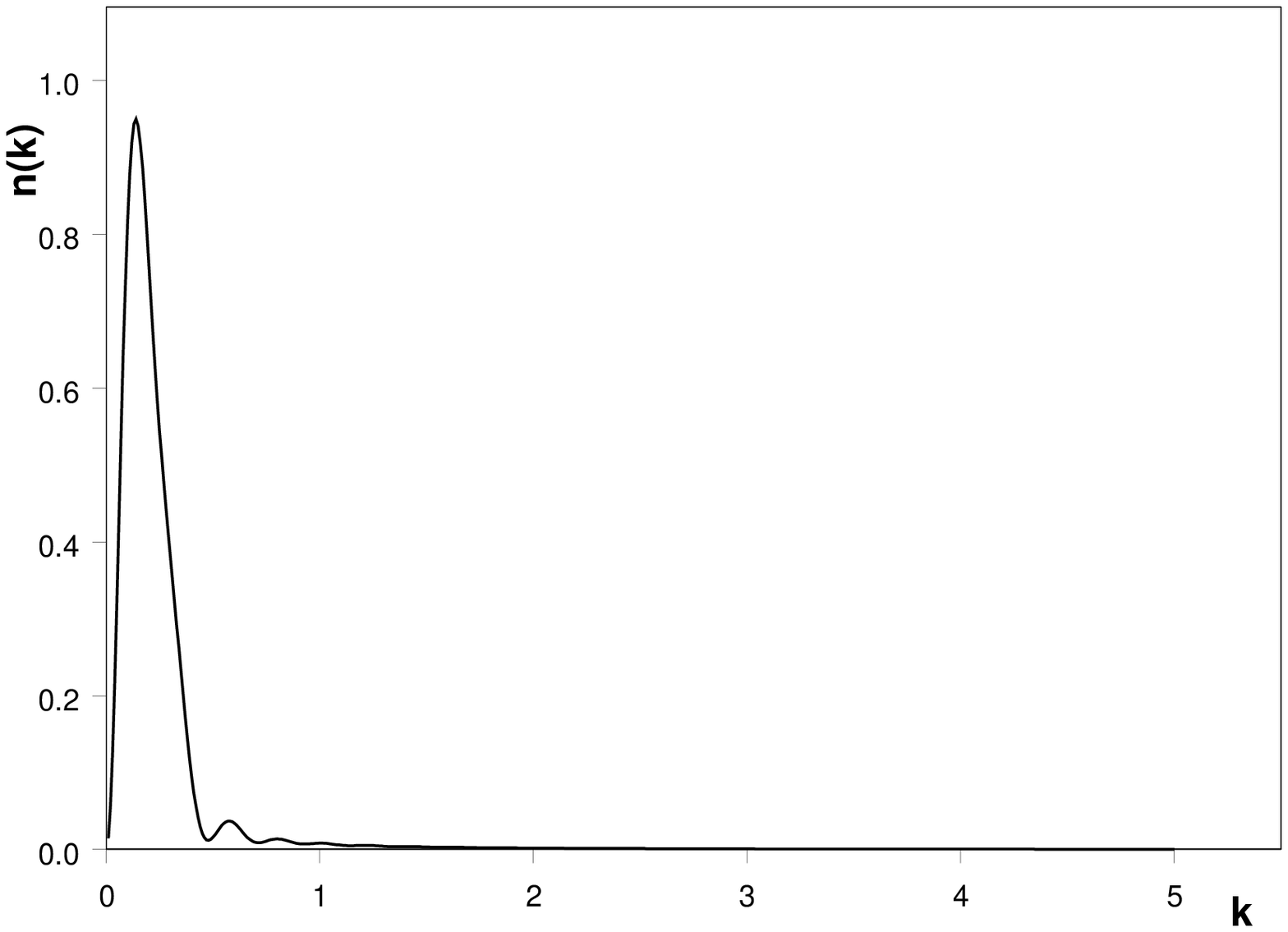,height=2.5in,width=2.5in,keepaspectratio=true}
\epsfig{file=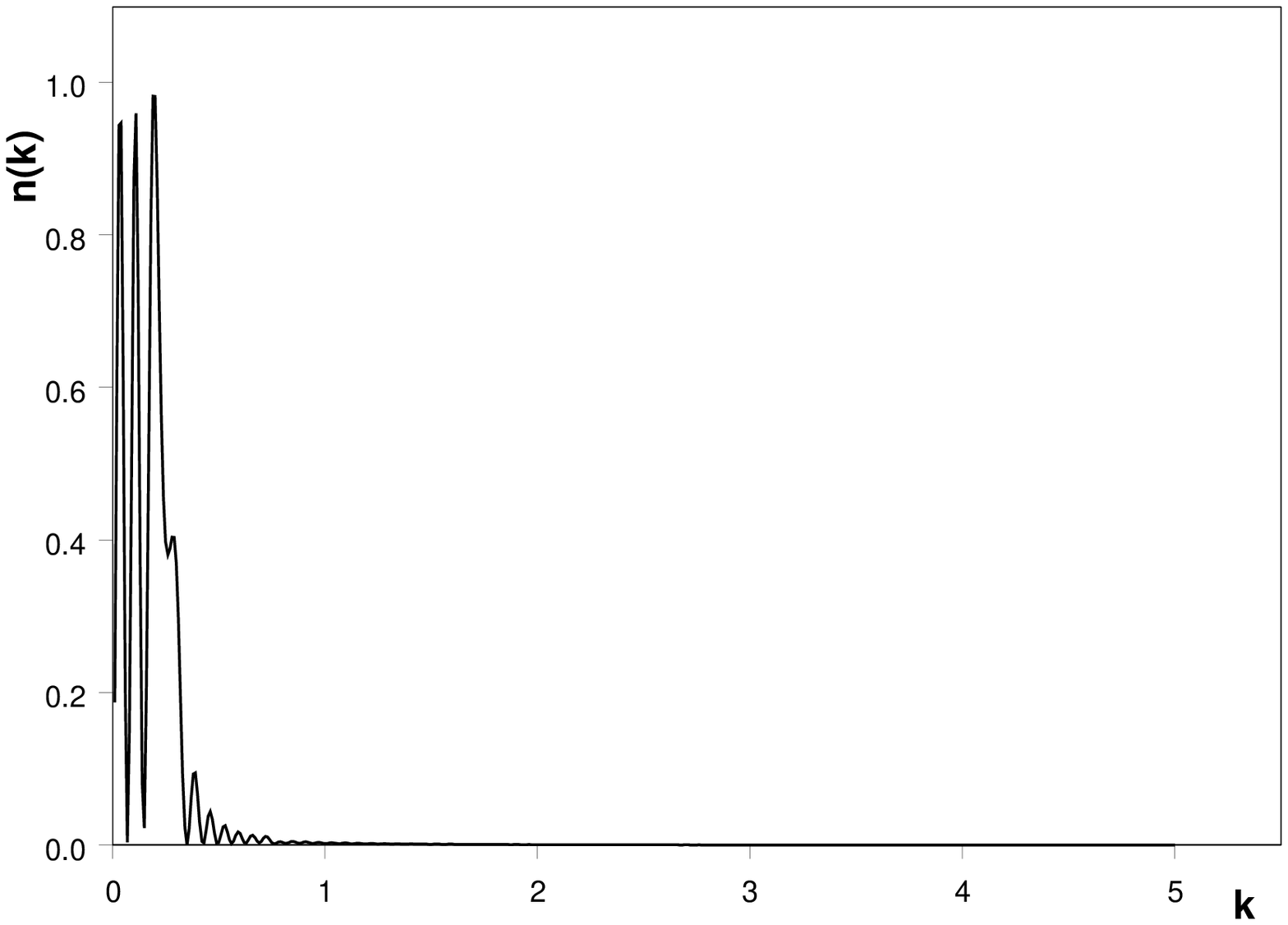,height=2.5in,width=2.5in,keepaspectratio=true}
\epsfig{file=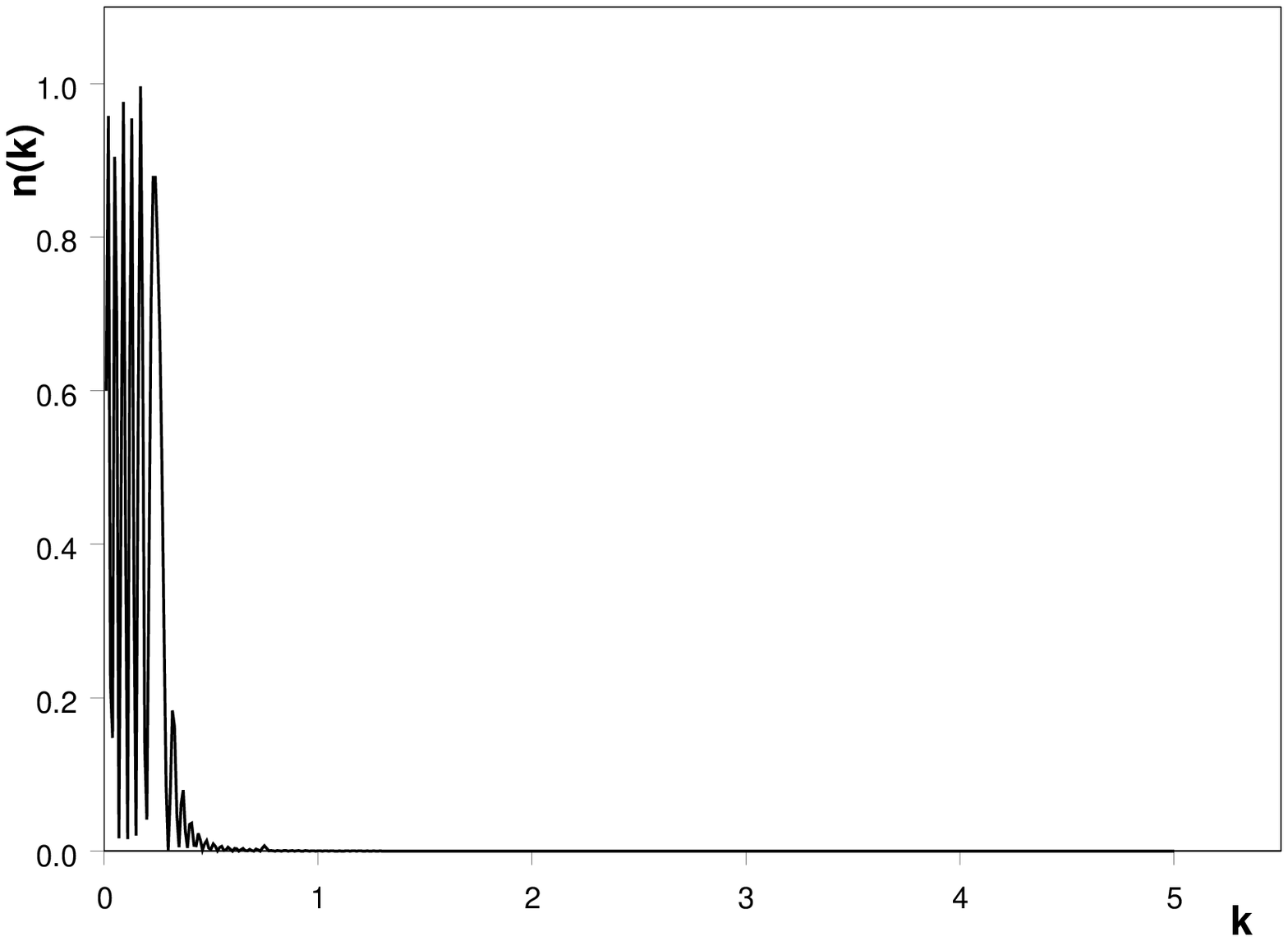,height=2.5in,width=2.5in,keepaspectratio=true}
\caption{$n(k)$ vs. $k$ at times $\tau=15.01;51.05;99.09$
respectively.} \label{fig:nofkosc}
\end{center}
\end{figure}

Figure (\ref{fig:chioft})  shows the phenomenon of ``catalyzed
regeneration'' or ``revival'' first observed in \cite{baackefermions}.
The mean field decays for a while and then regenerates itself. It is
important to note that it never regenerates back to the original value,
always to something less than that. The reason for this can be seen
from the evolution of the momentum distribution below (Figure
(\ref{fig:nofkosc})) which clearly shows an almost saturated
distribution of particles for momenta $k\lesssim \chi_0$.

The allowed band fills up to saturation very early on. After that,
the energy in the scalar cannot be transferred efficiently to
fermions and in fact the fermions, which only couple to the mean
field in our approximation, begin to transfer their energy back to
$\chi$. This depletes the band but not completely, which accounts
for incompleteness of the regeneration.

\subsection{Case 2: $\tilde{m}=0$ , $\Lambda<\bar{E}$ and $\left|
\chi _{0}\right|  <\chi_{\text{max}}$}

The massless case leads to a qualitatively similar dynamics of the
mean field and the fermionic fluctuations but with definite
quantitative differences in the time scale of oscillations damping
of the mean field and growth of the fermion fluctuation as
compared to the massive case. These are shown in
fig.(\ref{fig:chipsivst2}) below.

We note that now the only scale in the problem is completely
determined by the initial value of the mean field $\chi(0)$
whereas in the previous case there were two scales.

\begin{figure}[ht!]
\begin{center}
\epsfig{file=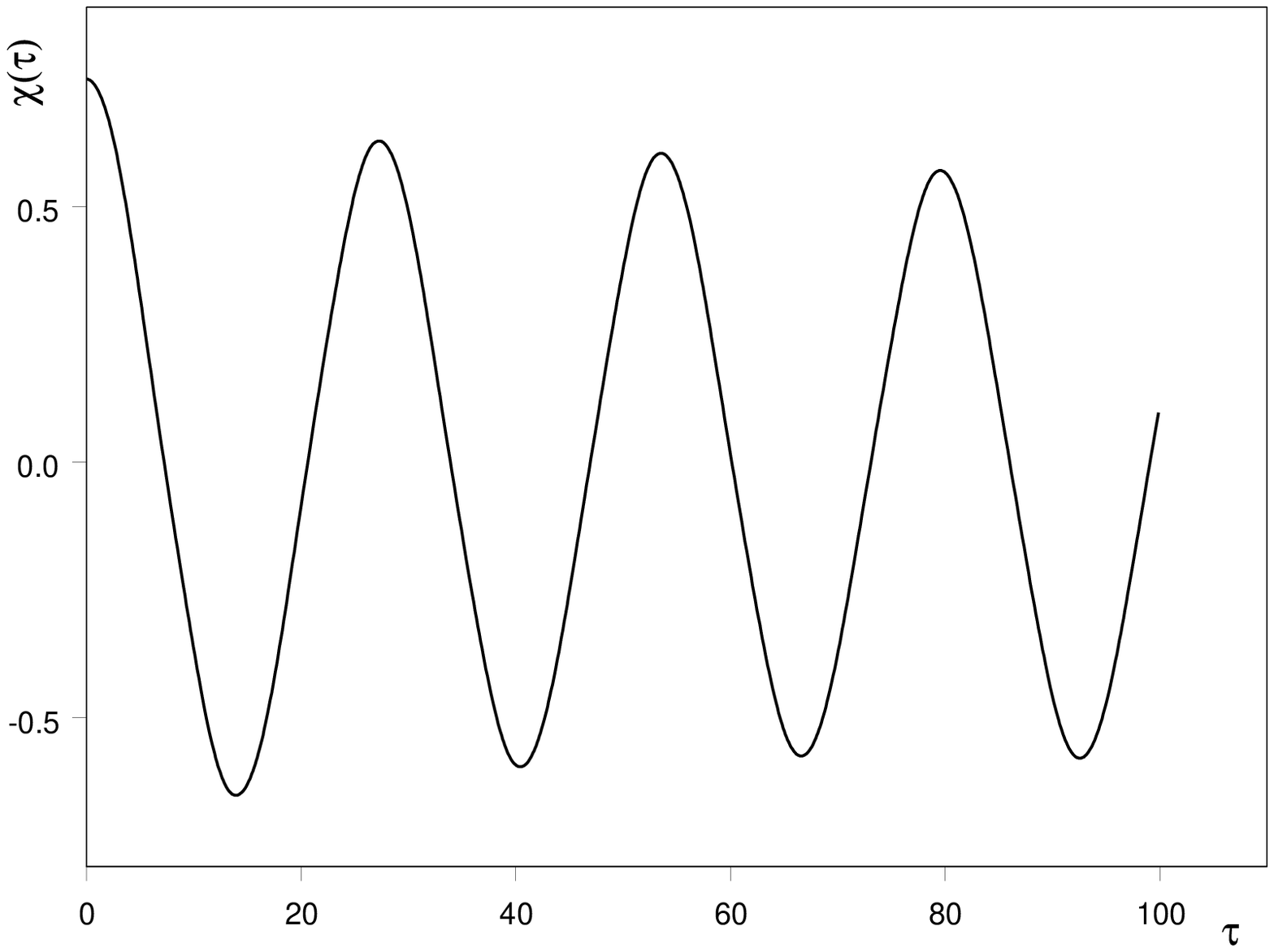,height=3.0in,width=3.0in,keepaspectratio=true}
\epsfig{file=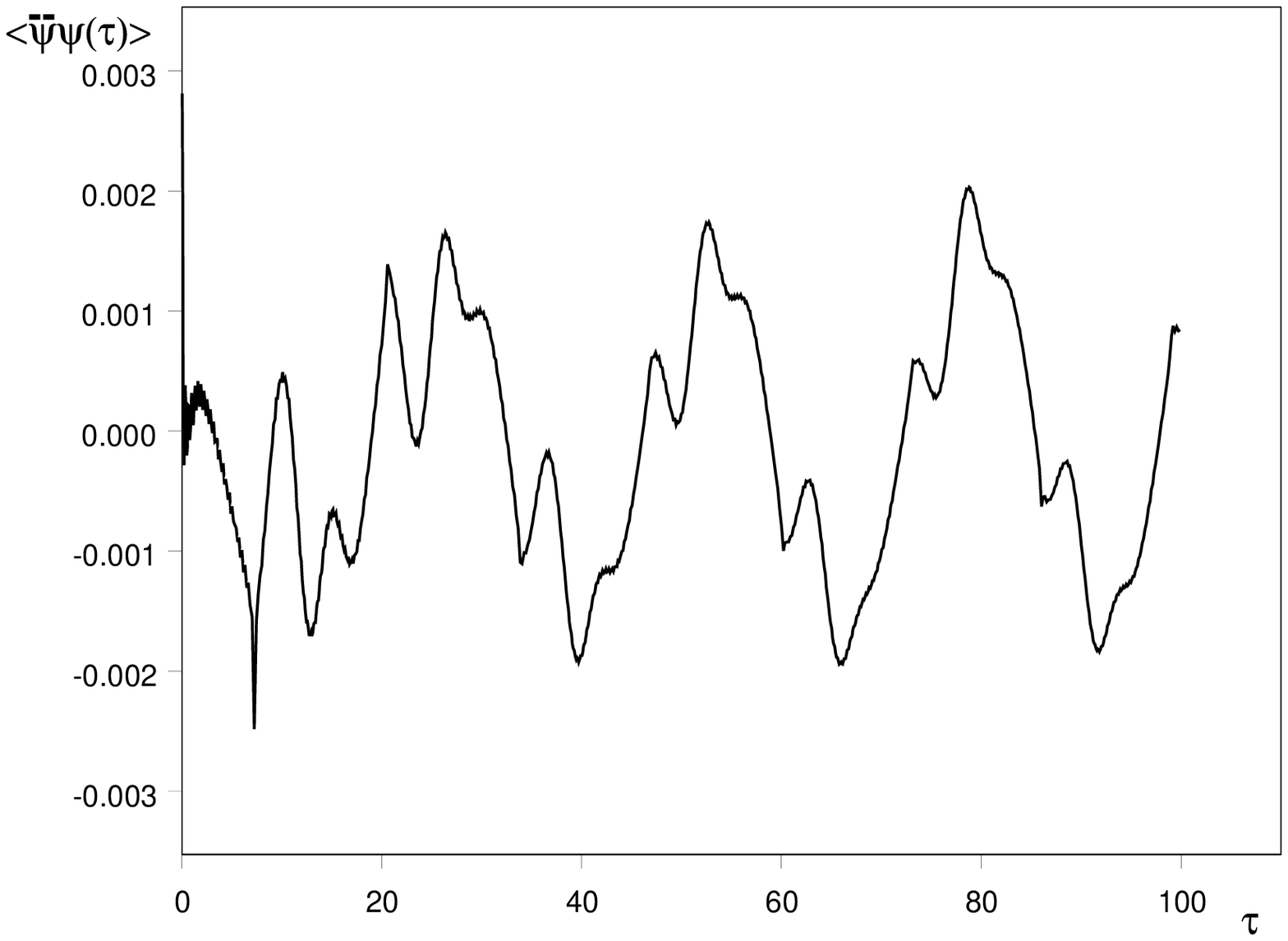,height=3.0in,width=3.0in,keepaspectratio=true}
\caption{$\chi(\tau)$ and $\langle\bar{\psi}\psi(\tau)\rangle_R^{>}$
vs. $\tau$. $\chi(0)=0.75$} \label{fig:chipsivst2}
 \end{center}
\end{figure}

\begin{figure}[ht!]
\begin{center}
\epsfig{file=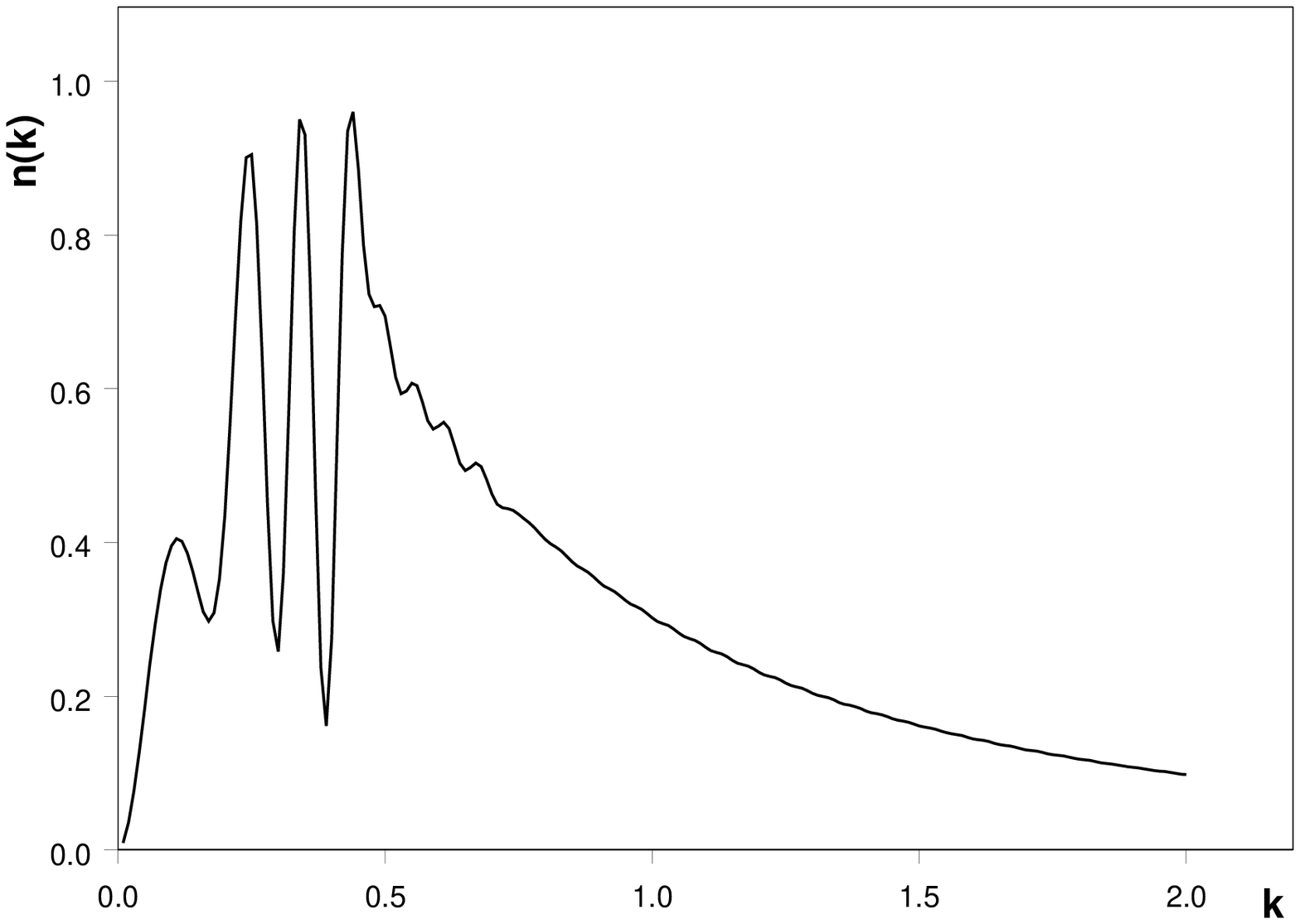,height=2.5in,width=2.5in,keepaspectratio=true}
\epsfig{file=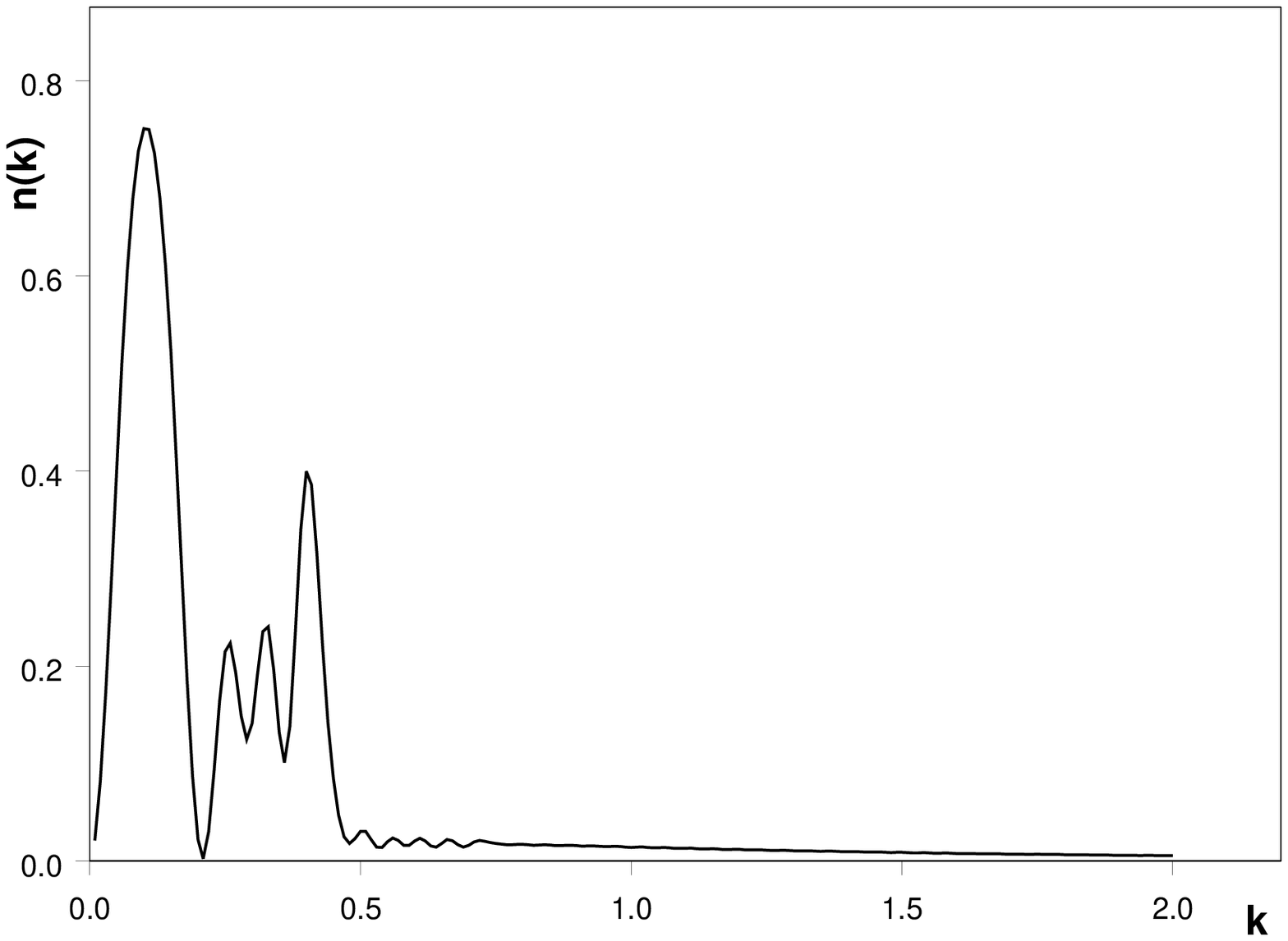,height=2.5in,width=2.5in,keepaspectratio=true}
\caption{$n(k)$ vs. $k$ at times $\tau=67.67;77.20$ respectively.}
\label{fig:nofk3}
\end{center}
\end{figure}

The figures for the dynamics of the mean field and fermion
fluctuations are qualitatively similar to those of the previous
case. Figure (\ref{fig:nofk3}) shows the momentum space
distribution of fermions.

The band structure is still present and is still set by the initial
value $\chi_{0}$. This is as it should be , especially since $\chi_{0}$
is now the only scale in the problem. Finally we again plotted
$\chi^{\prime\prime}$ versus $\chi\;$ with a result identical as that
shown in figure (\ref{fig:chippvschi}). We can again find a good fit to
$a \chi\left( \tau\right) +b \chi^{3}\left( \tau\right) \ln\left|
\chi\left( \tau\right)  \right| $ with the parameters $a,b$ very slowly
varying on the time scale of the oscillations. This fit shows that a
time dependent mass has been generated by the dynamics.
This is not so
surprising as scalar masses are not protected against radiative
corrections.

Again, just as in the previous case we find that new time scales
associated with damping in the amplitude of the mean field and the
renormalization of the parameters emerges.

\subsection{Case 3: $\tilde{m}=0$ , $\Lambda<\bar{E}$ and $\left|
\chi _{0}\right|  >\chi_{\text{max}}$}

\begin{figure}[ht!]
\begin{center}
\epsfig{file=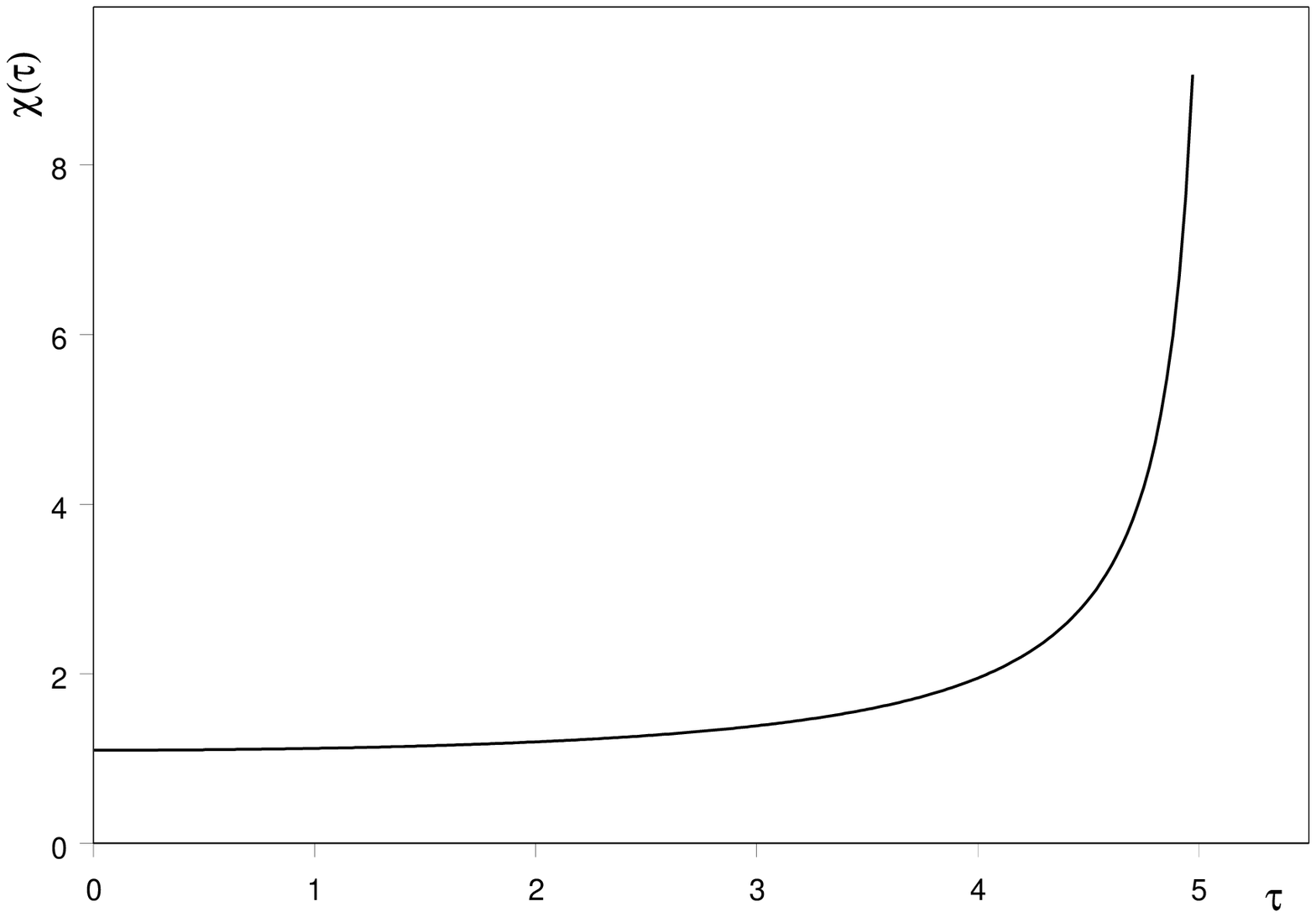,height=3.0in,width=3.0in,keepaspectratio=true}
\epsfig{file=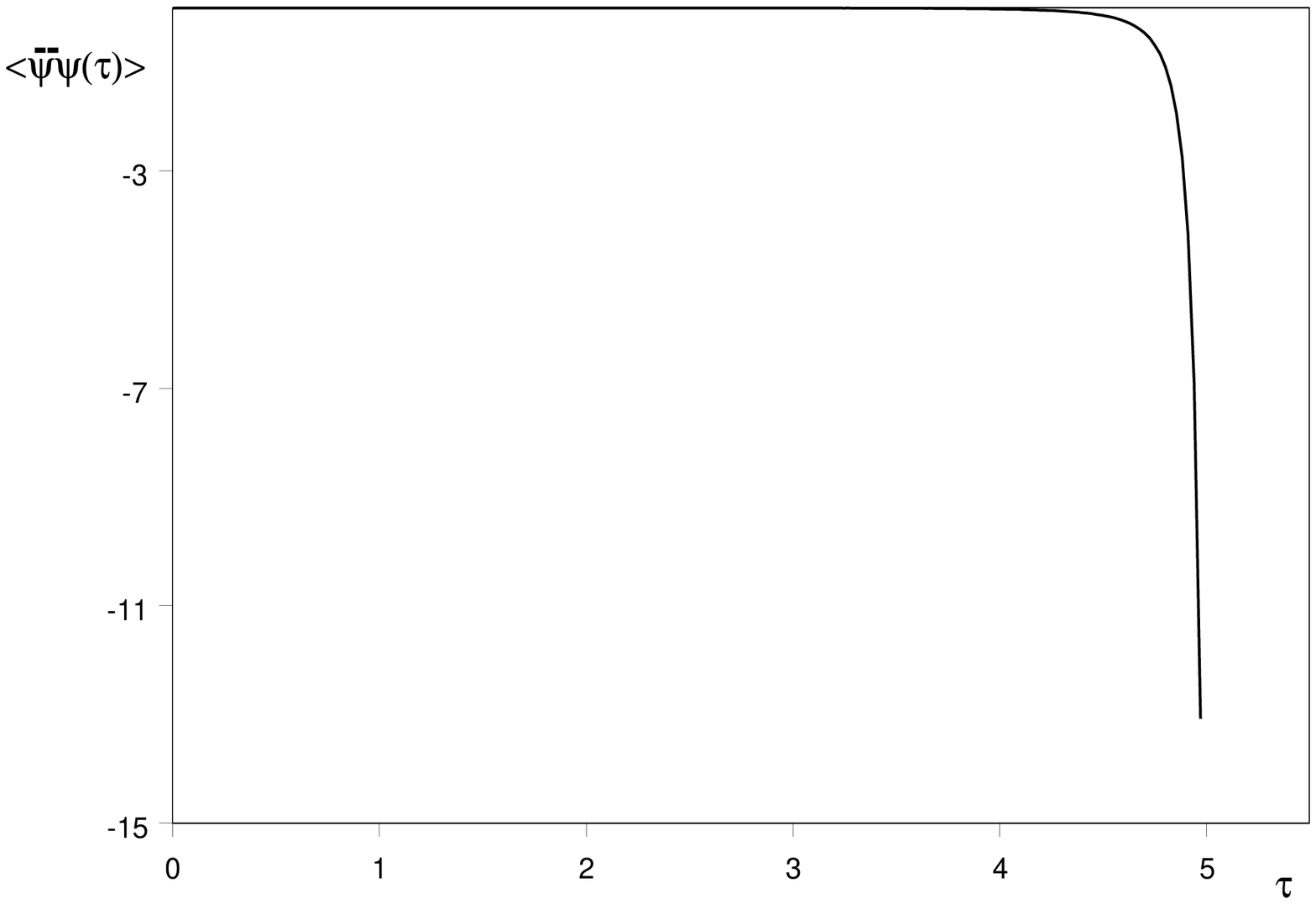,height=3.0in,width=3.0in,keepaspectratio=true}
\caption{$\chi(\tau)$ and $\langle\bar{\psi}\psi(\tau)\rangle_R^{>}$
vs. $\tau$ for $\chi(0)=1.1$. } \label{fig:runaway}
\end{center}
\end{figure}

In this case the mean field begins to the right of the maximum of
the effective potential and rolls down, running away up to the
scale of the cutoff, with a similar behavior for
$\langle\bar{\psi}\psi(\tau)\rangle$.
 This behavior is displayed in figure
(\ref{fig:runaway}). Once the mean field reaches an amplitude of
the order of the cutoff scale there is particle production on
this scale so the momentum integral over the fermionic fluctuations
is no longer accurate.  Therefore the dynamics is no longer
trustworthy and the evolution must be stopped.
In this case the dynamics is what would be expected from the
effective potential description: the mean field runs away as a
consequence of the effective potential being unbounded from below.

\bigskip

\subsection{Case 4: $\tilde{m}=0$ , $\Lambda>\bar{E}$ and $\left|
\chi _{0}\right|  >\chi_{\text{max}}$}

In the previous section we have discussed the possibility that the
presence of a Landau pole in the fully renormalized equations of
motion {\em might not} be a signal of discontinuities or
singularities in the evolution and that the dynamics will be
smooth when the mean field reaches {\em and crosses} the Landau
pole. There are two important hints that led us to this
conclusion: \textbf{a:}) the Landau pole does not appear
explicitly in the equation of motion written in terms of
$\mathcal{F}_{AF}$ (\ref{cutoffEOM}) (the logarithmic
divergence in $\mathcal{F}_{AF}$ is compensated by the logarithm
in $Z_{\Lambda}$).
\textbf{b:}) The self-consistent solution for
$\chi^{\prime\prime}(0)$ and for $\mathcal{F}(0)$ given by
(\ref{rightonLP}) when $\chi(0)$ is very near the Landau pole,
shows manifestly that the fermionic fluctuations cancel exactly
the contribution from the effective potential thus making the
``residue'' at the Landau pole vanish exactly. While this
remarkable cancellation has been gleaned in a particular case,
that in which the initial condition places the amplitude of the
mean field at the Landau pole, the combination of both arguments
are suggestive enough to conjecture that the dynamics will be
smooth in {\em all cases} as $\chi(\tau)$ crosses the Landau pole.

In order to probe this conjecture, we need to consider the case
$\Lambda \gg \bar{E}$ so that $\chi$ can can evolve past the
Landau pole but the dynamics should be reliable in such a way that
the amplitude of the mean field
must always be much smaller than the cutoff.

For $\Lambda \gg \bar{E}$ we see from the formulation
(\ref{cutoffEOM}) that $Z_{\Lambda} <0$ thus for very early time
when $\mathcal{F}_{AF} \sim 0$  the acceleration is negative and
the mean field climbs the potential hill instead of rolling down.
This a consequence of the fact that for a theory with a Landau
pole, if the cutoff is taken much larger than the Landau pole, the
``bare'' coupling becomes negative but very small. Thus in the
``almost finite'' formulation the only hint of the presence of a
Landau pole is in the opposite sign of the second derivative of
the mean field as compared to the case in which the cutoff is well
below the Landau pole.

Thus initially we expect that the second derivative is very small and
negative, the mean field will slowly {\em climb} up the potential hill
and the fermionic fluctuations will grow.

In what follows we study the cases with  $g=1.0;\chi _{0}=1.2, \Lambda
=200; E_L=2.7$ and $g=0.5;\chi _{0}=1.2, \Lambda =500; E_L=20.1$
respectively, to illustrate the main features of the dynamics.

\begin{figure}[ht!]
\begin{center}
\epsfig{file=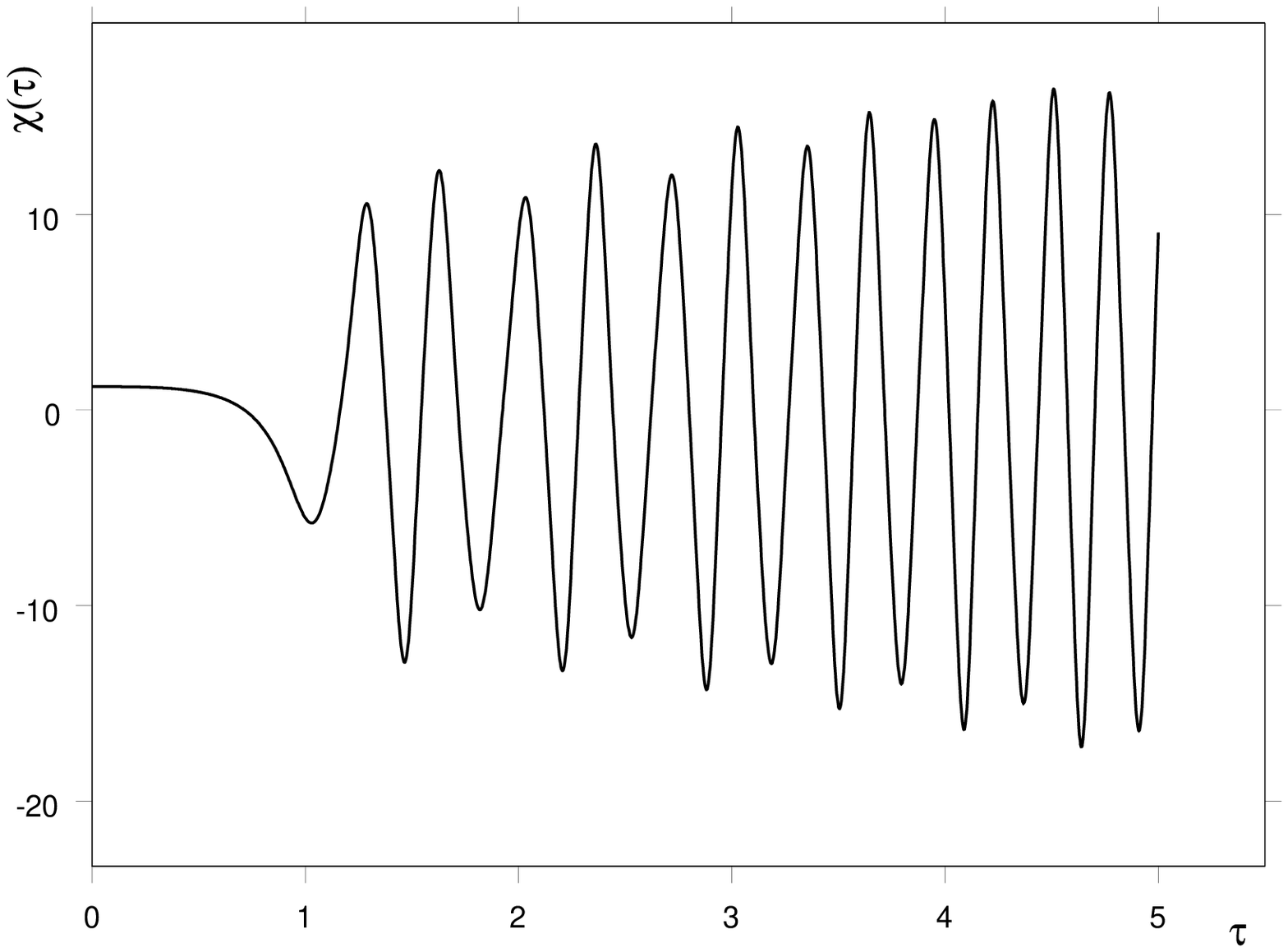,height=3in,width=3in,keepaspectratio=true}
\epsfig{file=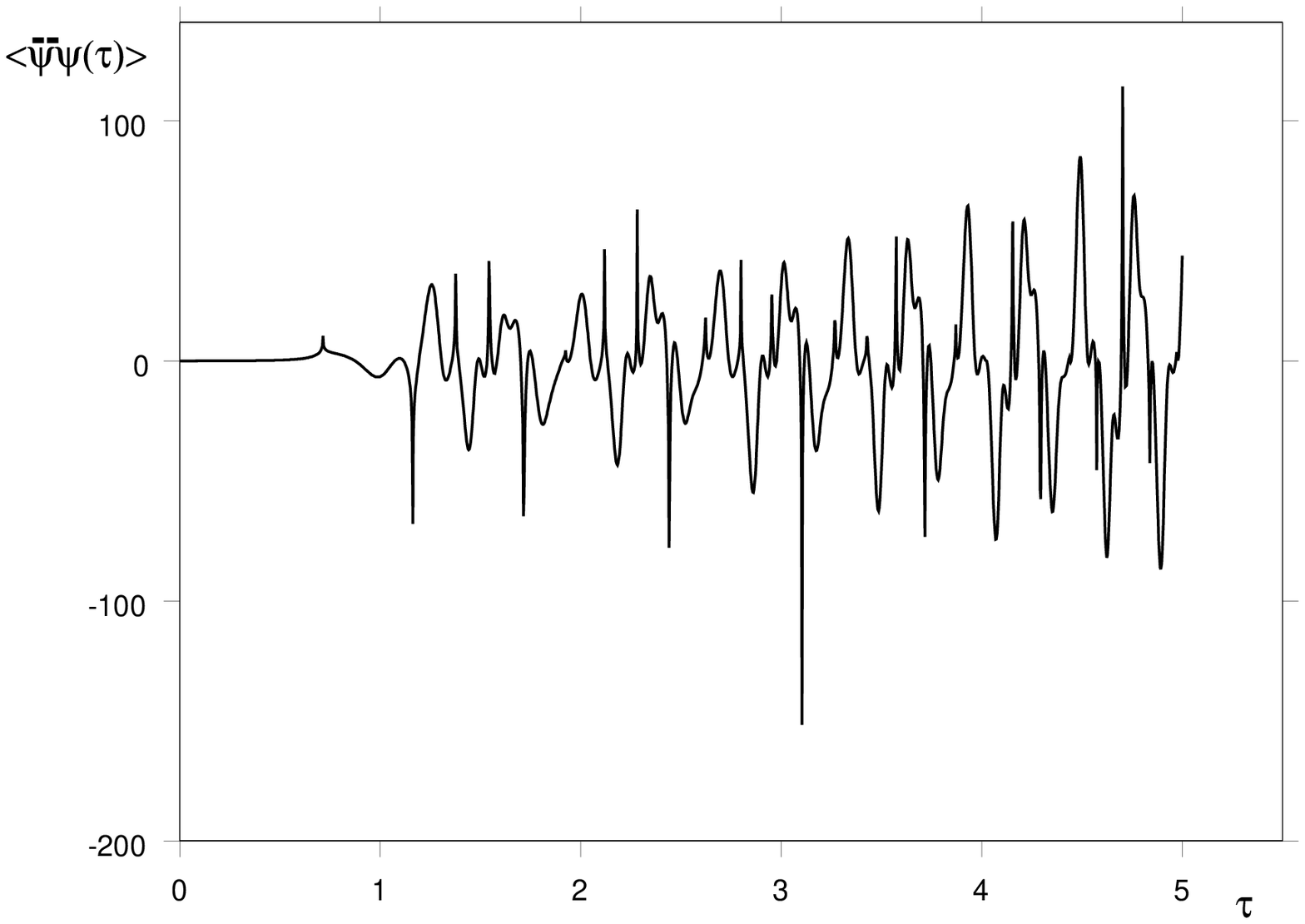,height=3in,width=3in,keepaspectratio=true}
\caption{$\chi(\tau)$ and $\langle\bar{\psi}\psi(\tau)\rangle_R^{>}$ vs.
$\tau$ for $\chi(0)=1.2; g=1.0; E_L =2.7$ } \label{fig:chipsibeyLP}
\end{center}
\end{figure}

\begin{figure}[ht!]
\begin{center}
\epsfig{file=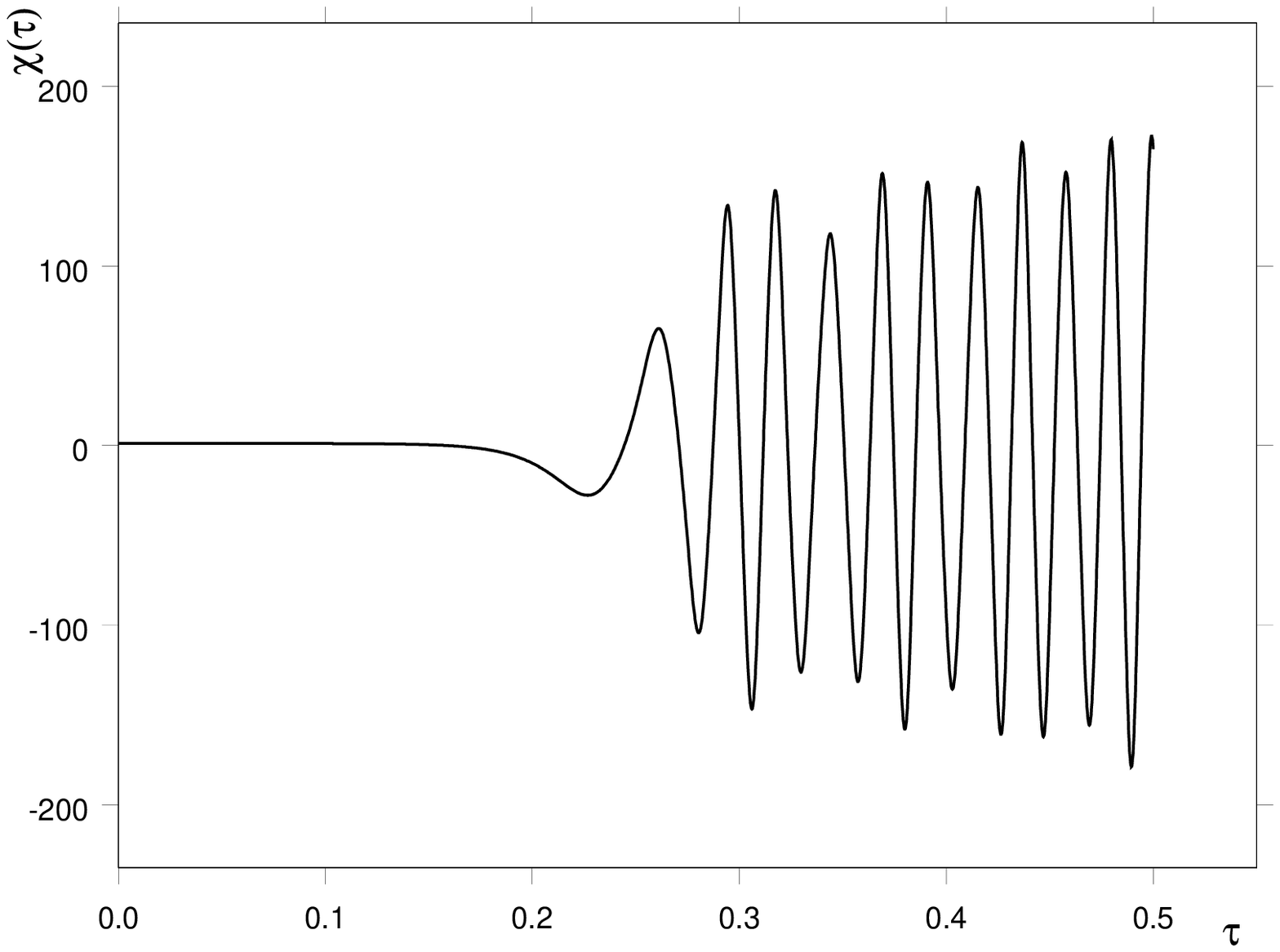,height=3in,width=3in,keepaspectratio=true}
\epsfig{file=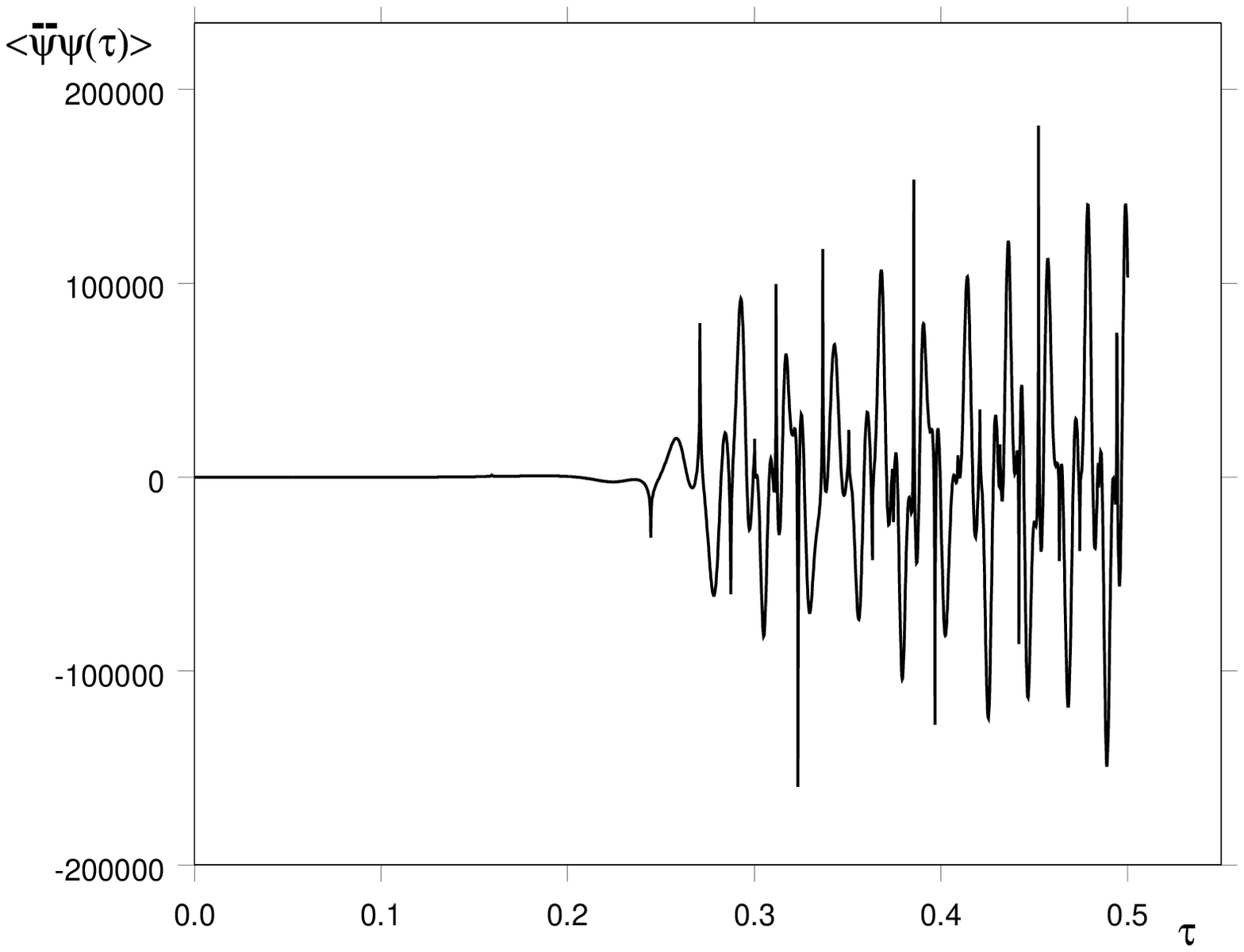,height=3in,width=3in,keepaspectratio=true}
\caption{$\chi(\tau)$ and $\langle\bar{\psi}\psi(\tau)\rangle_R^{>}$ vs.
$\tau$ for $\chi(0)=1.2; g=0.5; E_L =20.1$ } \label{fig:chipsibeyLP2}
\end{center}
\end{figure}

The full dynamics for the mean field and the renormalized chiral
condensate in these cases is displayed in figures
(\ref{fig:chipsibeyLP},\ref{fig:chipsibeyLP2}) for the mean field  and
the fully renormalized fermion fluctuation
$\langle\bar{\psi}\psi(\tau)\rangle_R^{>}$.

These figures show a remarkable behavior: $\chi$ initially climbs
up the potential, reaches the maximum and begins falling down
towards the mininum, overshoots, climbing up to the maximum on the
negative side and finally begins a plunge down the potential hill
on the negative side. As the amplitude of the mean field reaches
the Landau pole at a time $\tau_{LP}$ the (fully renormalized)
fermion condensate $\mathcal{F}(\tau)$ \emph{exactly} compensates
the contribution of the effective potential thus cancelling the
singularity of the running coupling at the position of the Landau
pole.

\begin{figure}[ht!]
\begin{center}
\epsfig{file=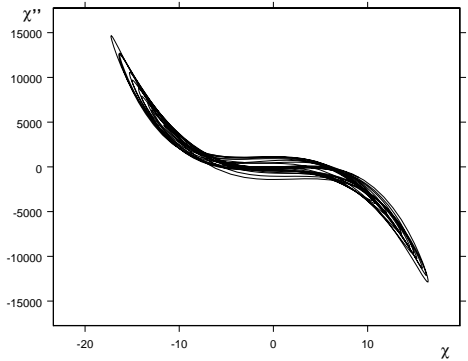,height=3in,width=3in,keepaspectratio=true}
\caption{$\chi^{\prime \prime}$ vs. $\chi$ parametrically for both
cases.} \label{fig:chippchibeyLP}
\end{center}
\end{figure}

When the amplitude becomes larger than $E_{L}$ the mean field
begins an {\em oscillatory} motion  about the origin with greater
and greater amplitude. The key to understanding this behavior lies
in the figure (\ref{fig:chippchibeyLP}) which depicts
$\chi^{\prime \prime}$ vs $\chi$ parametrically. Note that the two
different cases we are examining  are almost indistinguishable.

 A numerical fit to an excellent
accuracy reveals that   for $\tau >> \tau_{LP}$

\be \chi^{\prime \prime}(\tau) \sim b \chi^3(\tau) ~~;~~\label{chi3fit}
\ee

\noindent with $b$ slowly varying in time and saturating at long times
at a value $b \lesssim -2.0$.

Thus for $\tau >> \tau_{LP}$ the fermion condensate provides a
\emph{small} renormalization of the coefficient of $\chi^3 \ln|\chi|$.
Therefore when the amplitude of the mean field is much
larger than $E_L$, the $  \mathcal{F}\left(  \tau\right)  $ can be
neglected, the logs cancel, and 
the equation of motion takes the form:%
\begin{gather}
\chi^{\prime\prime}\left(  \tau\right)  +\frac{2}{2/g-\ln\left|  e\chi\left(
\tau\right)  \right|  }\left(  -\chi^{3}\left(  \tau\right)  \ln\left|
\chi\left(  \tau\right)  \right|  +\mathcal{F}\left(  \tau\right)  \right)
=0\text{ or}\nonumber\\
\chi^{\prime\prime}\left(  \tau\right)  \simeq b \chi^{3}\left(
\tau\right)~~;~~ \textmd{with} ~b \lesssim -2.0
 \label{openingup}%
\end{gather}

This shows that the mean field behaves at late time as if it were in a
\emph{quartic} potential; this would also be true in the massive case
since the mass term would eventually become subdominant. The
``potential'' appears to ``open up'' with time i.e. the coefficient of
the quartic term becomes smaller and smaller as the log in the
denominator dominates until it saturates.

\begin{figure}[ht!]
\begin{center}
\epsfig{file=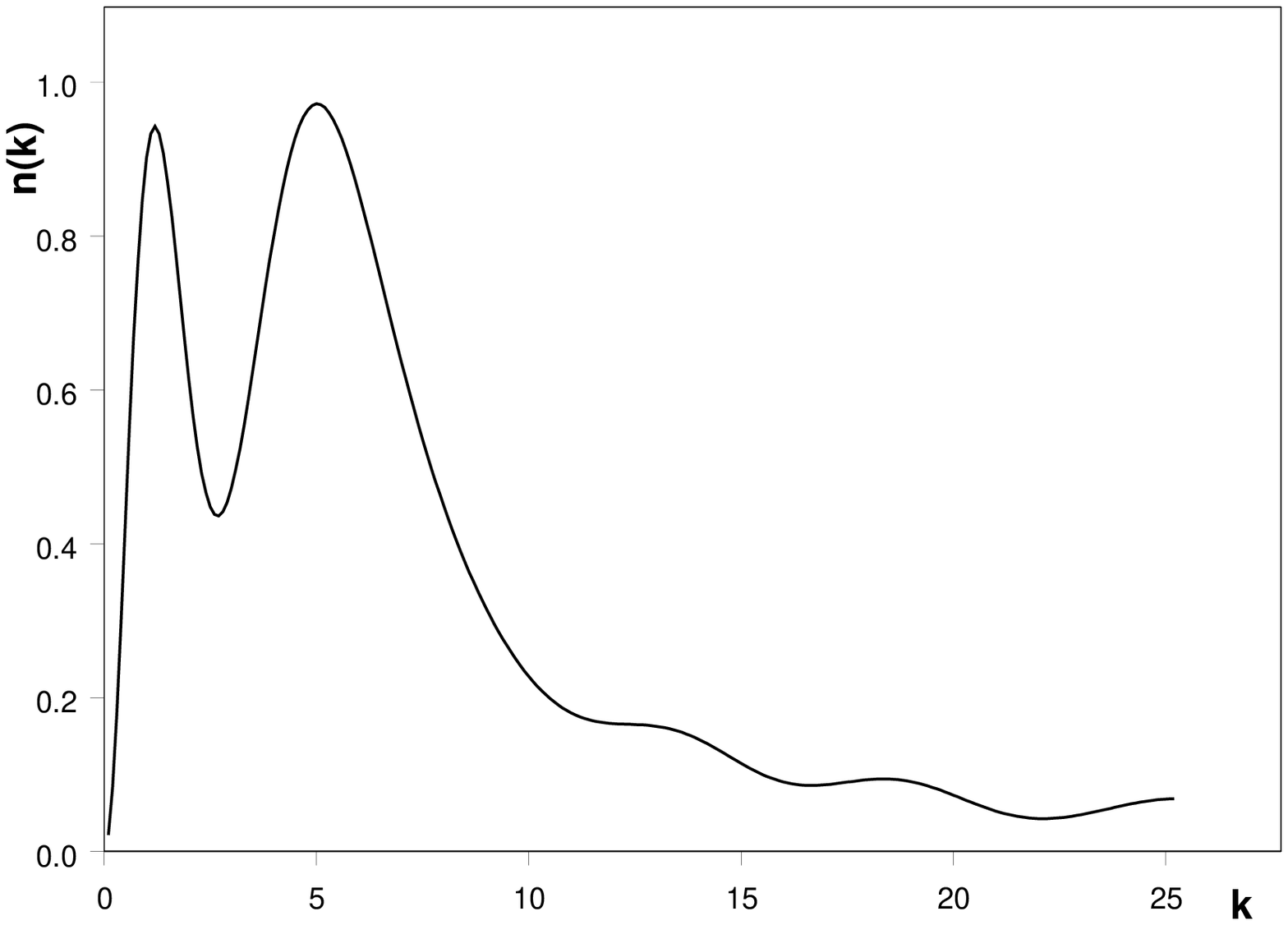,height=3in,width=3in,keepaspectratio=true}
\epsfig{file=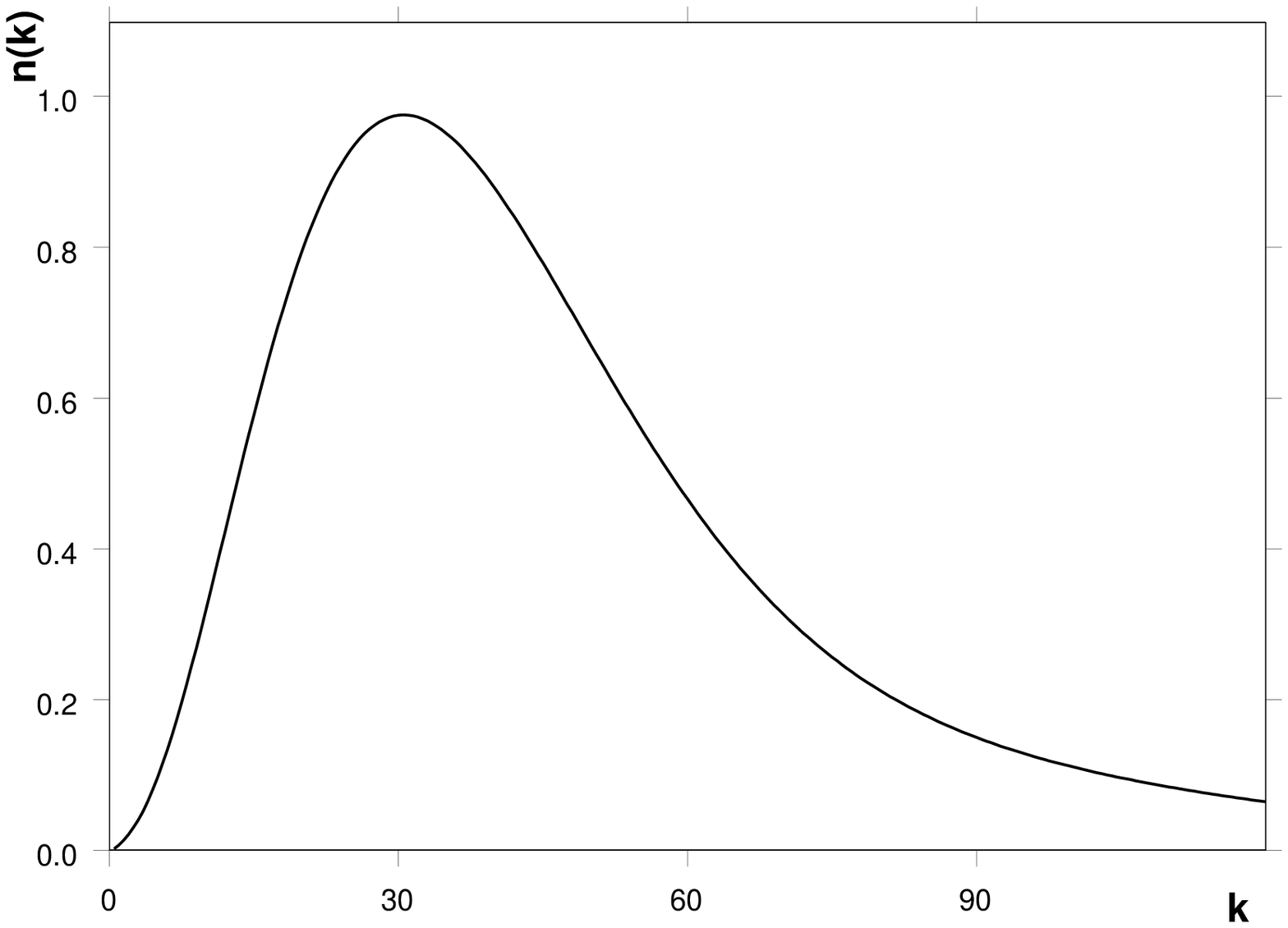,height=3in,width=3in,keepaspectratio=true}
\caption{$n(k)$ vs  $k$. Left figure is  for  $\tau=1.89,
g=1.0,E_L=2.7$, the right figure for $\tau=0.35,g=0.5,E_L=20.1$.}
\label{fig:nofkLP}
\end{center}
\end{figure}

Suppose that the initial value of the mean field were near the
maximum and that the coupling is such that $E_{L} >>1$. In this
case from the self-consistent solution (\ref{selfconsini}) we see
that $\mathcal{F} \sim \mathcal{O}(1)$ (for $\Lambda >> E_{L}$).
However we have argued and seen explicitly numerically that as the
mean field approaches and crosses the Landau pole, the fermion
condensate $\mathcal{F}(\tau)$ cancels the term $\chi^3(\tau)
\ln|\chi(\tau)|$.This implies that the fermion condensate becomes
of $\mathcal{O}(E_{L}^3)$. Since the renormalized fermionic
condensate is obtained by subtracting the terms from the effective
potential and the wave function renormalization, the only manner
in which this can actually happen, roughly speaking, is that the mode functions
$f_{1,2}$ evolve in time to almost saturation, i.e, that
$|f_{1,2}|^2 \sim 1$ for wave vectors up to the Landau pole scale
$\sim e^{2/g-1}$ (or beyond).  This will
lead to a fermion condensate of order $E^3_{L}$. This argument is
shown to be correct by figure (\ref{fig:nofkLP}) which displays
the occupation numbers of the produced fermions  as a function of
momentum for large $\tau$ when the mean field  oscillates.

Comparing the width of the band of occupied wavevectors with the
amplitude of the mean field in the oscillatory phase (see figures
\ref{fig:chipsibeyLP},\ref{fig:chipsibeyLP2})  reveals that the
width of the band is proportional to the amplitude of the mean
field, which, in turn is larger than the position of the Landau
pole. Thus states with wavevectors up to $E_L$ (or somewhat larger
as shown by fig.(\ref{fig:nofkLP})) are almost saturated with
occupation one (per spin). We interpret this as the formation of a
very dense fermionic plasma with a ``chemical potential'' of the
order of the Landau pole, since all states up to this scale are
filled by produced particles.

\textbf{Summary of the numerical analysis:} we want to highlight
several noteworthy features that emerge from the numerical
analysis: \textbf{a:} as presaged by the discussion above, the
mean field evolution is completely non-singular, even as it
crosses the Landau pole. The avoidance of singular behavior is
entirely due to the dynamical growth of the fermionic
fluctuations: they ensure that the quantity $-\chi^{3}\left(
\tau\right)  \ln\left|  \chi\left( \tau\right)  \right|
+\mathcal{F}\left(  \tau\right)  $ vanishes when $2/g-\ln\left|
e\chi\left( \tau\right)  \right|  $ does. \textbf{b:} The
backreaction of the fermions provide a small renormalization to
the effective potential after the mean field 
crosses the Landau pole, and as a result, the {\em
true} effective potential at large values of $\chi$ is upright and
\emph{quartic}.  The equation of motion is then of the approximate
form $\chi^{\prime \prime} \sim -2 \chi^3$ where the factor $-2$
receives small corrections from the fermion backreaction.
\textbf{c:} The non-equilibrium evolution
results in profuse particle production. The fermion occupation
number saturates generally up to momenta larger than  the
Landau pole, which in turn results in a very dense medium akin to
a fermi gas with chemical potential of the order of the Landau
pole.

\section{Pauli blocking vs. parametric amplification:}\label{sec:pauli}
An important question that we address in this section is the role
of parametric amplification in a fermionic theory.

In a {\em bosonic} theory, when the expectation value of the
scalar field oscillates around a minimum the fluctuation mode
functions obey an equation of motion with an effective oscillatory
mass squared term. The resulting Schrodinger-like equation for the
mode functions results in a spectrum with forbidden bands with
positions that depend on wave vector $k$ of the mode function and
the specific details of the oscillatory potential. For generic
initial conditions, the wavevectors $k$ in the forbidden bands
lead to an exponential growth of the mode functions with an
exponent, given by the imaginary part of the Floquet index, that
depends on the details of the potential. This is the phenomenon of
parametric amplification. The exponential growth of the mode
functions is associated with the build up of a non-equilibrium,
time dependent distribution function for the bosons, i.e, particle
production.

In the fermionic case, the Pauli exclusion principle restricts the
maximum occupation number of fermion states for a given wave
vector to be one. This is Pauli blocking, i.e, the quantum states
must have a finite occupation number. Obviously, even if the
fermionic fluctuation mode equations contain an oscillatory time
dependent mass, Pauli blocking should prevent any parametric
amplification and forbidden bands from occurring. This is somewhat
problematic, however. As our detailed numerical work shows, there
are cases in which the mean field is oscillatory, i.e, $\chi(t)
\sim \chi_0 \cos(\Omega t)$ and the fermion mode equations
(\ref{fmodeeqns}) take the form

\be \left[\frac{d^2}{dt^2}+p^2+\chi^2_0 \cos^2(\Omega t)\mp
i\Omega \chi_0 \sin(\Omega t) \right]f_{1,2 {\vec p}}(t) =0 \ee

At first glance it would appear that the oscillatory terms would
drive parametric amplification of the mode functions, which would
lead to an unbounded distribution function, contrary to Pauli
blocking. This is {\em not} the case, though. We now provide an
rigorous proof that fermionic mode functions {\em do not have
forbidden band structure} hence there is no parametric
amplification, and a perturbative argument that not only confirms
the exact proof but also illuminates the build-up of the
occupation number {\em in absence of parametric amplification}.

\subsection{No forbidden bands: formal proof}

The first order equations for the fermions (\ref{modof})
can be written in the following matrix form

\be \label{matrixform}  \left[
i\frac{d}{dt}-p\sigma_1-M(t)\sigma_3\right]f(t)=0 \ee

 \noindent with $\sigma_{1,3}$ the Pauli matrices and
 \be
 f(t)= \left(\begin{array}{c}
   f_1 \\
   f_2 \
 \end{array} \right).\ee

It is straightforward to check that \be f^{\dagger}(t)g(t) =
f^*_1(t)g_1(t)+f^*_2(t)g_2(t) = \mbox{constant} \label{consprob}\ee
\noindent for an arbitrary pair of solutions $ f(t) $ and
$ g(t) $ of the equation (\ref{matrixform}). This is a result of the
conservation of probability in the Dirac equation, or equivalently
the normalization condition on the Dirac spinors. Equation
(\ref{matrixform}) has two linearly independent spinor solutions
$h_1(t)~,~h_2(t)$ with initial conditions:

\be h_1(0)= \left(\begin{array}{c}
  1 \\
 0
\end{array} \right)~~;~~ h_2(0)= \left(\begin{array}{c}
  0 \\
1
\end{array} \right). \ee
Both solutions therefore obey

\be \label{const} h_1^{\dagger}(t)h_1(t)=1 ~~,~~
h_2^{\dagger}(t)h_2(t)=1 \ee

Furthermore it follows from the above identities that

\be h_1^{\dagger}(t)h_2(t) =0, \ee

\noindent i.e, the linearly independent solutions are orthogonal
at all times.

Since $M(t)$ is periodic with period T, i.e, $M(t+T)=M(t)$ then
$h_1(t+T)$ and $h_2(t+T)$ will also be solutions of
(\ref{matrixform}), but then they must be linear combinations of
the linearly independent solutions $h_1(t)~,~h_2(t)$:,

\bea
h_1(t+T) & = &  \alpha h_1(t) + \beta h_2(t) \nonumber \\
h_2(t+T) & = &  \gamma h_1(t) + \delta h_2(t) \nonumber \eea

\noindent where $\alpha,\beta,\gamma,\delta$ are complex
coefficients. The matrix of coefficients

\be M = \left( \begin{array}{cc}
  \alpha & \beta \\
  \gamma & \delta
\end{array}\right) \label{monodromy} \ee

\noindent is called the {\em monodromy} matrix and is the
important concept in Floquet theory. It represents an operator
that evolves the solution in time by one period.

The conditions

\be h^{\dagger}_1(t+T)h_1(t+T)=1~;~
h^{\dagger}_2(t+T)h_2(t+T)=1~;~h^{\dagger}_1(t+T)h_2(t+T)=0 \ee
\noindent obtained above, lead to the following conditions on the
coefficients

\be |\alpha|^2+|\beta|^2 =1 ~,~ |\gamma|^2+|\delta|^2 =1 ~,~
\alpha^*\gamma+\beta^*\delta =0, \ee

\noindent which in turn implies that the monodromy matrix is {\em
unitary}:

\be M^{\dagger}M =1. \ee

This unitarity property of the monodromy matrix can be traced back
to the conservation of probability of the solutions of the Dirac
equation (since it  is first order in time) as explicitly
determined by the condition (\ref{consprob}).

Floquet solutions are the eigenvectors of the monodromy matrix and
the (logarithm) of the eigenvalues are the Floquet exponents.
Floquet solutions therefore satisfy

\be F_{\pm}(t+T) = e^{i \varpi_{\pm}} F_{\pm}(t), \ee

\noindent where in general, the exponents $\varpi_{\pm}$ are
complex. Their imaginary parts determine the growth (or decay)
rate of the solutions and are responsible for parametric
amplification. However, the unitarity of the monodromy matrix
implies that the Floquet indices $\varpi_{\pm}$ are {\em real},
hence Floquet solutions develop a phase upon time evolution during
a period but their magnitude is constant. This then precludes
forbidden bands and hence there is no parametric amplification of
fermionic modes. Pauli blocking is a direct result of the Dirac
fields obeying a {\em first order} evolution equation in time.
This is also at the heart of the conservation of probability (or
normalization of the Dirac spinors), which in turn determines that
the monodromy matrix is unitary, hence Floquet indices are real.

\subsection{Perturbative argument:}

While the formal proof above unambiguously clarifies that there is
no parametric amplification of fermionic modes with an oscillatory
time dependent mass, we offer also a perturbative argument. We do
this both to highlight the main result of the exact proof above,
but also to illuminate why even when the Floquet indices are real
and there are no forbidden bands, time evolution with phases leads
to a build-up of the occupation number for some (resonant) values
of the wave vectors.

Consider the second order evolution equation for $f_{1,{\vec
p}}(t)$ (after rescaling by the scale ${\bar M}$)

\bea
&&\left[\frac{d^2}{dt^2}+p^2+M^2(t)+i\dot{M}(t)\right]f_{1,{\vec
p}}(t)=0 \nonumber \\
&& M(t)= \chi_0 \cos(\Omega t) \eea

A perturbative solution is obtained by considering $\chi_0 =
\epsilon <1$ and writing the formal expansion $f_{1,{\vec p}}(t)=
f^{(0)}_{1,{\vec p}}(t)+\epsilon f^{(1)}_{1,{\vec
p}}(t)+\epsilon^2 f^{(2)}_{1,{\vec p}}(t)+\cdots$ which leads to
the following hierarchy of equations

\begin{eqnarray}
\ddot{f}^{(0)}_{1,{\vec p}}(t)+ p^2 f^{(0)}_{1,{\vec p}}(t) & = & 0 \nonumber \\
\ddot{f}^{(1)}_{1,{\vec p}}(t)+ p^2 f^{(1)}_{1,{\vec p}}(t) & = &
-i\dot{M}f^{(0)}_{1,{\vec p}}(t) \nonumber \\
\ddot{f}^{(2)}_{1,{\vec p}}(t)+ p^2 f^{(2)}_{1,{\vec p}}(t) & = &
-i\dot{M}f^{(1)}_{1,{\vec p}}(t) -M^2(t) f^{(0)}(t) \nonumber \\
\quad \quad \vdots  \quad \quad & = & \quad \quad  \vdots \quad
\quad \label{hierarchy}
\end{eqnarray}

The zeroth order solution is of the form

\be f^{(0)}_{1,{\vec p}}(t) = A({\vec p}) e^{ipt}+B({\vec p})
e^{-ipt} \ee

\noindent with the coefficients $A({\vec p}),B({\vec p})$
determined by the initial conditions. The higher order
inhomogeneous equations are solved in terms of the retarded
Green's function

\be \label{Gret} G({\vec p},t-t') =
\frac{1}{p}\sin[p(t-t')]\Theta(t-t'). \ee

\noindent The general solution is given by

\be f^{(i)}_{1,{\vec p}}(t) = \int_0^{\infty} dt'G({\vec
p},t-t'){\cal I}^{(i)}(t') \ee

\noindent with ${\cal I}^{(i)}$ the inhomogeneity in the equation
for the i-th order contribution. We note that $f^{(i)}_{1,{\vec
p}}(0)=0;\dot{f}^{(i)}_{1,{\vec p}}(0)=0$, therefore the initial
conditions determine the constants $A({\vec p});B({\vec p})$ {\em
to all orders}.

The first order solution $f^{(1)}_{1,{\vec p}}(t)$ exhibits a
secular term, which grows linearly in time, when $\Omega = 2p$.
This obviously corresponds to the production of a fermion pair.
However this secular term is {\em purely imaginary}:
$f^{(1)}_{1,{\vec p}}(t) \propto i t \chi_0/p$. It corresponds to
a renormalization of the {\em phase} of the order solution. The
calculation to second order is lengthy but straightforward, with
the following remarkable result: purely imaginary secular terms
appear once again which grow in time, leading to a further
contribution to the phases. The {\em real} part of the secular
terms cancels out between the $M^2$ contribution (which involves
the zeroth order solution ) and the contribution from $i\dot{M}$
which appears {\em squared} thus always (in both cases for
$f_{1,2}$) with opposite sign as the $M^2(t)$ contribution.
Therefore, at least to second order, we find that the secular
terms generated from resonances when $\Omega \sim 2p$ are purely
imaginary, while the real part of the secular terms cancel
exactly. Thus there is {\em no exponential growth of the solution}
which would emerge from real secular terms in the perturbative
expansion. We have not attempted a higher order calculation, but
in view of the exact proof offered above, it is clear that the
result holds to all orders. The lowest order perturbative solution
also shows that $f_1$ and $f_2$ develop {\em opposite} phases so
that the occupation number given by eqn. (\ref{fermionnumber})
becomes of the form

\be n_q(\tau) \propto \sin(\varpi \tau) \ee

\noindent with $\varpi$ the real Floquet index. This explains the
behavior of the occupation number displayed in the figures above,
the oscillations and the saturation. Hence unlike the bosonic case
in which parametric amplification results in an exponential growth
of bosonic fluctuations, Pauli blocking at the level of the mode
functions is a consequence of real Floquet indices that result in
a bounded growth of the occupation number.

\section{\bigskip Conclusions, conjectures, implications and further questions}

We have used an expansion in $1/N$, where $N$ is the number of
fermion fields coupled to a scalar to understand the
non-perturbative dynamics of this system. In this limit, the
scalar fluctuations are suppressed relative to the fermionic ones
and we can consider mean field dynamics in a consistent
approximation that, at least in principle, can be improved on.

In the large $N$ limit, the theory exhibits dimensional
transmutation and symmetry breaking via the Coleman-Weinberg
mechanism with an effective potential that is unbounded below and
features a metastable minimum at the origin and two symmetric
maxima. The bulk of our work consisted in obtaining the equation
of motion for the expectation value of the scalar field, or mean
field and comparing the dynamics due to the \emph{full} equations
of motion, including the backreaction of the fermionic
fluctuations to that obtained from a renormalization group
improved effective potential.

The equation of motion obtained from the RG effective potential
\emph{alone} contains a Landau pole at a non-perturbative scale
$E_{L} \propto e^{2/g-1}$ where $g=y^2/2\pi^2$ and $y$ is the Yukawa
coupling. We should note that since we are working within the
large $N$ expansion, this Landau pole cannot be considered to be
an artifact of the perturbative expansion. The dynamics obtained
from the effective potential alone would yield singular behavior
when the amplitude of the mean field reaches the Landau pole.

The renormalization of the full equations of motion leads to a
running coupling constant that depends on \emph{time} through the
dependence on the amplitude of the mean field. This phenomenon is
akin to the dynamical renormalization found in
reference\cite{drg}. Furthermore we show that potential initial
singularities are self-consistently removed and that the dynamics
is smooth and free of Landau pole singularities, or discontinuities.

The fully renormalized equations of motion including the
backreaction of the fermionic fluctuations has two different
regimes depending on whether the  ultraviolet cutoff $\Lambda \ll
E_{L}$ or $\Lambda \gg E_{L}$. In the former case, if the initial
value of the mean field is between the origin and the maxima, the
mean field undergoes damped oscillations transferring energy to
the fermionic fluctuations. This energy transfer results in
fermion pair production within a band of wavevectors determined by
the mass of the scalar and the initial amplitude of the mean
field. It is found that the backreaction from the fermionic
fluctuations, encoded in the renormalized fermion chiral
condensate introduce a slowly varying renormalization of the
parameters of the effective potential.

When the absolute value of the initial value of the mean field is
larger than the maximum of the effective potential, both the mean
field and the fermion chiral condensate run away to the cutoff
scale, at which point the evolution must be stopped.

While it is commonly acknowledged that a theory that features a
Landau pole only makes physical sense when the cutoff is
\emph{below} the scale of the Landau pole, we decided to study the
dynamics even in the case in which $\Lambda \gg E_{L}$ to find out
if and how there are any singularities in the non-equilibrium
dynamics associated with the Landau pole. Typical statements on
the limit of validity of a theory with the Landau pole are mostly
based  on perturbative unitarity in S-matrix elements.

Our study is definitely non-perturbative as we study the evolution
for large amplitude mean field configurations. Whether
perturbative statements on S-matrix elements provide a limitation
to the non-perturbative approach is, to our knowledge, an open
question. We therefore proceed to study the theory in this regime
and to analyze the consequences without bias.

In this case we find  novel, remarkable behavior. When the
amplitude of the mean field reaches the Landau pole there is a
cancellation between the contribution from the effective potential
and that from the renormalized chiral condensate that prevents a
singularity in the dynamics at the Landau pole from developing.
The ensuing evolution is continuous. As the amplitude of the mean
field becomes larger than the Landau pole, it begins to
\emph{oscillate} with an amplitude that grows.  We find that the time
evolution of the mean field is well described by an effective
potential that is \emph{quartic and upright}.

We have seen that the fermion occupation number almost saturates
up to a wave vector of the order of the amplitude of the mean
field.  The resulting state is a very dense medium or plasm which
can be described as a cold, degenerate fermi gas with a 
chemical potential of the order of the energy scale of the
mean field.

We have also provided an exact proof of the fact that the Floquet
indices for the fermionic mode functions are \emph{real},
therefore preventing unbounded parametric amplification. This is the
manifestation of Pauli blocking at the level of the mode
functions. A perturbative proof of this result served to
illuminate the workings of the exact proof.

\bigskip

\textbf{Conjectures:} We conjecture, at this stage, that the
phenomenon of smooth dynamics beyond the Landau pole is somewhat
related to a recent observation that in a medium the actual
position of the Landau pole is actually shifted to much larger
scales\cite{hectorLP}.

While we have studied the theory in a regime in which perturbative
unitarity of S-matrix elements would suggest that the theory
breaks down, we have done so \emph{non-perturbatively} at the
mean-field level. It is conceivable, that while the dynamics is
smooth and there are no signals of pathologies, incompatibilities
may lurk in some (non-perturbative) matrix elements or physical
quantities that would cast doubt on the validity of the theory. To
assess this possibility requires a deeper study of correlation
functions beyond mean field and perhaps to next to leading order
in the large $N$. This is beyond the goal of this article and an
issue that deserves further scrutiny in its own right.

There has previously been a related conjecture that Landau poles
in some theories only appear in perturbative calculations or in
calculations of unobservables\cite{tarrach}.

\bigskip

\textbf{Implications:} There are many important cosmological
implications of our results. While previous studies of reheating
and preheating of fermionic theories have argued that fermions are
not very efficient for reheating because of Pauli blocking, in
this work we have seen that in the regime when the cutoff is much
larger than the Landau pole there is the possibility of abundant
particle production. After the mean field crosses the Landau pole,
the fermionic occupation numbers basically saturate up to a
momentum scale of the order of the Landau pole. This in turn
describes a dense medium with  a large chemical potential  of the
order of the Landau pole.

This could potentially be relevant in an extension of the Standard
Model with top-like quark sector with a Yukawa coupling of order
one. Since this theory will be part of an even larger theory
(perhaps Grand Unified) the cutoff should be taken to be of  the
order the GUT scale. On the other hand with a vacuum expectation
value of few hundred GeV's and a Yukawa coupling of order one, the
position of the Landau pole is at a scale of a few TeV's. Thus the
situation $\Lambda \gg E_{L}$ would naturally arise in these type
of scenarios.

Another potentially interesting implication in cosmology would be
the following: consider that after a phase transition the system
is trapped in the metastable vacuum. A bubble type configuration,
characterized by a collective
coordinate\cite{hectorinstantons,erickbubbles,cole} will tunnel
underneath the barrier of the effective
potential\cite{erickbubbles} and exit on the other side of the
maximum. The collective coordinate will then begin to roll down
the potential hill until it reaches the Landau pole at which point
the novel dynamics studied in this article will lead to fermion
pair production and the production of a dense medium with a large
chemical potential. We believe that these potential implications
deserve further study and we expect to address these in future
work.

\bigskip

\textbf{Further questions:} An important aspect that we have not
investigated is that of the new time scales that emerge from the
dynamics. For example when the mean field oscillates with small
amplitude in the first cases analyzed, there is a damping
associated with the energy exchange with the fermion fluctuations.
The time scale for damping is much larger than the typical
oscillation frequency, despite the fact that the coupling is
fairly strong. Another time scale is that associated with the
\emph{growth} of the amplitude of oscillations after the mean
field crosses the Landau pole, until the amplitude saturates.

We expect to apply the methods developed in\cite{drg,whit}to study
these questions and relegate this investigation to future work.

\bigskip

{\bf Acknowledgements} R.H. would like to thank Ira Rothstein for
valuable discussions. D.B. was supported in part by the NSF grants
 PHY-9988720 and  NSF-INT-9815064.
H. J. d. V. thanks the CNRS-NSF collaboration for support. R.H.
and \ M.R.M. were supported in part by the DOE grant number
DE-FG03-91-ER40682\ .

\end{document}